\documentclass[a4paper,11pt]{article}
\usepackage{jheppub} 
\usepackage{lineno}
\usepackage{slashed}
\usepackage{bbm}
\usepackage{tikz} 
 \usepackage{tikz-cd}
\usepackage{spectralsequences}
\usepackage{amsmath}
\usepackage{amssymb}
\usepackage{mathtools}
\usepackage{graphicx}

\DeclareSseqGroup \tower {} {
\class(0,0)
\DoUntilOutOfBounds {
\class(\lastx,\lasty+1)
\structline
}
}

\newcommand\tr{\mathrm{tr\,}}
\mathchardef\mhyphen="2D

\newcommand{\cA}{\mathcal {A}}

\newcommand*{\bC}{\ensuremath\mathbb{C}}

\newcommand*{\bR}{\ensuremath\mathbb{R}}

\newcommand*{\bZ}{\ensuremath\mathbb{Z}}

\newcommand*{\cL}{\ensuremath\mathcal{L}}
\newcommand*{\cM}{\ensuremath\mathcal{M}}
\newcommand*{\cN}{\ensuremath\mathcal{N}}

\newcommand*{\cR}{\ensuremath\mathcal{R}}

\newcommand{\ZZ}{\mathbb {Z}}

\newcommand{\tX}{\tilde X}

\newcommand{\tB}{\tilde B}

\newcommand{\tR}{\tilde R}
\newcommand{\tF}{\tilde F} 

\newcommand{\tA}{\tilde A}
\newcommand{\ta}{\tilde a}

\newcommand{\tome}{\tilde \omega}

\newcommand\beq{\begin{equation}}
\newcommand\eeq{\end{equation}}
\newcommand\bea{\begin{eqnarray}}
\newcommand\eea{\end{eqnarray}}
\newcommand\nn{\nonumber}
\newcommand{\vpc}{\varphi_{S^1}}
\def\dd{{\rm d}}

\def\a{\alpha}
\def\b{\beta}

\def\d{\delta}

\def\e{\epsilon}

\def\g{\gamma}
\def\G{\Gamma}

\def\m{\mu}
\def\n{\nu}
\def\o{\omega}

\def\p{\psi}

\def\S{\Sigma}

\newcommand{\cT}{\mathcal{T}}
\newcommand{\cE}{\mathcal{E}}

\newcommand{\cI}{\mathcal{I}}

\def\cL{\mathcal{L}}

\def\cA{{{\mathcal A}}}

\def\cM{\mathcal{M}} 
\def\cN{\mathcal{N}}

\def\6d{6\text{D}}
\def\5d{5\text{D}}
\def\7d{7\text{D}}

\title{\boldmath Integral cubic form of 5D minimal supergravities and non-perturbative anomalies in 6D (1,0) theories}

\author[a]{Peng Cheng,}
\author[b]{Michael N. Milam,}
\author[b]{Ruben Minasian}

\affiliation[a]{Arnold Sommerfeld Center, LMU München, Theresienstra{\ss}e 37, 80333 München, Germany}
\affiliation[b]{Institut de Physique Théorique, Université Paris-Saclay, CNRS, CEA, F-91191, Gif-sur-Yvette, France}

\abstract{A five-dimensional minimal supergravity theory coupled to vector and hypermultiplets is specified by a set of trilinear couplings, given by an intersection form $C_{IJK}$, and gravitational couplings specified by an integer-valued vector $a_I$ and is consistent when these data define an integral cubic form. For every Calabi-Yau threefold reduction of M-theory, this condition is satisfied automatically. Via suitable redefinitions of the basis of 5D vectors, this is also shown to be the case for the circle reductions of six-dimensional anomaly-free (1,0) theories. When the 6D theory has a $\mathbb{Z}_k$ gauge symmetry, we point out that the consistency of the circle reduction with nontrivial $\mathbb{Z}_k$ holonomy is closely related to 6D constraints derived by Monnier and Moore. These constraints are extended to semidirect products with continuous gauge groups $\mathbb{Z}_k \ltimes G$ and CHL-like circle compactifications. When $\mathbb{Z}_k$ acts on anti-self-dual tensor fields of 6D supergravity there should be a nontrivial action of holonomy on the topological Green-Schwarz terms.}

\begin{document}
\maketitle
\flushbottom

\section{Introduction}
\label{intro}
Since the dynamics of any quantum theory depends crucially on its spacetime dimension, it is often instructive to compare different theories related to each other via dimensional reduction/decompactification. Tracking how the salient and well-known aspects of dynamics manifest themselves in an adjacent dimension, where a priori very different quantum features are expected, can be illuminating.

One such example appears in the study of RG flows in quantum field theories in 2D/4D vs. 3D and their respective UV/IR fixed points:
\begin{itemize}
    \item In even dimensions (D $ = 2,4$) there exists a quantity $a$ which decreases along RG flow
    \begin{equation}
        a_{\text{UV}} > a_{\text{IR}},
    \end{equation}
    and can be computed by a local operator correlation function on the two- and four-dimensional sphere \cite{Zamolodchikov:1986gt,Komargodski:2011vj}:
    \begin{equation}
        \big\langle T_{2\text{D}/4\text{D}} \big\rangle  = a_{2\text{D}/4\text{D}}f(R) + \dots ,
    \end{equation}
    where $T$ is the trace of the energy-momentum tensor.
    \item In $\text{D}=3$, there is much evidence of the existence of a similar quantity \cite{Pufu:2016zxm}, given by the real part of the free energy on $S^3$, $F_{3\text{D}} = - \log |Z_{3\text{D}}|$:
    \begin{equation}
        F_{\text{UV}} > F_{\text{IR}}.
    \end{equation}
    \end{itemize}
    The key difference between even and odd dimensions is that unlike $a_{2\text{D}/4\text{D}}$ , $F_{3\text{D}}$ is a {\sl non-local} quantity. Finding relations between 3D quantities and 2D/4D ones in theories related by a circle compactification would be of interest. 
    
Perturbative anomalies  of continuous gauge or diffeomorphism symmetries (see e.g. \cite{Harvey:2005it}) provide another well-known example of dynamical differences in even- or odd-dimensional quantum field theories (QFT) or theories of quantum gravity (QG).\footnote{In this paper we mainly focus on supergravity examples, which are believed not to have any global symmetries at the quantum level.}
\begin{itemize}
 \item In even dimensions, dynamical chiral fields may break gauge/diffeomorphism symmetries at one-loop. In general, these anomalies cannot be removed by adding local counterterms. Hence, anomaly cancellation imposes strong conditions on the spectrum of an even dimensional QFT or QG.
    \item In odd dimensions, the non-invariance under continuous gauge/diffeomorphism transformations connected to identity can come only from local terms in the theory. The theory can be made well-defined by subtracting these ``bad" local  counterterms.
\end{itemize}
The above features of anomalies raise an interesting question and serve as the motivation for this paper: Consider an odd-dimensional QFT and QG that arise from an $S^1$ compactification of an even-dimensional QFT and QG:
\begin{equation}
    \label{schematical map of circle compactification}
    \varphi_{S^1}: \,\, \text{ Space of $\text{D} = 2n$ QFT and QG } \, \longrightarrow \, \text{ Space of $\text{D}= 2n-1$ QFT and QG }.
    \end{equation}
Given that perturbative anomaly cancellation imposes strong conditions on the space of even-dimensional QFT and QG, which apparently become void in odd dimensions, how does the reduction $\varphi_{S^1}$ work at the quantum level, and when is it invertible? How do perturbative anomalies of an even-dimensional theory manifest themselves upon $S^1$ compactification $\varphi_{S^1}$? 

As pointed out in \cite{Witten:1982fp},  when perturbative anomalies vanish the non-perturbative anomalies can give rise to further consistency conditions. A unified framework for the discussion of both perturbative and non-perturbative anomalies, via use of the $\eta$ invariant and the Atiyah-Patodi-Singer (APS) index theorem, is given by the Dai-Freed anomaly \cite{Dai_1994, Freed:2004yc, Garcia-Etxebarria:2018ajm, Monnier:2019ytc}. In this context the above question can be generalized:  What happens to anomalies (both perturbative and non-perturbative) of even dimensional theories after $S^1$ compactification $\varphi_{S^1}$?

In this paper, we study this question mainly in the context of 6D and 5D supergravities with 8 supercharges.  The anomaly cancellation in 6D $\mathcal{N} = (1,0)$ supergravity theories has a long history starting with \cite{Green:1984bx, Sagnotti_1992}. The non-perturbative $\bZ_k$ gauge anomalies in these theories were studied in \cite{Monnier:2018nfs}. This still leaves out cases where $\bZ_k$ symmetry acts on a continuous gauge group $G$ or on tensor multiplets of the $(1,0)$ theory, which we will discuss in Section \ref{sec:ano-NP}.\footnote{For a very partial list of works discussing aspects of anomalies in 6D $\mathcal{N} = (1,0)$ theories see
\cite{Erler_1994, Duff:1996rs, Berkooz:1996iz,
Kumar:2010ru,  Seiberg:2011dr, Park:2011ji, Monnier:2017oqd, Lee:2020ewl, Davighi:2020kok,
Dierigl:2022zll, 
Dierigl:2025rfn, Lockhart:2025lea}. For related global considerations in other dimensions, see \cite{ Dierigl:2022reg, Debray:2023rlx, Basile:2023knk}.}
 The results of the analysis of the image of these anomalies after $S^1$ compactification, i.e. the map $\varphi_{S^1}$ \eqref{schematical map of circle compactification},  can be summarized as:
 \begin{itemize}
     \item The perturbative anomaly in 6D will become gauge and diffeomorphism non-invariant local counterterms after $S^1$ compactification. These are generated by (KK towers of) charged fields at one-loop. When the parent 6D theory admits anomaly cancellation by the Green-Schwarz (GS) mechanism, the one-loop counterterms generated upon $S^1$ reduction cancel exactly the parts of the reduction of the 6D GS terms which are non-invariant in 5D.\footnote{This is not really something new, as this problem has been investigated in \cite{Corvilain:2017luj,Corvilain:2020tfb, Cheng:2021zjh}. Here we unify and streamline previous discussions.}
     \item When perturbative anomalies vanish, the non-perturbative anomalies have a nice topological description and are given by a cobordism invariant. In this case, the $\bZ_k$ non-perturbative anomaly again appears as a counterterm under the map $\vpc$. In this case, however, the anomaly becomes topological WZW-like terms, and the 6D anomaly-free condition translates into a quantization condition for these topological terms.   
 \end{itemize}
\noindent
Independently from the question about the fate of anomalies in the circle-reduced theories, one may wonder if there are intrinsically five-dimensional consistency conditions in QG. If so, how are the 6D and 5D conditions related?

5D pure supergravity with minimal supersymmetry has many similarities with the 11D theory. The bosonic content of the supergravity multiplet is given by  the graviton and a single field $A_{(\text{D}-2)/3}$ with a $\frac{(\text{D}+1)}{3}$-form field strength $F$. The Chern-Simons coupling, cubic in this field, makes the theory very interesting. The $1/3!$ coefficient in front of the CS terms renders the classical partition function ill-defined and necessitates a gravitational CS coupling $A_{(\text{D}-2)/3} \wedge X_{(2\text{D}+2)/3}(R)$, where $X(R)$ is a polynomial in the curvatures of appropriate degree (8 and 4 respectively).\footnote{Differently from 5D, in $\text{D}=11$ there is a $\bZ_2$ anomaly affecting the sign of the fermion determinant, due to the reality condition of the 11D gravitino.} This coupling is also at the origin of the chirality of the BPS objects charged under $A_{(\text{D}-2)/3}$ -- strings and M5-branes, respectively.  Hence the gravitational CS coupling is also required for cancellation of the chiral anomaly on the worldvolume of the BPS objects via inflow (see e.g. \cite{Freed:1998tg} for a discussion of both M5-branes and strings in 5D). 

While the quantum consistency of M-theory is both desirable and credible, the consistency of the minimal 5D pure supergravity is often questioned. This is  essentially due to the fact that it cannot arise from a Calabi-Yau (CY) compactification of M-theory or a circle reduction of a 6D anomaly-free theory, since either of these compactifications comes with some matter multiplets coupled to supergravity. Notably M-theory on CY3 will have at least one hypermultiplet. In general, less attention has been given to direct studies of the consistency conditions of 5D minimally supersymmetric gravity, in spite of their similarity to those in M-theory and their relative simplicity (for studies of the consistency of BPS objects, see e.g. \cite{Katz:2020ewz, Cheng:2021zjh}). 

In this paper, we take the 5D spacetime $M_5$ to be a spin manifold. This in general can be relaxed since M-theory is well-defined on a Pin$^+$ manifold $M_{11}$ \cite{Witten:2016cio, Freed:2019sco} with shifted quantization for $G_4$ fluxes. We will not address these cases here.

A 5D minimal supergravity theory is specified by the number of multiplets (vector and hyper) and by the Chern-Simons couplings at a generic point on the Coulomb branch 
\bea \label{eq:act1}
 \int_{M_5} \cL \supset  \int_{M_5} -\frac{1}{6}C_{I J K } A^{I} \wedge F^{J} \wedge F^{K}   - \frac{a_{I} }{48}A^{I} \wedge p_1(M_5), 
 \eea
 where $p_1(M_5)$ is the first Pontrjagin class of the bulk tangent bundle.
Assuming that the $F^I$ are integrally quantized ($F^I \in H^2(M_5, \bZ)$), the partition function of the theory is well-defined, provided that the triple-intersection form $C_{I J K }$ and vector ${a_{I} }$ satisfy
$$
\frac16 \Bigl[ C_{I J K} Q^{I}  Q^{J}  Q^{K}  +  \frac{a_{I}}{2} Q^{I} \Bigl] \, \in  \bZ \quad  \forall  \,Q^{I} \in \bZ {\qquad  \leftrightarrow \qquad  } 
\left\{ \begin{array}{l}   
C_{I J K } \in \bZ \qquad  \qquad \quad \forall \, I ,  J ,  K \\
C_{I I I} +  \frac{a_{I}}{2}  \in 6 \bZ  \qquad  \quad \forall I \\ 
 C_{I I J} + C_{I J J} \in 2 \bZ \qquad \forall \, I \neq J
\end{array}  \right.,
$$
defining the integral cubic form of the theory.

Two large overlapping sets of theories that are expected to satisfy these conditions are reductions of M-theory on CY manifolds and the simple circle reductions of 6D anomaly free $\mathcal{N} = (1,0)$ theories.

In the case of reductions of M-theory on a CY three-fold $X_3$, the vectors $A^{I}$ originate from the M-theory three-form, and the index $I$ runs over $H^{1,1}(X_3)$. The cubic form is then directly expressed in terms of topological data of $X_3$, and the divisibility conditions are easily verified. This is not new, but for completeness we give the proofs in subsection \ref{subsec:CY}.

Verifying the integrality of the 5D cubic form arising in the reduction of 6D theories, in spite of the latter having been studied quite extensively, is less straightforward. While a simple circle reduction is indeed expected to yield a consistent 5D theory, a  redefinition of the basis of 5D vectors is required. This redefinition has been studied in the cases when the respective 6D/5D theories are obtained from F/M-theory reductions,\footnote{A large, if partial, set of relevant examples in M/F-theory context see \cite{Morrison:2012ei, Braun:2014oya, Morrison:2014era, Anderson:2014yva, Klevers:2014bqa,Arras:2016evy, Anderson:2018heq,Raghuram:2018hjn,Anderson:2019kmx,Raghuram:2020vxm,Anderson:2023wkr,  Duque:2025kaa}. For an introduction to F theory, see \cite{Taylor:2011wt, Weigand:2018rez}.} i.e there is an underlying CY threefold (UV). The IR understanding of this redefinition in a generic supergravity theory (provided it exists) is not known. What we find is that the subtleties of the 6D GS couplings involving  (anti-)self-dual fields translate into much easier integrality conditions on the cubic form in a suitable basis. When the 6D theory has a $\bZ_k$ gauge symmetry, we show that the consistency of the reduction with holonomy {$\bZ_k$} allows to recover the 6D constraints of \cite{Monnier:2018nfs}. The consistency conditions on all possible $\bZ_k$ anomalies spelled out in \cite{Monnier:2018nfs} are satisfied in general F-theory compactifications.  These constraints are extended to semidirect products with continuous gauge groups $\bZ_k \ltimes G$ and CHL-like circle compactifications. When $\bZ_k$ acts on anti-self-dual tensor fields of 6D supergravity, the 5D conditions  will not be satisfied unless a nontrivial action of holonomy on the topological Green-Schwarz terms is accounted for. 

\vspace{.5cm}

\noindent
The paper is organized as follows. In Section \ref{sec:cubic} we discuss the quantization condition of $C_{IJK}$ and $a_I$ in \eqref{eq:act1} and show that the 5D supergravity theories from M-theory on a CY3 always satisfy these conditions. In Section \ref{sec:AT} we give a brief review of Dai-Freed anomaly theory, which takes both perturbative and non-perturbative anomalies into account. In Section \ref{sec:pert_an}, we discuss perturbative anomalies and how gauge and diffeomorphism non-invariant local terms get generated after $S^1$ compactification by one-loop calculation. We also use the adiabatic eta invariant to rederive the one loop 5D KK $U(1)$ CS term in \cite{Bonetti:2013ela}.  In Section \ref{sec:ano-NP}, we propose that non-perturbative $\bZ_k$ anomalies are mapped to quantization conditions of topological terms under $\vpc$. Assuming M/F-theory duality, we verify that the $\bZ_k$ anomaly conditions in \cite{Monnier:2018nfs} with topological Green-Schwarz terms are satisfied by a general F-theory compactification. We also extend these conditions to $\bZ_k\ltimes G$ gauge symmetry and discuss the $\bZ_k$ action on tensor multiplets. We conclude with a brief summary and a discussion of some open problems in Section \ref{sect: discussion}. 
 In Appendix \ref{app:reduce} we present the explicit forms for the dimensional reduction of many of the formal objects used in the main text, and in Appendix \ref{app:FDC} we perform a 5D loop computation for non-Abelian fields and verify the form of locally non-gauge invariant CS-terms. In Appendix \ref{Some bordism group calculations}, we collect some bordism calculations that are used in this paper. Appendix \ref{app:W} discusses the 3D reduction of the SU(2) global (Witten) anomaly.

\paragraph{Notational note} Notations introduced in the main text will be defined in due places. We collect here the main ones:
\begin{itemize}
    \item We follow standard 5D conventions \eqref{eq:act1}, and hence all the field strengths  have been divided by $2\pi$. For symmetry we do the same with the curvatures and gravitational couplings (e.g. in our notation  $ p_1(TM_5) = -\frac{1}{2}\tr R^2)$.
    \item All 6D quantities (other than constants) are tilded, while the corresponding 5D ones are without tildes.
    \item The numbers of 6D $\mathcal{N} = (1,0)$ multiplets are denoted by $T$ (the 2-form fields will be denoted as $B_2^{\alpha}$ with $\a = 1, ..., T+1$; here in addition to $T$ (anti-self-dual) tensor fields we included  the (self-dual) tensor field in the gravity multiplet), $V$ (with Abelian vectors $A^i$ with $i=1,..., V$), and $H$ for hypermultiplets. The self-dual tensor field in the gravity multiplet is denoted by $B_2^1$. The numbers of 5D vector and hypermultiplets are denoted by $n_V$  and $n_H$ respectively. When a 5D theory comes from a simple (untwisted) circle reduction of a 6D theory, $n_V = V + T + 1$. The KK vector field will be denoted by $A^0$. In the 5D basis, the Abelian vectors are denoted as  $A^I$ with $I=0,1,...,n_V$. 
    \item The trilinear and gravitational CS couplings in 5D, $C_{IJK}$ and $a_I$ respectively, are adapted to an integral basis of 5D vector fields $F^I \in H^2(M_5, \bZ)$. Analogous quantities computed from 6D reductions (with or without nontrivial $\bZ_k$ holonomy) are denoted by $c_{IJK}$ and $a^{\6d}_I$ respectively.
    \item The 5D local gauge and gravitational variations are denoted by $\delta_{\epsilon}$, while the large gauge transformations by $\delta_{\text{n}}$. Quantities in Fraktur font $\mathfrak{X}_2$ (see \eqref{eq:defX2}) and $\mathfrak{D}_6$ are invariant under  $\delta_{\epsilon}$ but not under $\delta_{\text{n}}$.
\end{itemize}


\section{Cubic form in 5D minimal supergravities}
\label{sec:cubic}
There are three different multiplets with eight supercharges in five dimensions -- the gravity multiplet, comprised of the graviton, a pair of gravitini, and a vector field, $n_V$ vector multiplets, each made up of a vector, a pair of spin-$\frac{1}{2}$ fermions, and a real scalar, and $n_H$ hypermultiplets, each made of four real scalars and a pair of spin-$\frac{1}{2}$ fermions. The classical theory is specified completely, in addition to the matter content, by an intersection matrix governing the Chern-Simons-like trilinear couplings of the theory $C_{IJK}$, where the indices run over all vectors including the one in the gravity multiplet ($I = 0,1, ..., n_V$). Even if the intersection matrix is integer-valued, the partition function of the theory can be ill-defined due to additional quantization conditions. This is not surprising, since the 5D minimal supergravity resembles the 11D one, where gravitational couplings are required in order to have a well-defined partition function. 

In fact, the M-theory construction of \cite{Witten:2016cio, Freed:2019sco} proving that the M-theory partition function is well-defined uses the notion of cubic refinement on a 12-manifold, which generalizes the most natural construction that applies to a 6-manifold $M_6$. We therefore look directly at the 6D case now. 

Given a finitely generated Abelian group $L$ with a trilinear form $\langle \cdot, \cdot, \cdot \rangle \, \rightarrow \, \ZZ$ \cite{Freed:2019sco}, postulate the existence of $\bar c \in L \otimes \ZZ_2$   such that
\beq
\langle {\bar c}, {\bar x}, {\bar x} \rangle = \langle {\bar x}, {\bar x}, {\bar y} \rangle + \langle {\bar x}, {\bar y}, {\bar y} \rangle \qquad (\mbox{mod} \,\, 2), \qquad \qquad  {\bar x}, {\bar y} \in L \otimes \ZZ_2.
\eeq
An element $c$ in $L$  such that  $c = {\bar c} \ \, (\mbox{mod} \,\, 2) $ is called characteristic, and $L_{\text{char}} \subset L$  is the set of all characteristic elements. On $M_6$ a function $\kappa_3(c)$ of characteristic class $c$ given by
\beq
\label{eq:kappa3}
\kappa_3(c) = \frac{c^3 - p_1(M_6) c }{48} \, ,
\eeq
(evaluated on $M_6$) is an integer independent of $c$. On $M_6$, with $L = H^2(M_6, \ZZ)/\mbox{torsion}$, ${\bar c} = w_2(M_6)$  and $\langle x,y,z \rangle = \langle x \cup y \cup z, [M_6] \rangle$. This is just a Dirac index 
$$ \mbox{ind}(D_{M_6}) = \int_{M_6} \widehat{A}(M_6)  e^{c/2}  \, .$$

An integer-valued cubic refinement of the symmetric trilinear form on  $L$  for a fixed characteristic class $c$ is given by a map $ L \, \rightarrow \ZZ$:
\beq \label{eq:cubic}
q_3^c(x) = \kappa_3(c + 2x) - \kappa_3(c) = \frac16 x^3 - \frac{1}{24} x p_1(M_6) + \cdots, \qquad \ \ x\in  H^2(M_6, \ZZ).
\eeq
Consider a 5D minimal supergravity theory with a single vector field (without fixing $n_H$) on $M_5$ with a bulk extension given by $M_6, \partial M_6 = M_5$, and add a gravitational coupling allowed by supersymmetry:
\beq\label{eq:pure5d+}
\cL_{5\text{D}}  \supset - \frac{C_{000}}{6} A \wedge F \wedge F - \frac{a_0}{48} A \wedge p_1(M_5) \, .
\eeq
Assuming that $F$ is a curvature of a $U(1)$ connection and is properly quantized (which is automatically true for CY compactifications), $F \in H^2(M_5, \bZ)$, the partition function of the theory will be well-defined, provided that \eqref{eq:pure5d+} is given by an integer-valued cubic refinement.

In parallel, one can consider BPS string sources given by  $\dd F = Q \delta_3 (\S)$, where $\delta_3 (\S)$ is the Poincaré dual to the string worldsheet $\S$, which should support a supersymmetric $\mathcal{N} = (0,4)$  theory. The central charges then are given by 
\bea
c_R = C_{000} Q^3 + \frac{a_0}{2}  Q \qquad \mbox{and} \qquad c_L = C_{000} Q^3 + a_0 Q.
\eea
The divisibility by 6 of $c_R$ for any value of $Q$, as required by (0,4) supersymmetry, is equivalent to \eqref{eq:act1} defining an integral cubic form and will be assured provided
\bea \label{eq:k0kr}
a_0 = 2 C_{000} (-1 + 6n ) \, , \qquad C_{000}, n\in \bZ.
\eea
Note that, differently from M5 anomalies,  neither the requirement of having a well-defined partition function nor the divisibility of $c_R$ by 6 fix the coefficient completely -- a priori it would be allowed to add extra integer-valued terms $\frac{n}{8} \int_{M_6} F \wedge \tr R^2 $.\footnote{For a 6D spin manifold $M_6$, $\frac{1}{2}p_1(M_6)= v_4(M_6)$ mod $2$ and the 4th Wu class $v_4(M_6) = 0$. Hence $\frac{1}{4}p_1(M_6) = -\frac{1}{8}\tr(R^2) \in H^4(M_6;\bZ)$ and $\int_{M_6} \frac{1}{8}F \wedge \tr R^2 \in \bZ$. }

The compactification of M-theory on a quintic (with $n_V=0$ and $n_H = 102$) has $C_{000} = 5$ and $a_0 = 50$ and is a special case of \eqref{eq:k0kr} with $n=1$.

In pure supergravity without matter, $C_{000}=1$ \cite{Cremmer:1980gs}, and one would expect $c_R = Q^3 - Q + 6n Q $ and $c_L = Q^3 +12nQ - 2Q$, i.e. $(c_R,c_L) = (6,11)$ for $n=1$ and $Q=1$. A $Q=1$ string worldsheet is expected  to have a single $\mathcal{N} = (0,4)$ multiplet possibly coupled to a free CFT. Note  that the string center of mass would have central charges $(c_R, c_L) = (6,3)$. Before any discussion about potential implications, however, we should pause and remark that, differently from CY compactifications, here we do not have an argument about integer quantization of $F$. It would be interesting to return to this question.

\subsection{Consistency conditions on CS coefficients of  minimal 5D supergravities}
\label{subsec:CY}
A generic supergravity theory with $n_V$ vector and $n_H$ hypermultiplets will have a CS sector governed by an intersection matrix $C_{IJK}$ (with $I,J,K = 0,1,...,n_V$), which determines the trilinear couplings of the vector fields, and a vector $a_I$, determining the gravitational CS terms. The relevant part of the action is given by
\bea \label{eq:5dL}
 \int_{M_5} \cL &&\supset  \int_{M_5} -\frac{1}{6}C_{I J K } A^{I} \wedge F^{J} \wedge F^{K}   + \frac{a_{I} }{96}A^{I} \wedge \tr R^2 \nn \\
&& = \int_{M_6} - \frac{1}{3!} C_{I J K } F^{I} \wedge F^{J} \wedge F^{K}   - \frac{a_{I} }{48}F^{I} \wedge p_1(M_6),
\eea
where $M_6$ is a manifold bounded by $M_5$: $\partial M_6 = M_5$, and we assume that both manifolds are spin. 

$F^{I} = \dd A^{I}$  determines the first Chern class of line bundles corresponding to $U(1)$ gauge groups of a 5D theory at a generic point of the Coulomb branch.\footnote{$\Omega^{\text{Spin}}_{5\text{D}}(BU(1)) = 0$ so the extension is not obstructed.} Once more, all  explicit factors of $2 \pi$ are suppressed -- each curvature $F$ or $R$ comes with $\frac{1}{2\pi}$.

The partition function  $Z_{5\text{D}} \propto \mbox{Det} (\mbox{\text Fermi}) \exp \{i\int_{M_5} \cL\}$, and in order for it  to be well-defined and notably independent from the bulk extension $M_6$, it is required that 
\beq \label{eq:PF}
\text{exp} \bigg\{ i\int_{M_5} \cL \bigg\} = 1 \ \ \ \Leftrightarrow \ \ \ 
I = \int_{M_6}-\frac{1}{3!} C_{I J K } F^{I} \wedge F^{J} \wedge F^{K}   - \frac{a_{I} }{48}F^{I} \wedge p_1(M_6)  \, \in \, \bZ.
\eeq
This requirement in turn leads to a set of consistency conditions on a generic 5D minimal supergravity that can be summarized as a set of conditions on the intersection matrix $C_{IJK}$ and the vector $a_I$:
\begin{subequations}
\label{5d consistency condition}
\bea
&& C_{I J K } \in \bZ \qquad  \qquad \quad \forall \, I ,  J ,  K, \label{eq:CC1}  \\
 && C_{I I I} +  \frac{a_{I}}{2}  \in 6 \bZ  \qquad  \quad \forall I,  \label{eq:CC2}\\
&&   C_{I I J} + C_{I J J} \in 2 \bZ \qquad \forall \, I \neq J. \label{eq:CC3} 
\eea
\end{subequations}
Note that  the above conditions imply also $a_I \in 2\bZ$. These conditions can be seen as the  generalization of the cubic form construction for a single vector field \eqref{eq:cubic}.

Note that analysis of the partition function will give constraints up to an ad hoc addition of independently integer-valued contributions, i.e. would be expected to be weaker than \eqref{eq:CC1} -- \eqref{eq:CC3}. On the other hand, we may recall that  in the case of  M-theory, the set of conditions coming from the M5 anomalies \cite{Freed:1998tg, Harvey:1998bx} does not have these ambiguities and implies the conditions derived from the partition function. So now as well we can analyze the integrality and  divisibility properties of the central charges of $\mathcal{N} = (0,4)$ BPS strings and then verify that \eqref{eq:PF} is indeed satisfied.

In the presence of string sources, the invariance of the CS couplings in \eqref{eq:5dL} is broken and localizes on the string worldsheet. Requiring that these are canceled against the string anomalies, i.e. the anomaly inflow mechanism, determines the  central changes of the BPS strings  as (see e.g. \cite{Freed:1998tg, Harvey:1998bx})
\bea \label{eq:CeCh}
    c_R &=&  C_{I J K} Q^{I}  Q^{J}  Q^{K}  +  \frac{a_{I}}{2} Q^{I}, \nn \\
     c_L - c_R &=&  \frac{a_{I}}{2} Q^{I},
\eea
where $Q^{I} \in \bZ$ is the charge of the string with respect to $F^{I}$ for any $I$. The $\mathcal{N} = (0,4)$ worldsheet supersymmetry requires 
$$
c_R \, \in \, 6 \bZ \qquad  \forall  \,Q_{I} \, .
$$
Note that $c_L - c_R \in \bZ$, and hence $\frac{1}{2}a_I\in \bZ$ is also required.

Considering a BPS string with a single non-zero charge $Q^I$, one can rewrite $c_R$ as
$$
c_R = C_{III} \Big[ (Q^I)^3 -Q^I \Big] + \left(C_{III} + \frac{a_I}{2} \right) Q^I \quad \forall Q^I \in \bZ.
$$
Given that the first term is always divisible by 6, the condition  \eqref{eq:CC2} follows. The integrality of $\frac{1}{2}a_I$  in turn implies the integrality of $C_{III}$ for any value of the index $I$.

Now consider a string with two non-zero charges $Q^I$ and $Q^J$ ($I \neq J$). Subtracting $\frac13\left(C_{I I I} +  \frac{1}{2}a_I\right) (Q^I)^3 $ from $\frac{1}{3}c_R$, one obtains $ C_{IIJ}(Q^I)^2 Q^J + C_{IJJ}Q^I (Q^J)^2  \in 2 \bZ$. Using that $(Q^I)^2 Q^J \pm Q^I (Q^J)^2 \in 2 \bZ$, this can be rewritten as:
\begin{equation}
\begin{split}
&(C_{IIJ} + C_{IJJ} ) (Q^I)^2 Q^J + C_{IJJ} \Big[Q^I (Q^J)^2  - (Q^I)^2 Q^J  \Big]  \in 2 \bZ \quad \\& \Rightarrow \quad C_{IIJ} \in \bZ , \,\,\,\,  C_{IIJ} + C_{IJJ} \in 2 \bZ, \,\,\,\,\, \ \ \ \ \forall I \neq J. 
\end{split}
\end{equation}
 
Finally, using \eqref{eq:CeCh}, a three charge configuration with $Q^I$, $Q^J$, and $Q^K$ ($I \neq J \neq K$) yields
$$
\frac{c_R}{6} - \frac16\left(C_{I I I} +  \frac{a_{I}}{2}\right) (Q^I)^3 - \frac12 \Big[ C_{IIJ}(Q^I)^2 Q^J + C_{IJJ}Q^I (Q^J)^2 \Big] = C_{IJK}Q^{I}  Q^{J}  Q^{K}   \, \in \, \bZ,
$$
completing the condition \eqref{eq:CC1}.

Before turning to the partition function and \eqref{eq:PF}, recall that a Dirac index is given by 
\begin{equation}
\label{6d index theorem 1}
   \mbox{ ind} (D_{M_6}) = \int_{M_6}\frac{1}{3!}F^3 - \frac{1}{24}F \wedge p_1(M) \in \bZ,
\end{equation}
where $F= \sum n_{I} F^{I}$ is a linear combination of $U(1)$ field strengths with integer coefficients. One immediate implication of \eqref{6d index theorem 1} is that $\frac{1}{4}p_1(M_6) \in H^{4}(M_6;\bZ)$, which for a 6D spin manifold can also be derived by noticing that the fourth Stiefel-Whitney (SW) class $w_4(M_6) = 0$. Equation \eqref{eq:PF} then yields 
\bea \label{eq:PF1}
I &\cong&  \int_{M_6} \frac16 \left( C_{I I I} +  \frac{a_{I}}{2}  \right) F^I \wedge F^I \wedge F^I   \\
&+& \int_{M_6} \frac12  \sum_{I \neq J} \left( C_{IIJ}F^I \wedge F^I \wedge F^J + C_{IJJ}F^I \wedge F^J \wedge F^J \right) + \sum_{I \neq J \neq K} C_{IJK} F^I \wedge F^J \wedge F^k, \nn
\eea
where $\cong$ denotes equality mod $\bZ$, which comes from using \eqref{6d index theorem 1} for $F=F^I$ in the first term. Using \eqref{eq:CC1} -- \eqref{eq:CC3}, every single term in \eqref{eq:PF1} is integer-valued.

We note that the integer-valued invertible transformation $F^I \,\mapsto \, N^I_{\, \, J} F^J$, where $N \in \text{SL}(n_V, \bZ)$ is a matrix with unit determinant, preserves the consistency conditions \eqref{eq:CC1} -- \eqref{eq:CC3}. This can be proven using arguments similar to those proving the conditions themselves.

\subsection{Consistency conditions in CY3 compactifications of M-theory}
Calabi-Yau threefold compactifications of M-theory provide a large class of consistent minimal supergravities, and hence the expectation is that the consistency conditions \eqref{eq:CC1} -- \eqref{eq:CC3} are satisfied due to general basic properties of the CY geometry. The terms in \eqref{eq:5dL} descend from 11D couplings:
\bea \label{eq:11dL}
 \int_{M_{11}} \cL && \supset \int_{M_{11}} -\frac{1}{6} C_3 \wedge G_4 \wedge G_4   - \frac{1}{192} C_3 \wedge \left[ p_1(TM_{11})^2 - 4 p_2  (TM_{11}) \right]  \nn \\
&& = \int_{M_{12}}- \frac{1}{3!} G_4 \wedge G_4 \wedge G_4  -  G_4 \wedge X_8 (TM_{11}),
\eea
where $\partial M_{12}=M_{11}$. Compactifying on a Calabi-Yau threefold $X$  gives rise to \eqref{eq:5dL} with 
\begin{equation}
    \label{5d supergravity CS from CY3}
    C_{IJK} = \int_{X}\o_I \cup \o_J \cup \o_K \, \qquad \mbox{and} \qquad  a_I = \int_{X}\o_I \cup c_2(TX),
\end{equation}
where $I,J,K = 1,..., h^{1,1}(X) = n_V +1$.

The integrality of the intersection matrix $C_{IJK}$ is clear, and only the conditions
\begin{subequations}
\bea
 \int_{X}\o_I\cup \o_J\cup \o_J + \o_I\cup \o_I\cup \o_J &\in&  2\bZ \,\,\, \ \ \ \ \forall \, I \neq J, \label{eq:CCCY1}\\
\int_{X}\o_I \cup \o_I \cup \o_I +  \frac12  \o_I \cup c_2 (TX)  &\in&  6\bZ \ \,\,\, \ \ \ \forall \, I,  \label{eq:CCCY2}
\eea
\end{subequations}
need to be verified. In fact these statements are the content of Theorem 5 in the classical paper by Wall using differential topology \cite{wall1966classification}. Below we present a different proof.

Start by recalling that the form $\o_I$ is the dual of the divisor $D_I$, and the CY condition (triviality of the canonical bundle) gives $c_1 (TD_I \oplus N) =0$, implying that   $\o_I = c_1(TD_I) = - c_1(N)$ (where $N$ is the normal bundle). From the other side, 
$c_1(TD_I) = w_2(D_I)$ mod $2$, and the (integral lift of) the second Stiefel-Whitney class $w_2(D_I)$ is the characteristic vector in $H^2(D_I;\bZ)$, i.e.
    \begin{equation}
        x\cup x + x\cup w_2(D_I) = 2\bZ \qquad  \forall x \in H^2(D_I;\bZ).
    \end{equation}
    Hence, dropping the pullback signs,  we obtain:
\begin{equation}
\begin{split}
    \int_X \o_I\cup \o_J\cup \o_J + \o_I\cup \o_I \cup \o_J &= \int_{D_I}\o_I\cup \o_J +\o_J\cup \o_J \\& = \int_{D_I}w_2(D_I)\cup \o_J +\o_J\cup \o_J = 0 \, \ \text{mod} \,\, 2,
\end{split}
\end{equation}
proving  \eqref{eq:CCCY1}. Using the same relations, equation  \eqref{eq:CCCY2} can be rewritten as 
\bea
\int_X \o_I \cup \o_I \cup \o_I +  \frac12  \o_I \cup c_2 (TX) &=& \int_{D_I} c_1(TD_I)^2 + \frac12 c_2(TD_I\oplus N) \nn \\
& = & \int_{D_I} \frac{c_1(TD_I)^2  + c_2(TD_I) }{2}.
\eea
In the last equality we have used $c_2(TD_I\oplus N) = c_2(TD_I) + c_1(TD_I) c_1(N)  = c_2(TD_I) - c_1(TD_I)^2$. 

Assuming the divisor $D_I$ can be represented by a complex surface, using the Hirzebruch-Riemann-Roch theorem gives
\begin{equation}
    \sum_{i}(-)^ih^{0,i}(D_I) = \int_{D_I} \mbox{td}(TD_I) = \int_{D_I}\frac{c_1(TD_I)^2 + c_2(TD_I)}{12} \in \bZ,
\end{equation}
from which  $ \int_X \o_I \cup \o_I \cup \o_I  +  \frac{1}{2}\o_I \cup c_2(TX) \in 6\bZ$ follows.

\vspace{.5cm}
\noindent
There is a caveat to the above argument --  we have assumed that (multiple copies of) every effective divisor $D_I$ can be represented by a smooth complex surface inside $X$. So far we do not know whether this is a true statement or not for a general CY3. Here, we fill this gap by the following argument:
\begin{itemize}
    \item Choose a set of effective divisors $\{D_I\} = (D_1,..,D_{h^{1,1}})$ that generate $H^{2}(X;\bZ)$; here $D_I$ is the Poincaré dual of $\omega_I$ in \eqref{eq:CCCY1}, \eqref{eq:CCCY2}.
    \item Choose a very ample divisor $D_0$ inside the Kähler cone of the CY3, and then the above proof works for $D_0$, as it can be represented by a smooth complex surface, i.e.:
    \begin{equation}
        \label{intersection number for D0}
        D_0 \cdot D_0 \cdot D_0 + \frac{1}{2} D_0 \cdot c_{2}({TX}) = 0 \ \text{mod} \ 6.
    \end{equation}
    \item For every $D_I$, choose a very large $a\in 2\bZ_{>0}$ such that $D_0 + \frac{1}{a}D_I$ is also inside the Kähler cone of the CY3; then $aD_0 + 2 D_I$ is also a very ample divisor, and hence the above proof gives
    \begin{equation}
        \label{intersection of very ample divisor}
        (aD_0 +  D_I)\cdot (aD_0 +  D_I) \cdot (aD_0 +  D_I) + \frac{1}{2}(aD_0 + D_I)\cdot c_2({TX}) = 0 \ \text{mod} \ 6.
    \end{equation}
    Combining \eqref{intersection number for D0} and \eqref{intersection of very ample divisor} gives
    \begin{equation}
    \label{intersection of effective divisors 1st}
         D_I \cdot D_I \cdot D_I + \frac{1}{2} D_I \cdot c_{2}({TX}) = 0 \ \text{mod} \ 6.
    \end{equation}
    Hence we have established \eqref{eq:CCCY2}.
    \item  Applying \eqref{intersection of effective divisors 1st} to the effective divisor $D_I+D_J$ with $I\neq J$ gives
    \begin{equation}
    \label{intersection of effective divisors 2nd}
        D_I \cdot D_I \cdot D_J + D_I \cdot D_J \cdot D_J = 0 \ \text{mod} \ 2,
    \end{equation}
    and equation \eqref{eq:CCCY1} is established.
\end{itemize}

\section{Anomaly theories}
\label{sec:AT}
Local and global anomalies in D dimensions are captured by a (D+1)-dimensional (invertible) field theory called the anomaly theory. This framework will prove to be very useful for our analysis and will be reviewed in this section. For convenience some of the explicit considerations here will be restricted to fermions. 

Consider a theory $\cT$ with gauge group $G$; then the anomaly of the gauge group $G$ is described by a phase ambiguity $\cA(g)$ of the partition function $Z_{\cT}$ under a gauge transformation $g\in G$:
\begin{equation}
    \label{anomaly phase 1st}
    g: Z_\cT \rightarrow Z_{\cT^g} = \cA(g) \cdot Z_\cT, \qquad \qquad \cA(g)  \in U(1).
 \end{equation}
$\cA \neq 1$ signals the quantum inconsistency of the theory $\cT$. Luckily, given a theory $\cT$, there are ways to study the anomaly $\cA$ without knowing the full partition function $Z_\cT$, as we review below.
\subsection{Perturbative anomalies}
Perturbative anomalies only appear in even-dimensional theories with chiral fields \cite{Harvey:2005it}. In this case, the (consistent) anomaly $\cA$ is given by a solution of the Wess-Zumino consistency condition.
\paragraph{Wess-Zumino consistency condition} Given a theory $\cT$ on  a D-dimensional spacetime $M_{\text{D}}$ with gauge group $G$ (with lie algebra $\mathfrak{g}$), then under the gauge transformation $A_G \to A_G + D_\mu \e_1$ with $\e_1 \in C^{\infty}(M_{\text{D}};\mathfrak{g})$ and $A_G$ the gauge fields of gauge group $G$, the effective action of the theory $\cT$, $\G_{\cT}(A_G)$, transforms as:
\begin{equation}
    \G_{\cT}(A_G) \to  \G_{\cT}(\cA) + \delta_{\e_1}\G_{\cT}(A_G),
\end{equation}
 and satisfies the Wess-Zumino consistency condition:
 \begin{equation}
     \label{WZ consistency condition}
     \delta_{\e_1}\delta_{\e_2}\G_{\cT}(A_G)-\delta_{\e_2}\delta_{\e_1}\G_{\cT}(A_G) = \delta_{[\e_1,\e_2]}\G_{\cT}(A_G), \quad \e_{1,2} \in C^{\infty}(M_{\text{D}};\mathfrak{g}).
 \end{equation}

 The descent procedure provides the solution to (\ref{WZ consistency condition}):
 \begin{equation}
 \label{Descendant procedure}
    \begin{aligned}
I_{\text{D}+2}(F_G) & = \text{d} I_{\text{D}+1}^{(0)}(A_G,F_G), \\
\delta_\e I_{\text{D}+1}^{(0)}(A_G,F_G) & = \text{d} I_{\text{D}}^{(1)}(\e,A_G,F_G),\\
    \cA &= \exp\{i I_{\text{D}}^{(1)}(\e,A_G,F_G)\}.
\end{aligned}
\end{equation} 
The subscripts here indicate the rank of the differential form, while the superscripts denote the number of variation parameters for local quantities.
The ambiguities in determining the anomaly $\cA$ (or $I_{ \text{D}}^{(1)}(\e,A_G,F_G)$) in \eqref{Descendant procedure} correspond to the local counterterms that can be added to the theory $\cT$. The above discussion can also be generalized to include gravity. In that case, $I_{\text{D}+2}$ will be  a $(\text{D}+2)$-form polynomial of both gauge field strength $F_G$ and the spacetime curvature $R$. Hence, in general, the anomaly $\cA$ is determined by a $(\text{D}+2)$-form polynomial in  $F_G$ and $R$. For a D-dimensional theory with chiral fermions (for  D  even) in representation $\cR$ of gauge group $G$, $I_{\text{D}+2}$ is given by the index density \cite{Alvarez-Gaume:1983ihn,Alvarez-Gaume:1984zst,Alvarez-Gaume:1984zlq}:
\begin{equation}
    \label{anomaly index density}
     I_{\text{D}+2} = \widehat{A}(T M_{\text{D}}) \, \text{ch}_{\cR}(F_G)\big|_{\text{D}+2} \ ,
\end{equation}
where 
\begin{equation}
   \widehat{A}(T M_{\text{D}}) = 1 + \frac{\text{tr}(R^2)}{48}+ \cdots, \ \ \ \ \ \ \text{ch}_{\cR}(F) = \text{Tr}_{\cR_G}(e^{F_G}).
\end{equation}
Vanishing of the perturbative anomaly, $\cA =1$, means
\begin{equation}
    \label{vanishing of perturbative anomaly}
     I_{\text{D}+2} = 0.
\end{equation}
The right hand side of (\ref{anomaly index density}) contains only even forms, and in odd-dimensional theories, $I_{\text{D}+2} =0$ automatically. In other words, in odd dimensions, the perturbative anomaly only comes from gauge non-invariant local couplings and can be removed by the addition of local counterterms. 

\paragraph{Example: 4D QED with fermions} Consider 4D $U(1)$ gauge theory with two chiral fermions with gauge charges $q_1,q_2$. Then (\ref{anomaly index density}) in this case is
\begin{equation}
    I_{6} = (q_1^3 + q_2^3)\frac{F^3}{6} +(q_1+q_2)F\wedge \frac{\text{tr}(R^2)}{48},
\end{equation}
where the first term is the pure gauge anomaly and the second term is the mixed gauge-gravitational anomaly. Vanishing of anomalies requires
\begin{equation}
    q_1^3 + q_2^3 = q_1+q_2 =0,
\end{equation}
and a set of solutions is $q_1 = -q_2 = Q \in \bZ$.

\subsection{Dai-Freed theory of anomalies}
 Given a theory $\cT$ on a D-dimensional spacetime $M$, the Dai-Freed anomaly (see \cite{Dai_1994, Garcia-Etxebarria:2018ajm, Lee:2020ojw} for more details) associates a bulk $(\text{D}+1)$-dimensional manifold $X$ with $M$ (i.e. $\partial X= M$), where all the relevant bundle structures (gauge bundles, spin structure, etc.) on $M$ are extended to $X$ (see Figure \ref{X_{12}}).\footnote{We assume this is always the case. The obstructions to extend relevant structures from $M$ to $X$ are related to generalized $\theta$-terms of the theory $\cT$ and will be ignored when studying the anomalies of $\cT$. }

Denoting the bulk partition function $\a(X)$, Dai-Freed theory roughly says that the combined theory is well-defined under a gauge transformation $g$:
\begin{equation}
\label{Well defined combine system}
Z_{\cT}(M) \a(X) = Z_{\cT^g}(M) \a^g (X) ,
 \end{equation}
with
\begin{equation}
    \label{bulk partition function of X}
    \a(X) = \exp\{ 2 \pi i \eta_X \},
\end{equation}
where $\eta_X$ is the eta invariant bulk Dirac operator $D_X$ with suitable boundary conditions \cite{atiyah1975spectral1,atiyah1975spectral2,Dai_1994}.

\paragraph{$\eta$ invariant} The $\eta$ invariant is defined in \cite{atiyah1975spectral1} in terms of the spectrum of the Dirac operator $D_X$ on a manifold $X$.\footnote{Certain boundary conditions need to be imposed to make $D_X$ self-adjoint if $X$ has a boundary.} For the eigenvalues $\{\lambda_i,...\}$ of $D_X$, $\eta(X)$ is defined as
\begin{equation}
    \label{definition of eta}
    \eta(X) = \lim_{s\to 0} \sum_{\lambda_i}\frac{\text{sgn}(\lambda_i)}{2|\lambda_i|^{-s}}.
\end{equation}
$\text{sgn}(\lambda_i)$ denotes the signature of $\lambda_i$ with the choice $\text{sgn}(0) =1$. $\eta$ is a geometric invariant and in general depends on the geometric properties of $X$, such as the metric.

If the theory $\cT$ is well defined, it should not depend on the extension $X$. Together with (\ref{Well defined combine system}), this means that there should not be a difference between two different bulk partition functions, i.e. $\frac{\a(X_1)}{\a(X_2)}=1$. Thanks to the gluing property of $\eta$ proven in \cite{Dai_1994}, this difference is given by $\exp\{2 \pi i \eta_{Y}\}$ of the closed ($\text{D}+1$)-dimensional manifold $Y$, obtained by gluing $X_{1}$ and $X_2$ along their shared boundary $M$: $Y = X_1 \sqcup_{M}\bar{X}_2$,  where the bar indicates that for $X_2$ the orientation is reversed (see Figure \ref{X_{12}}): 
\begin{equation}
    \label{gluing property of eta inv}
    \exp\{2 \pi i \eta_{Y}\} = \exp\{2 \pi i (\eta_{X_1}-\eta_{X_2}) \} = \frac{\a(X_1)}{\a(X_2)}.
\end{equation}
To summarize, there exists the following equivalence:
\begin{equation}
    \label{DF eta for anomaly}
    \text{D-dim theory $\cT$ has no chiral anomaly} \ \ \iff \ \ \exp\{2\pi i \eta_Y\} = 1,
\end{equation}
for all $(\text{D}+1)$-dimensional manifolds $Y$ with corresponding structures.\footnote{Notice that since the structure of $X_{1,2}$ and the Dirac operators $D_{X_{1,2}}$ is determined by $\cT$, the same applies to $Y$.}
\begin{figure}
    \centering
    \includegraphics[width=0.9\linewidth]{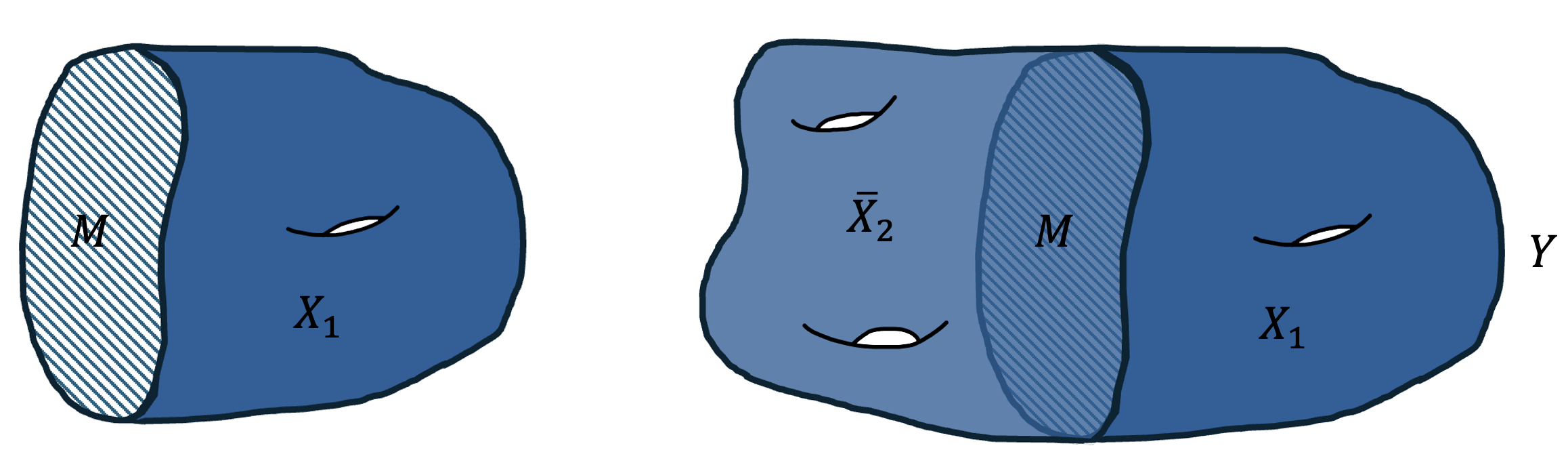}
    \caption{$Y$ from gluing $X_1,\bar{X}_2$}
    \label{X_{12}}
\end{figure}

\paragraph{APS index theorem} Assuming a ($\text{D}+1$)-dimensional manifold $Y$ (with the gauge bundle structures, etc., extended) can be bounded by a $(\text{D}+2)$-dimensional manifold $Z_{\text{D}+2}$, then
\begin{equation}
    \label{APS index theorem}
    \eta_{Y} = \int_{Z_{\text{D}+2}} \widehat{A}(TZ_{\text{D}+2}) \, \text{ch}_{\cR}(F_G) \ \ \text{mod }\bZ \,.
\end{equation}
Using the APS index theorem \cite{atiyah1975spectral1,atiyah1975spectral2}, the difference of the LHS and the RHS is the index of the Dirac operator on $Z_{\text{D}+2}$ with suitable boundary conditions. 
As before, $\cR$ is the representation of the fermions charged under the gauge group $G$, as determined by the D-dimensional theory $\cT$.

Combining (\ref{DF eta for anomaly}),(\ref{APS index theorem}), and (\ref{anomaly index density})  with the fact that the right-hand side of (\ref{APS index theorem}) is in general not an integer ($Z_{\text{D}+2}$ is a $(\text{D}+2)$-dimensional manifold with boundary), we see how the Dai-Freed theory captures the ordinary description of perturbative anomalies:
\begin{equation}
    \label{perturbative anomaly from df}
    \exp\{2\pi i \eta_Y\} = 1 \iff \widehat{A}(TZ_{\text{D}+2}) \, \text{ch}_{\mathcal{R}}(F_G)\big|_{\text{D}+2} = 0,
\end{equation}
when the ($\text{D}+1$)-dimensional manifold $Y$ with all relevant structures extended can be bounded by $Z_{\text{D}+2}$.

\paragraph{Non-perturbative anomalies} In fact,  as (\ref{DF eta for anomaly}) also applies to a manifold $Y$ that cannot be bounded by a $(\text{D}+2)$-dimensional manifold $Z_{\text{D}+2}$, the Dai-Freed theory contains more than just the perturbative anomalies. This type of manifold $Y$ is characterized by the bordism group $\Omega_{\text{D}+1}^{\text{Spin}}(BG)$:
\begin{equation}
    \label{def of bordism group}
    \Omega_{\text{D}+1}^{\text{Spin}}(BG) = \{Y: Y \neq \partial Z_{\text{D}+2}\}/\text{\{bordant equivalence\}}.
\end{equation}
Here $Y_1$ and $Y_{2}$ are bordant equivalent if $\exists Z_{d+2}$ ${\text{s.t.}} \,  \partial Z_{d+2} = Y_{1}\sqcup \bar{Y}_{2}$, where $\bar{Y}_{2} $ is $Y_2$ with the orientation reversed. $BG$ contains the bundle structure information regarding the gauge groups of the theory, and the superscript Spin indicates that we consider $(\text{D}+1)$-dimensional manifolds with spin structure. There will be new anomalies from the torsion part of $\Omega_{\text{D}+1}^{\text{Spin}}(BG)$,\footnote{The free part of $\Omega_{\text{D}+1}^{\text{Spin}}(BG)$ does not give rise to new anomalies; instead they characterize the generalized Wess-Zumino-Witten terms of the theory $\cT$. We will discuss this below.} which consists of those manifolds $Y_{\text{D}+1}$ (up to bordant equivalence) that cannot be bounded by any $(\text{D}+2)$-dimensional $Z_{\text{D}+2}$, but some copies of $Y_{\text{D}+1}$ can be bounded. What's more, \cite{atiyah1975spectral2} tells us that although $\eta_Y$ is a geometric invariant, it can be promoted to a cobordism invariant when the perturbative anomaly vanishes (i.e. when the right-hand side of (\ref{APS index theorem}) is zero). 

To be more precise, 
\begin{equation}
    \label{non perturbative anomaly 1st}
    \exp\{2 \pi i \eta_Y\} \in \text{Hom}_{\bZ}\big(\Omega_{\text{D}+1,\text{tor}}^{\text{Spin}}(BG), U(1)\big), 
\end{equation}
when the perturbative anomaly vanishes. Then the condition that the theory $\cT$ is anomaly free \eqref{DF eta for anomaly} becomes
\begin{equation}
    \label{non perturbative anomaly 1st-}
    \exp\{2 \pi i \eta_Y\} = 1 \in \text{Hom}_{\bZ}\big(\Omega_{\text{D}+1,\text{tor}}^{\text{Spin}}(BG), U(1)\big). 
\end{equation}
To summarize, the Dai-Freed theory characterizes the anomaly in the following way:
\begin{itemize}
    \item The perturbative anomaly is given by $\widehat{A}(TZ_{\text{D}+2})\, \text{ch}_{\cR}(F_G)\big|_{\text{D}+2}$.
    \item When the perturbative anomaly vanishes ($\widehat{A}(TZ_{\text{D}+2})\, \text{ch}_{\cR}(F_G){\big|_{\text{D}+2}} = 0$), the non-perturbative anomaly is described by $\exp\{2\pi i \eta_Y\} \in \text{Hom}_{\bZ}\big(\Omega_{\text{D}+1,\text{tor}}^{\text{Spin}}(BG), U(1)\big)$.
\end{itemize}

\paragraph{Cobordism invariants} Dai-Freed theory characterizes anomalies in terms of cobordism invariants. Note $\widehat{A}(TZ_{\text{D}+2}) \,  \text{ch}_{\cR}(F_G)\big|_{\text{D}+2}$ detects (the free part of) the (D$+2$)-dimensional bordism class:
\begin{equation}
    \label{index density is bordism inv}
    \int_{N_{\text{D}+2}} \widehat{A}(TZ_{\text{D}+2}) \, \text{ch}_{\cR}(F_G) \neq 0 \in \bZ \ \ \Rightarrow \ \ \nexists P_{\text{D}+3} \text{ s.t. } \partial P_{\text{D}+3} = N_{\text{D}+2}, 
\end{equation}
for a closed $(\text{D}+2)$-dimensional manifold $N_{\text{D}+2}$. Together with (\ref{non perturbative anomaly 1st}), we see that the anomaly $\cA$ of the D-dimensional theory $\cT$ is classified by $\text{Inv}^{\text{Spin}}(BG)$ defined by the sequence:
\begin{equation}
    \label{bordism description of anomalies 1st}
    0 \to \text{Hom}_{\bZ}\big(\Omega_{\text{D}+1,\text{tor}}^{\text{Spin}}(BG), U(1)\big) \to {\text{Inv}^{\text{Spin}}(BG)}  \to \text{Hom}_{\bZ}\big(\Omega_{\text{D}+2}^{\text{Spin}}(BG), \bZ\big) \to  0.
\end{equation}
 
\paragraph{Example: Global $SU(2)$ (Witten) anomaly} In \cite{Witten:1982fp}, a global anomaly for 4D $SU(2)$ gauge theories with $n_f$ fundamental Weyl fermions, due to $\pi_4(SU(2)) = \bZ_2$, is discussed. Vanishing of this global $SU(2)$ anomaly requires $n_f$ to be even. This example fits well into the Dai-Freed theory framework, i.e. \eqref{bordism description of anomalies 1st}:
 \begin{itemize}
     \item The perturbative anomaly vanishes:
     \begin{equation}
         I_{6}^{\text{pert}} = \widehat{A}(TX)\, \text{ch}_{\text{fund}}(F_{SU(2)})\big|_{6} = 0.
     \end{equation}
     \item $\Omega_{5\text{D},\text{tor}}^{\text{Spin}}\big(BSU(2)\big) = \bZ_2 \,\, \Rightarrow \,\, \text{Hom}_{\bZ}\big(\Omega_{5\text{D},\text{tor}}^{\text{Spin}}(BSU(2)), U(1)\big) = \bZ_2$ suggests the existence of a potential non-perturbative anomaly.
     \item Putting the theory with $SU(2)$ instanton number $1$ on $S^4$, and using the large gauge transformation $g \neq \text{id} \in \pi_{4}(SU(2))$ of the $SU(2)$ gauge potential $A$, as well as  the transformed gauge potential $A^g = gAg^{-1} + g \text{d} g^{-1}$, \cite{Witten:1982fp} constructed a 5D ``mapping torus" $M_{5\text{D}}$:
     \bea
         (M_{5\text{D}},A) &= (S^4\times [0,1],A = (1-t) A\big|_{S^4} + t A^{g}\big|_{S^4})/(x\in S ^4 \times \{0\},A_{t=0}) \nn \\
         &\sim  (x\in S ^4 \times \{1\},A_{t=1}),
     \eea
     
    where $(M_{5\text{D}},A)$ is the generator of $\Omega_{5\text{D},\text{tor}}^{\text{Spin}}\big(BSU(2)\big)$. Using the adiabatic ansatz, \cite{Witten:1982fp} shows that the partition function of  4D $SU(2)$ gauge theories with $n_f$ fundamental Weyl fermions $Z_{n_f}$ is an element\footnote{{Up to a factor $|Z_{n_f}|$ that is irrelevant for the anomaly.}} of $\text{Hom}_{\bZ}\big(\Omega_{5\text{D},\text{tor}}^{\text{Spin}}(BSU(2)), U(1)\big)$, i.e.
     \begin{equation}
         Z_{n_f}\big((M_{5\text{D}},A)\big) \propto (-)^{n_f},
     \end{equation}
     and that the vanishing of the $SU(2)$ global anomaly requires $n_f$ to be even.
 \end{itemize}

\subsection{Physical interpretation of bordism groups}
\label{Physical interpretation of other bordism groups}

In this paper, we refer to the anomaly  encoded in $\text{Hom}_{\bZ}\big(\Omega_{\text{D}+1,\text{tor}}^{\text{Spin}}(BG), U(1)\big)$, determined  by \eqref{bordism description of anomalies 1st}, as the non-perturbative  anomaly.\footnote{According to this definition, although the 4D Witten anomaly is a non-perturbative anomaly, its 6D analogues are actually captured by a perturbative anomaly \cite{Lee:2020ewl,Davighi:2020kok}.}  When perturbative anomalies vanish,  the vanishing of non-perturbative anomalies puts further non-trivial constraints on consistent theories.

The free part of $\Omega_{\text{D}+1 }^{\text{Spin}}(BG)$ is also physically relevant. $\text{Hom}_{\bZ}\big(\Omega_{\text{D}+1,\text{free} }^{\text{Spin}}(BG), \bZ\big)$ characterizes the generalized WZW terms that can be added to the D-dimensional theory \cite{Lee:2020ojw}. To illustrate this point, we shall just use one pertinent example.

\paragraph{Example: 5D $\cN=1$ supergravity with a $U(1)$ field}
\label{example 5d minimal supergravity}
 For 5D $\cN=1$ supergravity with a $U(1)$ field with curvature $F$, the relevant bordism groups are (see Appendix \ref{Some bordism group calculations} for the bordism group calculations):
\begin{equation}
    \label{Omega7dU(1)}
    \Omega_{7\text{D}}^{\text{Spin}}\big(BU(1)\big) = 0,
\end{equation}
\begin{equation}
    \label{Omega6dtorU(1)}
    \Omega_{6\text{D},\text{tor}}^{\text{Spin}}\big(BU(1)\big) = 0.
\end{equation}
Hence there are no anomalies for the theory we are considering. At the same time
\begin{equation}
    \label{Omega6dfreeU(1)}
     \Omega_{6\text{D},\text{free}}^{\text{Spin}}\big(BU(1)\big) = \bZ\oplus \bZ,
\end{equation}
and
\begin{equation}
    \label{dualOmega6dfreeU(1)}
     \text{Hom}_{\bZ}\big(\Omega_{6\text{D},\text{free}}^{\text{Spin}}(BU(1)),\bZ\big) = \bZ\oplus \bZ \,. 
\end{equation} 
The two generators of (\ref{dualOmega6dfreeU(1)}) are
\begin{equation}
    F^3, \quad \frac{1}{6}F^3 + \frac{1}{48}F\wedge \text{tr} (R^2).
\end{equation}
Hence the element $\cI \in \text{Hom}_{\bZ}\big(\Omega_{6\text{D},\text{free}}^{\text{Spin}}(BU(1)),\mathbb{Z}\big)$ takes the form:
\begin{equation}
\label{elementofdualOmega6dfreeU(1)}
    \cI = a F^3 + b F\wedge \text{tr}(R^2) \ \   \text{ s.t. }  \int_{N_6} \cI \in \bZ \ \,\,\, \ \forall \text{ 6D closed spin manifolds $N_6$}.
 \end{equation}
 On the other hand, there is a topological sector of 5D $\cN=1$ supergravity, given by the 5D CS term \eqref{eq:pure5d+}. Extending the relevant structure to a 6D bulk $M_6$ with $\partial M_6 = M_5$\footnote{This can be done, since $\Omega_{5\text{D}}^{\text{Spin}}\big(BU(1)\big) = 0$} and demanding that the integral be independent of the bulk extension is equivalent to the following condition:
\begin{equation}
    \label{WZW quantization condition}
     \int_{N_6} -\frac{C_{000}}{6} F \wedge F \wedge F   + \frac{a_0}{96}F \wedge \tr R^2  \in \bZ  \  \quad \forall \text{ 6D closed spin manifolds $N_6$} \,.
\end{equation}
Comparing (\ref{WZW quantization condition}) and (\ref{elementofdualOmega6dfreeU(1)}), we conclude that well-defined topological terms in 5D $\cN=1$ supergravity with $U(1)$ gauge group are characterized by $ \text{Hom}_{\bZ}\big(\Omega_{6\text{D},\text{free}}^{\text{Spin}}(BU(1)),\bZ\big)$.

\paragraph{Remark} In the above example we used the special property $\Omega_{5\text{D}}^{\text{Spin}}(BU(1)) = 0$. But in some cases there can exist obstructions to extending the D-dimensional theory to a $(\text{D}+1)$-dimensional bulk,  $\Omega_{\text{D}}^{\text{Spin}}(BG) \neq 0$. When this happens, generalized $\theta$ terms can be added to the theory given by $\text{Hom}_{\bZ}\big(\Omega_{\text{D}}^{\text{Spin}}(BG),U(1)\big)$.

\subsection{6D Green-Schwarz mechanism and isometries}
\label{sec:ano6d}

If the 6D anomaly (and the corresponding anomaly theory) factorizes, one can find couplings in supergravity, depending locally on the fields of the theory, such that when exponentiated and tensored with the partition function of the theory, the product is well-defined. This is the (generalized) Green-Schwarz mechanism \cite{Sagnotti_1992}, and the couplings are the GS counterterms. In this subsection we review the relevant details of the 6D GS mechanism and then present its dimensional reduction on the circle. Looking ahead to this, we use the notation that fields with tildes are 6-dimensional, while those without are 5-dimensional after circle compactification.

The situation for 6D $\mathcal{N} = (1,0)$ supergravities is the following: in order to cancel anomalies via the generalized GS mechanism, the anomaly polynomial must factorize as
\begin{equation}
    \label{6d perturbative anomaly polynomial}
    \tilde{I}_{8}^{\text{pert}} = \tfrac{1}{2}\Omega_{\a\b} \tilde{X}_4^\a \wedge \tilde{X}_4^\b,
\end{equation}
where $\a,\b$ run over the $(T+1)$ (anti-)self-dual tensors in the theory, and $\Omega_{\alpha \beta}$ is a constant symmetric inner product of signature $(1,T)$. Similarly to the situation in 10D, if the $B$-fields transform under general gauge and local Lorentz transformations as
\begin{equation}\label{eq:deltaB}
    \delta_{\tilde{\e}} \tilde{B}_2^\a = \text{d} \tilde{\lambda}_1^{\alpha}-\tilde{X}_{2}^{(1) \alpha\, },
\end{equation}
for the usual 2-form $\tilde{X}_2^{(1)}$ defined via descent ($\tilde{X}_4 = \text{d}\tilde{X}_3^{(0)}, \delta \tilde{X}_3^{(0)} = \text{d} \tilde{X}_2^{(1)}$), then adding the GS term to the action:
\begin{equation}
\label{6d GS term}
    \mathcal{L}_{\text{GS}} = \tfrac{1}{2} \Omega_{\a \beta} \tilde{B}_2^{\a}\wedge \tilde{X}_4^\b,
\end{equation}
cancels the anomaly (\ref{6d perturbative anomaly polynomial}) and makes the 6D $\mathcal{N} = (1,0)$ supergravity well-defined at the perturbative level. When we study the non-perturbative anomaly, we will assume that the perturbative anomaly vanishes after implementing the GS mechanism. In this case the gauge-invariant 3-form field strength for the $B$-fields is now modified to
\begin{equation}
\label{anomB}
	\tilde{H}^{\alpha}_3 = \text{d} \tilde{B}^{\alpha}_2 + \tilde{X}_3^{(0)\alpha}.
\end{equation}
In 6D $\mathcal{N} = (1,0)$ supergravity, one has four kinds of multiplet:
\begin{itemize}
    \item One supergravity multiplet $(\tilde{g}_{\m\n}, 2\tilde{\psi}^+_\m, \tilde{B}_{\m\n}^{+})$, where $\tilde{g}_{\m\n}$ is the graviton, $\tilde{\psi}^+_\m$ is a symplectic Majorana-Weyl (SMW) gravitino,  and $ \tilde{B}_{\m\n}^{+}$ is a self-dual 2-form.
    \item $H$ hypermultiplets $(2\tilde{\psi}^{-},4 \tilde{\phi})$, where the hyperini, $\tilde{\psi}^{-}$, are 6D SMW with negative chirality and $\tilde{\phi}$ is a real scalar.
    \item $V$ vector multiplets $(\tilde{A}_\m,2\tilde{\psi}^{+})$, with $\tilde{A}_{\m}$ being the gauge fields and $\tilde{\psi}^{+}$ being a 6D SMW gaugino with positive chirality.
    \item $T$ tensor multiplets $(\tilde{B}_{\m\n}^{-},2\tilde{\chi}^{-},\tilde{\phi})$, with $\tilde{B}_{\m\n}$ an anti-self-dual 2-form field, $\tilde{\chi}^{-}$ a 6D SMW tensorino with negative chirality, and $\tilde{\phi}$ a real scalar.
\end{itemize}
   The scalars in the tensor multiplets parametrize a $SO(1,T)/SO(T)$ moduli space. The hypermultiplets contain scalars in a quaternionic representation. We consider a gauge group $G$ which is a direct product of a semi-simple Lie group $\mathcal{G}$ and $n$ $U(1)$ factors:
\begin{equation}
\label{G}
	G = \mathcal{G} \times U(1)^n = \prod_{m} \mathcal{G}_m \times U(1)^n,
\end{equation}
such that $V = \sum_m \dim \mathcal{G}_m + n$. We use $\text{Tr}_{\mathcal{R}} [ \, \cdots \,]$ to denote the trace over a specific representation $\mathcal{R}$ and $\text{tr}[\, \cdots \, ]$ to denote the trace in the fundamental representation (or in the vector representation for gravitational terms). We denote the field-strengths of the simple non-abelian gauge factors by $\tilde{\mathcal{F}}_m$ and the $U(1)$ field-strengths by $\tilde{F}^i$. 

The total perturbative anomaly 8-form for a generic gauge group of the form (\ref{G}) is given by  \cite{Sagnotti_1992, Erler_1994}
\begin{equation}
\begin{split}
	\tilde{I}_8 = &\ \tfrac{1}{5760}(H - V + 29T - 273)\big[\text{tr}\tilde{R}^4 + \tfrac{5}{4}[\text{tr}\tilde{R}^2]^2\big] + \tfrac{1}{128}(9 - T) [\text{tr} \tilde{R}^2]^2 \\& - \tfrac{1}{96} \sum_m \big(A_{\text{adj}_m} -\sum_{\mathcal{R}_m} x^m_{\mathcal{R}_m} A_{\mathcal{R}_m}\big) \text{tr}\tilde{R}^2 \,  \text{tr} \tilde{\mathcal{F}}_m^2 \\& + \tfrac{1}{24} \sum_m \big(B_{\text{adj}_m} - \sum_{\mathcal{R}_m} x^m_{\mathcal{R}_m} B_{\mathcal{R}_m}\big)\text{tr} \mathcal{\tilde{F}}_m^4 \\&  + \tfrac{1}{24} \sum_m \big(C_{\text{adj}_m} - \sum_{\mathcal{R}_m} x^m_{\mathcal{R}_m} C_{\mathcal{R}_m}\big) [\text{tr}\mathcal{\tilde{F}}_m^2]^2 \\&
    - \tfrac{1}{6} \sum_{i, m} \sum_{q_i, \mathcal{R}_m} x^{i,m}_{q_i, \mathcal{R}_m} q_iE_{\mathcal{R}_m} \tilde{F}^i \text{tr} \tilde{\mathcal{F}}^3_m- \tfrac{1}{4} \sum_{m,n} \sum_{\mathcal{R}_m, \mathcal{S}_n} x^{m,n}_{\mathcal{R}_m, \mathcal{S}_n} A_{\mathcal{R}_m} A_{\mathcal{S}_n} \text{tr} \tilde{\mathcal{F}}_m^2 \text{tr}\tilde{\mathcal{F}}^2_n\\&
    - \tfrac{1}{2} \sum_{m,i,j} \sum_{\mathcal{R}_m, q_i, q_j} x^{m, i,j}_{\mathcal{R}_m, q_iq_j}q_i q_j A_{\mathcal{R}_m} \tilde{F}^i \tilde{F}^j\text{tr} \tilde{\mathcal{F}}^2_m \\& - \sum_{i,j,k,l} \sum_{q_i,q_j,q_k,q_l} x^{i,j,k,l}_{q_i, q_j, q_k, q_l} q_i q_j q_k q_l \tilde{F}^i \tilde{F}^j \tilde{F}^k \tilde{F}^l,
\end{split}
\end{equation}
where the representation theoretic coefficients are defined by
\begin{equation}
	\text{Tr}_{\mathcal{R}} \tilde{F}^2 = A_{\mathcal{R}} \text{tr} \tilde{F}^2, \ \ \ \ \  \ \ \text{Tr}_{\mathcal{R}} \tilde{\mathcal{F}}^4 = B_{\mathcal{R}} \text{tr} \tilde{\mathcal{F}}^4 + C_{\mathcal{R}} (\text{tr} \tilde{\mathcal{F}}^2)^2, \ \ \ \ \  \text{Tr}_{\mathcal{R}} \tilde{\mathcal{F}}^3 = E_{\mathcal{R}} \text{tr} \tilde{\mathcal{F}}^3,
\end{equation}
and the $x^{m,i,j}_{\mathcal{R}_m,q_i,q_j}$ coefficients label the number of hypermultiplets in the representation $\mathcal{R}_m$ of gauge factor $m$ and with charges $q_i$ and $q_j$ under the $i$th and $j$th $U(1)$ factors respectively. Similarly,  $x^{m,n}_{\mathcal{R}_m,\mathcal{S}_n}$ is the number of hypermultiplets both in representation $\mathcal{R}_m$ of gauge factor $m$ and $\mathcal{S}_n$ of gauge factor $n$, with the other $x$'s defined analogously. The irreducible terms in $\text{tr}\tilde{R}^4$, $\text{tr} \tilde{\mathcal{F}}_m^4$, and $\text{tr} \tilde{F}_m^3 \tilde{F}^i$,  cannot be canceled via the GS mechanism. Cancellation of such terms requires the following three constraints on the field-content and the representations to be satisfied\footnote{The coefficient of $ \text{Tr} \tilde{\mathcal{F}}^3_m \tilde{F}_i$ need not be zero for a more general GS mechanism to work, which will make the $U(1)_{i}$ gauge field massive \cite{Berkooz:1996iz}. Here we will not consider this case.}:
\begin{equation}
\begin{split}
	&\text{tr}\tilde{R}^4 \ \, \ \ \ : \ \ \  H - V + 29 T - 273 = 0, \\&
	\text{Tr} \tilde{\mathcal{F}}_m^4 \ \ \ : \ \ \ B_{\text{adj}_m} - \sum_{\mathcal{R}_m} x_{\mathcal{R}_m}^i B^i_{\mathcal{R}_m} = 0, \\&
	 \text{Tr} \tilde{\mathcal{F}}^3_m \tilde{F}_i : \ \ \ \sum_{\mathcal{R}_m, q_i} x_{\mathcal{R}_m, q_i}^{m,i} q_i E_{\mathcal{R}_m} = 0.
\end{split}
\end{equation}
If these three constraints are satisfied, the anomaly indeed takes the form of (\ref{6d perturbative anomaly polynomial}), where the 4-form $\tilde{X}_4^{\alpha}$ is given in terms of the curvatures by
\begin{equation}\label{eq:tX4}
	\tilde{X}_4^{\alpha} = \tfrac{1}{8} \ta^{\alpha} \text{tr} \tilde{R}^2 + \tfrac{1}{2} \sum_{m} b_m^{\alpha} \text{tr} \tilde{\mathcal{F}}_m^2 + \tfrac{1}{2}\sum_{i,j} b^{\alpha}_{ij} \tilde{F}^i \tilde{F}^j,
\end{equation}
for some $\mathbb{R}^{1,T}$ vectors of coefficients, $\tilde{a}^{\alpha}$ and $b^{\alpha}$.\footnote{All 6D quantities have tildes. Not to clutter the notation too much we do not put tildes on the anomaly polynomial coefficients, except for $\ta^{\alpha}$. In the reduction to 5D, $\ta^{\alpha}$ becomes a part of $a_I$, as discussed in subsection \ref{subsex:TI}.}
Comparison with the anomaly polynomial shows that the inner products of the anomaly coefficients satisfy the following anomaly cancellation relations (see e.g. \cite{Taylor:2011wt} and references therein): 
\begin{equation}
\begin{split}
	& \ \ \ \ \ \ \ \ \ \ \ \ \ \ \ \ \ \ \ \ \ \ \ \ \ \ (\text{tr}\tilde{R}^2)^2: \ \ \ \ \ \ \ \, \ta \cdot \ta = 9 - T, \\&
	\ \ \ \ \ \ \ \ \ \ \ \ \ \ \ \ \ \ \ \ \ \ \ \ \ \ \text{tr}\tilde{R}^2 \, \text{Tr} \tilde{\mathcal{F}}_m^2: \ \ \ \ta \cdot b_m = \tfrac{1}{6} \big[A_{\text{adj}_m} - \sum_{\mathcal{R}_m} x^m_{\mathcal{R}_m} A_{\mathcal{R}_m} \big], \\&
	 \ \ \ \ \ \ \ \ \ \ \ \ \ \ \ \ \ \ \ \ \ \ \ \ \ \ \text{tr} \tilde{R}^2 \tilde{F}_i \tilde{F}_j: \ \ \ \ \ \ta \cdot b_{ij} = - \tfrac{1}{6} \sum_{q_i q_j}x^{i,j}_{q_i, q_j} q_i q_j, \\&
	\ \ \ \ \ \ \ \ \ \ \ \ \ \ \ \ \ \ \ \ \ \ \ \ \ \ (\text{tr}\tilde{\mathcal{F}}_m^2)^2: \ \ \ \, \ \ \ b_m \cdot b_m = -\tfrac{1}{3}\big[C_{\text{adj}_m} - \sum_{\mathcal{R}_m} x^m_{\mathcal{R}_m} C_{\mathcal{R}_m}\big], \\&
	\ \ \ \ \ \ \ \ \ \ \ \ \ \ \ \ \ \ \ \ \ \ \ \ \ \ \text{tr}\tilde{\mathcal{F}}_m^2 \text{tr} \tilde{\mathcal{F}}_n^2: \ \ \ \ b_m \cdot b_n = \sum_{\mathcal{R}_m, \mathcal{S}_n} x^{m,n}_{\mathcal{R}_m, \mathcal{S}_n} A_{\mathcal{R}_m} A_{\mathcal{S}_n}  \ \ \  (m \neq n), \\&
	\ \ \ \ \ \ \ \ \ \ \ \ \ \ \ \ \ \ \ \ \ \ \ \ \ \ \text{tr}\tilde{\mathcal{F}}_m^2 \tilde{F}_i \tilde{F}_j: \ \ \ \ \, b_m \cdot b_{ij} = \sum_{\mathcal{R}_m q_i, q_j} x^{m,i,j}_{\mathcal{R}_m, q_i, q_j} q_i q_j A_{R_m}, \\&
	\ \ \ \ \ \ \ \ \ \ \ \ \ \ \ \ \ \ \ \ \ \ \ \ \ \ \tilde{F}_i \tilde{F}_j \tilde{F}_k \tilde{F}_l: \ \ \ \ \ b_{ijkl} = \sum_{q_i, q_j, q_k, q_l} x^{i, j, k, l}_{q_i, q_j, q_k, q_l} q_i q_j q_k q_l,
\end{split}
\end{equation}
where we have defined the symmetrized $b_{ijkl} = b_{ij} \cdot b_{kl} \, + \, b_{ik} \cdot b_{jl} + b_{il} \cdot b_{jk}$. One has that the inner-products corresponding to the non-Abelian factors are integers \cite{Kumar:2010ru}:
\begin{equation}
	\ta \cdot b_m, \ b_m \cdot b_n \in \mathbb{Z}  \ \ \ \ \ \forall m,n.
\end{equation}
A number of further important global anomaly constraints have been derived for the inner products of the vectors $\ta^{\alpha}$, $b^{\alpha}_m$, and $b^{\alpha}_{ij}$ \cite{Seiberg:2011dr, Monnier:2017oqd}, which we list below:
\begin{equation}
\begin{split}
	&\ta \cdot b_m + b_m \cdot b_m \in 2 \mathbb{Z},  \\&
	6 \ta \cdot b_{ij} \in \mathbb{Z}, \\&
	b_m \cdot b_{ij} \in \mathbb{Z}, \\&
	b_{ijkl} \in \mathbb{Z} \ \ \ \ \ \forall i,j,k,l, \\&
	b_{ii} \cdot b_{ii} + 2 \ta \cdot b_{ii} \in 4 \mathbb{Z}, \\&
	b_{ii} \cdot b_{ij} + 2 \ta \cdot b_{ij} \in 2\mathbb{Z}, \\&
	b_{ii} \cdot b_{jj} + 2 b_{ij}\cdot b_{ij} + 6\ta \cdot b_{ij} \in 2\mathbb{Z}.
        \label{3.13}
\end{split}
\end{equation}
There is the further constraint
\begin{equation}
	b_{m} \cdot b_{ii} \in 2 \mathbb{Z} \text{  for  } \mathcal{G}_m = SU(N\geq3),\
	 SO(N \geq 6), \  G_2, \ F_4, \ E_{6,7,8}.
     \label{3.14}
\end{equation}
The analogous constraint for $\mathcal{G}_m = Sp(2N)$ holds if the generalized completeness hypothesis is assumed. Finally, using string charge quantization and the generalized completeness hypothesis, one finds
\begin{equation}
	\ta, \, b_i, \, \tfrac{1}{2}b_{ii}, \, b_{ij} \in \Lambda_S.
        \label{3.15}
\end{equation}
Moreover, a necessary condition for a construction of the Green-Schwarz term in topologically non-trivial backgrounds is that $\tilde{a}$ is characteristic \cite{Monnier:2018nfs}, i.e.
\begin{equation}
	\ta \cdot x + x \cdot x \in 2 \mathbb{Z} \ \ \,\,\,\,  \forall x \in \Lambda_S.
        \label{3.16}
\end{equation}
In this case, the conditions \eqref{3.13} and \eqref{3.14} actually follow directly from \eqref{3.15} and \eqref{3.16}. For 6D $(1,0)$ supergravity with $U(1)$ gauge groups discussed later in this paper, we assume the string charge quantization condition \eqref{3.15}.

\paragraph{Some remarks about GS mechanism} 
\begin{itemize}
    \item The anomaly contribution from GS terms can be characterized {as a mixed anomaly between electric and magnetic higher-form symmetries of a  bulk theory in one dimension higher (with boundary given by the spacetime of the theory under consideration) \cite{Gaiotto:2014kfa, Hsieh:2020jpj}.}
    \item    The (anti-)self-duality of tensor fields in the CS terms leads to additional complications \cite{Monnier:2018nfs}. The full treatment usually requires a quadratic refinement to, roughly speaking, take care of the factor of $\frac{1}{2}$ in (\ref{6d GS term}).
    \item In theories with $\bZ_k$ symmetries the $\tilde{X}_4^{\alpha}$ factors in the anomaly polynomial can be modified and shifted by a topological torsion class \cite{Monnier:2018nfs}.  This possibility is related   to the fact that $H^4(B\bZ_k;\bZ) = \bZ_k$, and will play an important role in our considerations, notably in subsection \ref{subsec:TGS}.
    \item The anomalous $B$-field transformations \eqref{anomB} lead to a modified Bianchi identity: 
    \begin{equation}
        \label{GS Bianchi identity}
        \dd \tilde{H}_3^{\alpha} = \tilde{X}_4^{\alpha}(\tilde{R}, \mathcal{\tilde{F}}_m, \tilde{F^i}),
    \end{equation}
    which imposes constraints on the gauge bundle and tangential structures of spacetime manifolds -- i.e. $\tilde{X}_4$ has to be trivial with a trivialization $\tilde{X}_4 = \text{d} \tilde{H}_3$. The resulting tangential structure of the spacetime manifold is called a (twisted) string structure. The relevant bordism group should be $\Omega_{*-\text{dim}}^{\text{(twisted) string}} (BG)$. See \cite{Lee:2020ewl} for relevant discussions. In this paper, we will not consider (twisted) string tangential structures. 
\end{itemize}

\subsubsection{Circle reduction} \label{sec:circle}
The study of the relations between 6D $\mathcal{N} = (1,0)$ theories and 5D minimal supergravity, and notably GS vs. CS terms, goes back to the early days of F-theory \cite{Morrison:1996na,Morrison:1996pp,Ferrara:1996wv, Intriligator:1997pq} and continues up through more recent works \cite{Bonetti:2011mw, Bonetti:2012fn, Bonetti:2013cza} and \cite{Corvilain:2017luj, Corvilain:2020tfb}.
Consider the 6D spacetime to be a circle-fibered manifold, $U(1) \hookrightarrow \tilde{M}_6 \stackrel{\pi}{\longrightarrow} M_5$, with globally defined one-form $e$ whose exterior derivative is the pullback of an element of the second integral cohomology of the 5D base: $\text{d}e = \pi^*T,$ with $T \in H^2(M_5,\mathbb{Z})$.\footnote{In the context of reducing characteristic classes, we use $T$ for the curvature of the circle fibration (the graviphoton field-strength). By raising indices, it appears also as a one-form and zero-form (see Appendix \ref{app:reduce}). $F^0 = \dd A^0$ always appears as a two-form.} 

For comparison with 5D the loop calculations, it is worthwhile to keep track of the possible ambiguities.
Note that ambiguity in the descent procedure coming from the anti-derivative $\tilde{X}_3^{(0)} = \text{d}^{-1}[\tilde{X}_4]$ results in the ambiguity in the transformation of $\tB$ which is present already in 6D.  Clearly one can add any closed form to the definition of $\tilde{X}_3^{(0)}$ without modifying $\tilde{X}_4$. Similarly, $\tilde{X}_2^{(1)} = \text{d}^{-1}[\delta \tilde{X}_3^{(0)}]$ is ambiguous.

We now consider the dimensional reduction.
The first step is to understand how $\tX_4$ (we suppress the index $\alpha$ here) reduces. In the circle fibered geometry, it can be decomposed as $\tilde{X}_4 = X_4 + X_3 \wedge e$. 
As noted above, symbols with tildes will represent fields on the total space, whereas those without will represent fields on the base space. The pull-back signs have been suppressed here; the explicit expressions will be given in Appendix \ref{app:reduce}. The characteristic class $\tilde{X}_4$ is closed, which translates to the lower-dimensional relations
\begin{equation}\label{eq:h-v}
	\text{d} X_4 - X_3 T = 0, \ \ \ \ \ \ \ \ \ \ \ \text{d} X_3 = 0.
\end{equation}
Notably, $X_4$ is not closed. On the other hand, decomposing the relation $\text{d} \tilde{X}_3^{(0)} = \tilde{X}_4$, gives the equations
\begin{equation}
	X_4 = \text{d} X_3^{(0)} + X_2^{(0)} T, \ \ \ \ \ \ \ \  X_3 = \text{d}X_2^{(0)}. \label{2.6}
\end{equation}
The combination $[X_4 - X_2^{(0)}T]$ is closed, and $X_4$ can be written  as
\begin{equation}
	\tilde{X}_4 = \big[X_4 - X_2^{(0)}T\big] + \text{d}\big[X_2^{(0)} \wedge e\big].
\end{equation}
There is, in fact, a unique choice of $X_2^{(0)}$ such that $[X_4 - X_2^{(0)}T]$ is globally well-defined and is gauge-invariant under 5D local gauge transformations and diffeomorphisms \cite{Liu:2013dna}. We denote this specific $X_2^{(0)}$ by $\mathfrak{X}_2$, as it plays an important role in the following. We see from equation (\ref{2.6}) that any other choice of $X_2^{(0)}$ will differ from $\mathfrak{X}_2$ by a closed term. We will write this closed term locally as $\text{d} \text{x}_1$. A generic $\tilde{X}_3^{(0)}$ is given by
\begin{equation}\label{eq:defX2}
	\tilde{X}_3^{(0)} = X_3^{(0)} + (\mathfrak{X}_2 + \text{d} \text{x}_1) \wedge e,
\end{equation}
where $X_3^{(0)}$ is defined by the equation
\begin{equation}
	X_3^{(0)} = \text{d}^{-1}[X_4 - \mathfrak{X}_2 T - \text{d} \text{x}_1 T] \,.
\end{equation}
There is still some remaining ambiguity, since, once again, any closed form can be added to a given choice of $X_3^{(0)}$. We leave this additional ambiguity implicit in the following. 

There is an important subtlety about the reduction of gauge transformations with non-trivial Wilson lines. We can discuss this by simply examining the abelian fields in the theory. Consider the following explicit decomposition of a 6D abelian gauge field:
\begin{equation}
	\tilde{A} = A + \phi e = A + \phi (\text{d}\theta + A^0),
\end{equation}
where a factor of the circle radius  has been absorbed in the definition of $\phi$. The reduction of a 6D gauge transformations $\delta_{\tilde{\epsilon}}\tilde{A} = \text{d} \tilde{\epsilon}$ gives rise to two different types of 5D gauge transformations, which we refer to as 5D local gauge transformations, $\delta_{\epsilon}$, and 5D large gauge transformations (LGTs), $\delta_{\text{n}}$. For a general gauge transformation combining both types, we use $\delta_{\tilde{\epsilon}}$. A general 6D gauge transformation reduces to a 5D local gauge transformation of the vector field $A$, as well as a discrete integer shift of the scalar mode (accompanied by a compensating shift of $A$):
\begin{equation}
	\delta_{\tilde{\epsilon}} A = \text{d} \epsilon - \text{n} A^0, \ \ \ \ \ \ \ \ \delta_{\tilde{\epsilon}} \phi = \text{n}.
\end{equation}
In the following formal discussion we do not specify which type of gauge transformation we are performing. Now we examine the dimensional reduction of the variation of the $B$-field:
\begin{equation}
	\delta B_2 =  \text{d} \lambda_1 + \lambda_0 T -X_2^{(1)} , \ \ \ \ \ \ \delta B_1 =  \text{d}\lambda_0-X_1^{(1)}.
\end{equation}
We see that the lower-dimensional one-form $B_1$ does not transform as a bona fide vector field for a generic choice of anti-derivative $\tilde{X}_3^{(0)}$. One can either choose $\tilde{X}_2^{(1)}$ in the descent, such that its vertical component, $X_1^{(1)}$, is closed, or one must perform a shift of $B_1$ to find an object with the correct transformation properties in 5D. By reducing the descent equations, we know that $\text{d} X_1^{(1)} = \delta \mathfrak{X}_2 + \text{d} \delta \text{x}_1$. For local gauge transformations of the 5D field-strengths, $\delta_{\epsilon} \mathfrak{X}_2 = 0$, and so $\delta_{\epsilon} \text{x}_1 = X_1^{(1)}$ up to a closed term. For any choice of the anti-derivative,
\begin{equation}\label{eq:newA1}
	A_1 \equiv B_1 + \text{x}_1,
\end{equation}
transforms as a normal 5D vector field. The associated gauge-invariant 5D field-strength is $F_2 \equiv \text{d}A_1 = \text{d}[B_1 + \text{x}_1]$. Now we examine the reduction of the gauge-invariant 6D field strength:
\begin{equation}
\label{H3}
	H_3 = \text{d} B_2 - B_1 T + X_3^{(0)}, \ \ \ \ \ \ \ H_2 = \text{d} B_1 + X_2^{(0)}.  
\end{equation}
By substitution, we see that $H_2$ is indeed gauge-invariant under $\delta_{\epsilon}$:
\begin{equation}
\label{H2}
	H_2 = F_2 - \text{d} \text{x}_1 + X_2^{(0)} = F_2 + \mathfrak{X}_2.
\end{equation}

\noindent 
Now we turn to the reduction of the Green-Schwarz term. It becomes
\begin{equation}
	S_{\text{GS}} = - \tfrac{1}{2}\int \text{d}\tilde{B}_2 \cdot \tilde{X}_3^{(0)} = - \tfrac{1}{2}\int (\tilde{H}_3 - \tilde{X}_3^{(0)}) \cdot \tilde{X}_3^{(0)}.
\end{equation}
$\tilde{X}_3^{(0)} \cdot \tilde{X}_3^{(0)} = 0$ by the symmetry of the dot-product combined with the anti-symmetry of the wedge product, and contracting with the isometry vector, we find
\begin{equation}
-\iota_v (S_{\text{GS}}) = \tfrac{1}{2}\int H_3 \cdot (\mathfrak{X}_2 + \text{d}\text{x}_1) - H_2 \cdot {X}^{(0)}_3.
\end{equation}
Using \eqref{H2} and \eqref{H3} and integrating out $B_2$ \cite{Bonetti:2013cza}, one finds the following top-form 5D couplings:
\begin{equation}
\label{NI}
   -\tfrac{1}{2}\int \text{x}_1 \cdot [X_4 - \mathfrak{X}_2 T] - \mathfrak{X}_2 \cdot X^{(0)}_3 - 2 A_1 \cdot [ X_4 - \mathfrak{X}_2 T]  + A^0 F_2 \cdot F_2.
\end{equation}
The first two terms are not invariant under 5D local transformations $\delta_{\epsilon}$ and must clearly be canceled in order to have a consistent theory. The latter terms are invariant under 5D gauge and local Lorentz transformations; however, they do vary under $\delta_{\text{n}}$. We will return to this point below.

\paragraph{Comments on reduction of heterotic-like BI} The existence of a locally invariant two-form $\mathfrak{X}_2$ is crucial for a seemingly trivial fact that a circle reduction of a D-dimensional supergravity with 16 supercharges  yields a (D$-1$)-dimensional supergravity. The former has a duality group $O(10-\text{D}, N)$ which acts on the vector fields in the theory, $(10-\text{D})$ of which reside in the gravity multiplet and $N$ of which reside in the vector multiplets. The coupling of the Yang-Mills multiplets to gravity necessitates a Bianchi identity of the form
\begin{equation}
    \dd \tilde{H}_3  = \tilde{X}_4 \qquad \Rightarrow \qquad \d_{\tilde{\epsilon}} \tilde{B}_2 = -\tilde{X}_2^{(1)}.
\end{equation}
Depending on the details of the D-dimensional theory, $\tilde{X}_4$ may or may not contain a gravitational contribution, but always has a pure gauge part proportional to $\tilde{F}^2$.\footnote{At the classical level $N$ is not fixed, but is, in general, subject to consistency constraints.} 

The reduction of this theory is expected to produce a 16 supercharge supergravity in one dimension lower, with a symmetry group $O(11-\text{D}, N+1)$ acting on $(12 - \text{D} +N$) vectors. The two new vector fields come from the off-diagonal part of the D-dimensional metric and the $\tilde{B}_2$ field. However, $\tilde{B}_2$ with one leg on the circle is not a vector for a generic choice of $\tilde{X}_2^{(1)}$. $A_1$ defined in \eqref{eq:newA1} is. A shift by $\text{x}_1$ is needed in order to obtain a vector field ($\d_{\epsilon} A_1 = \dd \epsilon$). Note that this vector is not invariant under the large gauge transformations:
\begin{equation}
    \delta_{\text{n}} F_2 = - \delta_{\text{n}} \mathfrak{X}_2 = - 2\text{n}F + \text{n}^2  T.
\end{equation}
This transformation is, however, within the $O(11- \text{D}, N+1)$ lattice and is given by the so-called Wilson line shift \cite{Fraiman:2018ebo, Israel:2023tjw}. 

\section{Reduction of perturbative anomaly}
\label{sec:pert_an}
In this section, we discuss anomalies at the perturbative level and their dimensional reduction and the eta invariant as a means to compute lower dimensional CS coefficients, as well as regularization ambiguities and cancellation of the anomaly in LGTs. We explain how this cancellation allows us to extract (most of) the CS coefficients and show that most of the 5D coefficients obtained via one-loop calculation seem to fail the 5D consistency conditions (\ref{5d consistency condition}). The solution to this puzzle will be presented in Section \ref{sec:ano-NP}.
\subsection{Circle reduction of anomalies}\label{sec:cirle2} 
The 6D (sum-factorized) anomaly can be seen by inflow from a 7D term:
\begin{equation}
	S_{\mathcal{A}_7} = \int_{M_7} \tilde{\mathcal{A}}_7 = \tfrac{1}{2}\int_{M_7} \tilde{X}_3^{(0)} \cdot \tilde{X}_4,
\end{equation}
where $\partial M_7 = M_6$. 
Its variation 
\begin{equation}
\delta S_{\mathcal{A}_7} =  \tfrac{1}{2}\int_{M_7} \text{d} \tilde{X}_2^{(1)} \cdot \tilde{X}_4 = \tfrac{1}{2}\int_{M_6} \tilde{X}_2^{(1)} \cdot \tilde{X}_4,
\end{equation}
is the 6D (consistent) anomaly, that is canceled by GS terms. Assuming that there is a circle isometry (generated by a vector $v$) and dimensionally reducing $\tilde{\mathcal{A}}_7 = \tfrac{1}{2} \tilde{X}_3^{(0)} \cdot \tilde{X}_4$ directly gives
\begin{equation}
\label{contracted 7 form}
	\iota_v \tilde{\mathcal{A}}_7 =  \tfrac{1}{2}\iota_v [\tilde{X}_3^{(0)} \cdot \tilde{X}_4] = \tfrac{1}{2} \big[ X_3^{(0)} \cdot X_3 + X_2^{(0)} \cdot X_4 \big],
\end{equation}
which can be further rewritten as
\begin{equation}\label{eq:isoD}
	\iota_v \tilde{\mathcal{A}}_7 = \mathfrak{D}_6 + \text{d} \Delta_5,
\end{equation}
with
\begin{equation}
\label{NI2}
	\Delta_5 = \tfrac{1}{2}\big[-X_3^{(0)}\cdot \mathfrak{X}_2 + \text{x}_1 \cdot  [X_4 -  \mathfrak{X}_2 T]\big], \ \ \ \ \ \ \ \ \ \mathfrak{D}_6 = X_4 \cdot \mathfrak{X}_2 - \tfrac{1}{2}\mathfrak{X}_2^2 T.
\end{equation}
Clearly $\text{d}\Delta_5$ descends via Stokes' theorem to 5D and gives the negative of the local non-invariant terms found via circle reduction of the GS term which were derived in the last section. $\mathfrak{D}_6$ comprises  two terms that are invariant under local gauge transformations but not under large gauge transformations.
While $\mathfrak{D}_6$ is not a total derivative, its variation under LGTs is, and it is not hard to check that $\text{d}^{-1}[\delta_{\text{n}} \mathfrak{D}_6]$ is indeed equal to the negative of the variation of the final two terms in \eqref{NI} coming from the reduction of the GS term.

 General study of anomaly reduction \cite{Corvilain:2017luj, Cheng:2021zjh, Corvilain:2020tfb} for $\tX_4 = \tF \wedge \tF$ (for abelian $\tF$) suggests that the terms not invariant under $\delta_{\epsilon}$ in \eqref{NI2} are generated at one-loop by summing over the massive KK tower after circle reduction and that these terms should be canceled by the reduction of the GS term. The one-loop three-vector CS coefficients, including non-invariant terms, were calculated in \cite{Corvilain:2020tfb}, and it was confirmed that they do indeed cancel the reduction of the GS term if the 5D loop calculation is performed by means of a reduced 6D Pauli-Villars (PV) regulator. The cancellation of these non-invariant terms is expected, as the 6D $\mathcal{N} = (1,0)$ theory is free of perturbative anomalies with the inclusion of GS terms.
\paragraph{Remark} As noted in \cite{Corvilain:2017luj} and \cite{Corvilain:2020tfb}, the non-invariant terms calculated using zeta function regularization and the dimensionally-reduced PV regulator are not the same, and only the reduced 6D PV regulator correctly cancels the reduction of the GS term. We note that this discrepancy is a result of the specific choice of anti-derivative made in the definition of the GS term, namely that of the usual CS term $\tilde{\omega}_3^{(0)} = \tilde{A}\tilde{F}$. For such a term, $\omega_2^{(0)} = \mathfrak{X}_2 + \text{d}\text{x}_1$ with $\text{x}_1 = -\phi A$.  The result found via zeta function regularization, which we will use in the following, however, corresponds to a different choice of anti-derivative, namely one which is manifestly invariant under LGTs: $X_2^{(0)} = \mathfrak{X}_2 - \text{d}[2\phi A + \phi^2 A^0] = -2 \text{d}\phi(A + \phi A^0)$. We find that $\Delta_5(\tilde{F})$, calculated using $\text{x}_1 = -2 \phi A - \phi^2 A^0$, matches the non-invariant terms found in \cite{Corvilain:2020tfb}. The full expressions for multiple $U(1)$ factors are given in Appendix \ref{app:reduce}.
 
 Now we focus on the effect of large gauge transformations. The 6D theory is invariant also under these transformations, so the full 5D theory must be as well. As noted above, the reduction of the GS term generates CS couplings which are invariant under local gauge transformations, but which vary under LGTs. This variation under $\delta_{\text{n}}$ must also be canceled by one-loop effects, and as can be verified from the explicit expressions in \cite{Corvilain:2020tfb}, the floor-function terms in the CS coefficients cancel the anomaly in LGTs. In fact, requiring the good CS couplings to cancel the variation under LGTs of the reduction of the GS term allows one to constrain the coefficients which involve the abelian fields.
\begin{itemize}
    \item LGT of the GS term after circle compactification: Since it is only the abelian fields which vary under the LGTs, it is useful to introduce the following splitting of the formal $X$'s:
\begin{equation}
	\tilde{X}^{\alpha}_4 = \tilde{X}^{\alpha}_4 (\tilde{R}, \tilde{\mathcal{F}}) + \tfrac{1}{2}b^{\alpha}_{ij} \tilde{F}^i \tilde{F}^j.
\end{equation}
Similarly
\begin{equation}
	\mathfrak{X}^{\alpha}_2 = \mathfrak{X}^{\alpha}_2(\tilde{R},\tilde{\mathcal{F}}) + \tfrac{1}{2}b^{\alpha}_{ij} [2\phi^i F^j T + \phi^i \phi^j T^2].
\end{equation}
The variation of $\mathfrak{D}_6$ under the LGTs is
\begin{equation}
\label{LGT OF D6 1st}
	\delta_{\text{n}}\mathfrak{D}_6 = \big\{ [X_4  - \mathfrak{X}_2 T] - \tfrac{1}{2}\delta_{\text{n}}[ \mathfrak{X}_2(F)]T \big\} \cdot \delta_{\text{n}} [\mathfrak{X}_2 (F)],
\end{equation}
where
\begin{equation}
\label{LGT of X2}
	\delta_{\text{n}} [\mathfrak{X}_2(F)]  = \tfrac{1}{2} b_{ij} [ 2 \text{n}^i F^j  - \text{n}^i \text{n}^j T].
\end{equation}
Substituting \eqref{LGT of X2} into \eqref{LGT OF D6 1st} gives
\begin{equation}
\begin{split}
	\delta_{\text{n}} \mathfrak{D}_6 \ \ = \ & \tfrac{1}{16}(\ta \cdot b_{kl}) \big[ \text{tr} R^2 + \text{tr} [RT^2_1] + \text{tr}[\nabla T_1 \nabla T_1 ] \big][2 \text{n}^k F^l - \text{n}^k \text{n}^l T] \\& \, \tfrac{1}{4} (b_m \cdot b_{kl}) \sum_m \text{tr} \mathcal{F}^2_m [2 \text{n}^k F^l - \text{n}^k \text{n}^l T]  \\&
	\tfrac{1}{24} b_{ijkl} \big[4 \text{n}^i F^j F^k F^l - 6 \text{n}^i \text{n}^j F^k F^l T + 4 \text{n}^i \text{n}^j \text{n}^k F^l T - \text{n}^i \text{n}^j \text{n}^k \text{n}^l T^3 \big]. \label{LGT of D6 2nd}
\end{split}
\end{equation}
Clearly these can be written as CS terms on the 5D boundary, e.g. the term $ F^j F^k F^l$ in \eqref{LGT of D6 2nd} gives $A^jF^kF^l$.
 \item LGT of one-loop generated CS terms: 
  the one-loop generated 5D CS terms (with KK towers of fermions running in the loop \cite{Bonetti:2013ela}) can be found in \cite{Corvilain:2017luj} and \cite{Corvilain:2020tfb} for theories with purely abelian fields. In those works, the coefficients were found via one-loop computation. However, just requiring the one-loop terms to cancel the variation of the reduced GS terms, $-\text{d}^{-1}[\delta_{\text{n}} \mathfrak{D}_6]$ allows to determine the coefficients except for the constant part of $k_{0mm}$, $k_{0RR}$, and $k_{000}$:
\begin{equation}
\begin{split}
\label{one loop U(1) CS}
	S_{\text{CS}} = & \ \ \, \tfrac{1}{96}\big[  k_{iRR} A^i + k_{0RR} A^0\big] \big(\text{tr} R^2 + \text{tr}[RT_1^2] + \text{tr}[\nabla T_1 \nabla T_1] \big) \\& - \tfrac{1}{4}\big[ k_{imm} A^i + k_{0mm} A^0 \big] \sum_m \text{Tr}_{\mathcal{R}} \mathcal{F}_m^2  \\&
	-\tfrac{1}{6}\big[ k_{ijk} A^iF^jF^k + 3k_{ij0} A^iF^jT + 3k_{i00} A^iT^2 + k_{000} A^{0} T^2 \big],
\end{split}
\end{equation}
with coefficients (up to a constant)
\begin{equation}
\begin{split}
\label{kgrav}
	&k_{iRR} = -\sum_{\text{h}} q^{\text{h}}_i\big(-1 + 2\lfloor \mu^{\text{h}} \rfloor \big), \\&
    k_{0RR} = -\sum_{\text{h}} \lfloor \mu^{\text{h}} \rfloor \big(\lfloor \mu^{\text{h}} \rfloor - 1 \big) + a^{6\text{D}}_0,\\& 
     k_{imm} = -\sum_{\text{h}} q^{\text{h}}_i\big(1 + 2\lfloor \mu^{\text{h}} \rfloor \big), \\&
    k_{0mm} =  -\sum_{\text{h}} \lfloor \mu^{\text{h}} \rfloor \big(\lfloor \mu^{\text{h}} \rfloor + 1 \big) + b^m_0,
\end{split}
\end{equation}
where $\mu^{\text{h}} = \sum_i q^{\text{h}}_i \phi^i$, $\lfloor {\mu}^{\text{h}} \rfloor$ is the floor function of $\mu^{\text{h}}$, and the index $\text{h}$ runs over all of the hyperini. The purely abelian fields' CS coefficients are given by\footnote{assuming the perturbative anomaly cancellation and ignoring the gauge non-invariant terms.}
\begin{equation}
\begin{split}
\label{kgauge}
	&k_{ijk} = -\sum_{\text{h}} q^{\text{h}}_i q^{\text{h}}_j q^{\text{h}}_k \big[\tfrac{1}{2} + \lfloor \mu^{\text{h}} \rfloor \big], \\&
	k_{ij0} =  -\sum_{\text{h}} q^{\text{h}}_i q^{\text{h}}_j \big[\tfrac{1}{12} + \tfrac{1}{2}\lfloor \mu^{\text{h}} \rfloor ( \lfloor {\mu}^{\text{h}} \rfloor + 1) \big], \\&
	k_{i00} =  -\tfrac{1}{6}\sum_{\text{h}} q^{\text{h}}_i \lfloor \mu^{\text{h}} \rfloor ( \lfloor \mu^{\text{h}} \rfloor + 1)  \rfloor ( 2\lfloor \mu^{\text{h}} \rfloor + 1), \\&
	k_{000}  = -\tfrac{1}{4} \sum_{\text{h}} \lfloor \mu^{\text{h}} \rfloor^2  \big(\lfloor \mu^{\text{h}} \rfloor + 1 \big)^2 + c_{000}.
\end{split}
\end{equation}
\end{itemize}
Requiring the cancellation fixes the coefficients $k$, except for the constant terms $a^{\text{6D}}_0$, $b^m_0$, and $c_{000}$. It can be checked explicitly that the LGTs of \eqref{one loop U(1) CS} cancel the LGTs of the final two terms in \eqref{NI}.  Thanks to this fact, we can set $\lfloor {\mu}^{\text{h}} \rfloor = 0$, as is commonly done in the literature, and we make this choice in the following.\footnote{The functions of $\lfloor \mu^{\text{h}} \rfloor$ appearing in  \eqref{kgauge} as well as in $k_{0mm}$ and $k_{imm}$ agree with the expressions that come from regularizing the loop integrals. Two different solutions  canceling \eqref{LGT OF D6 1st} are allowed for $k_{0RR}$ and $k_{iRR}$. Eq. \eqref{kgrav} gives the form that after fixing LGTs agrees with the loop calculations and leads to correct 5D consistency conditions, as discussed in Sections \ref{subsex:TI} and \ref{sec:ano-NP}. }

\paragraph{Isometric descent} The 5D non-invariant terms in the reduction of the anomaly \eqref{NI2} 
\begin{equation}\label{eq:redGS-total}
	\Delta_5 = \text{d}^{-1} \big[\iota_v \text{d}^{-1} \tilde{I}_{\text{D}+2} - \mathfrak{D}_6 \big],
\end{equation}
are generated in 5D by integrating out massive KK modes of the chiral (anomalous) 6D fields running in the loop. Hence it is convenient to rewrite it as a sum of individual terms. 

The reduction of the anomaly along an isometry is given by a variation of a local coupling. This reduction is given by 
\begin{equation}
\label{reduce}
 I^{(1)}_{\text{D}-1} = 	-\iota_v \text{d}^{-1} \delta_{\tilde{\epsilon}} \text{d}^{-1} \tilde{I}_{\text{D}+2} =\text{d}^{-1} \delta_{\tilde{\epsilon}} \iota_v \text{d}^{-1} \tilde{I}_{\text{D} + 2}.
\end{equation}
The first equality is a direct statement of the horizontal-vertical decomposition of the anomaly ($
	\text{d}^{-1} \delta_{\tilde{\epsilon}} \text{d}^{-1} \tilde{I}_{\text{D}+2} = \tilde{I}_{\text{D}}^{(1)} = I^{(1)}_{\text{D}} + I^{(1)}_{\text{D}-1}\wedge e$), while the second follows from $v$ being an isometry generator (Lie derivative w.r.t. $v$ annihilates everything). The equality holds for the specific choice of $\delta_{\tilde{\epsilon}} = \delta_{\epsilon}$. For this choice, however, $\mathfrak{D}_6$ is annihilated, and the RHS of \eqref{reduce} is just $\Delta_5$. Since the operators $\delta_{\tilde{\epsilon}}, \iota_v,$ and $\text{d}^{-1}$ all act linearly on $\tilde{I}_{\text{D}+2}$,  splitting the total anomaly into the sum of the components coming from each type of chiral field, we find
\begin{equation}\label{eq:Delta-total}
	\Delta_5 (\tilde{R},\tilde{\mathcal{F}},\tilde{F}) = (H - V + T) \Delta^{\psi}_5 (\tilde{R})- \Delta_5^{\psi_{\mu}} + (T-1)\Delta^{B_{\mu \nu}}_5 + \Delta_5^{\psi}(\tilde{R},\mathcal{\tilde{F}},\tilde{F}),
\end{equation}
where $\Delta_5^{\Phi}$ for the individual fields $\Phi = (\psi, \psi_{\mu}, B_{\mu \nu})$ are the isometric descendants of the respective anomalies:
\begin{equation}\label{eq:deltaPhi}
	\Delta_5^{\Phi} = \text{d}^{-1} [\iota_v \text{d}^{-1}\tilde{I}_8^{\Phi}  - \mathfrak{D}^{\Phi}_6 ] \,,
\end{equation}
and where we have split the spin-$\tfrac{1}{2}$ fermion anomaly into a pure-gravitational portion and a portion which depends on the gauge curvatures. When the 6D field content is such that the anomaly $\tilde{I}_6^{(1)}$ factorizes, so do the 5D non-invariant terms, and we recover
$$\Delta_5 = \tfrac{1}{2}\big[-X_3^{(0)}\cdot \mathfrak{X}_2 + \text{x}_1 \cdot  [X_4 -  \mathfrak{X}_2 T]\big] \,.$$
Note that   the appearance of $ \text{x}_1 $ in the non-invariant terms is non-physical -- in the anomaly-free theories, these cancel. It is associated with the ambiguities in taking inverse derivatives, and is matched by ambiguities in the choices of regularization schemes in the loop calculations. 

Note that this situation is not just specific to 6D. In the 10D type-IIB theory, the individual anomaly contributions sum up to zero, and hence no 10D counterterms are needed and no non-invariant terms arise after circle reduction. Hence, the loop contributions from integrating out the massive modes of the individual chiral fields should also sum up to zero. Similarly in the circle reduction of the 10D heterotic string, the GS term gives rise to 9D top-form couplings that are not invariant. They are canceled by integrating out the massive KK modes of chiral fields, provided their individual contributions agree with the nine-forms $\Delta_9^{\Phi}$ obtained via isometric descent.

\subsection{$\eta$ invariant and 5D topological couplings}
In 6D $\mathcal{N} = (1,0)$ supergravity, the phase contribution to the partition function from chiral fields with fixed background (i.e. gauge bundle $A_G,F_G$, metric $g$, etc.) is given by $\exp\big\{2\pi i \, \eta_{7\text{D}}(A_G,F_G, g,...)\big\}$ (see Section \ref{sec:AT}), the exponentiated $\eta$ invariant of the 7D bulk whose boundary is the 6D spacetime (together with the associated tangential and gauge bundle structures). The anomaly free condition becomes
\begin{equation}
    \label{6d anomaly free condtion from DF}
    \eta_{7\text{D}, \text{grav}} + \eta_{7\text{D}, V} +\eta_{7\text{D}, H} +\eta_{7\text{D}, T} + \cA_{\text{GS}} = 0 \,\,\,\, {\text{mod}} \  \bZ,  
\end{equation}
for any 7D closed spin manifold with associated structures. In (\ref{6d anomaly free condtion from DF}),
\begin{itemize}
    \item $\eta_{7\text{D}, \text{grav}}$ corresponds to the symplectic Majorana-Weyl (SMW) gravitini $\p^{+}_{\m}$ in the gravity multiplet and the self-dual tensor $B_{\m\n}^{+}$.
    \item $\eta_{7\text{D}, V}$ corresponds to the SMW gaugini $\lambda^{+}$ in the vector multiplets.
    \item $\eta_{7\text{D}, H}$ corresponds to the SMW fermions $\p^{-}$ in the hypermultiplets.
    \item $\eta_{7\text{D},T}$ corresponds to the SMW fermions $\chi^{-}$ and the anti-self-dual tensors $B_{\m\n}^{-}$ in the tensor multiplets.
    \item $\cA_{\text{GS}}$ denotes the contribution from the $\text{GS}$ terms in the 6D theory, as reviewed in Section \ref{sec:AT}.
    \end{itemize}
    
Compactifying a 6D supergravity theory on $S^1$ gives a 5D theory with a canonical KK $U(1)$ vector $A^0$, with field strength $F^0$. Hence we are in the situation discussed in subsection \ref{Physical interpretation of other bordism groups}. As mentioned there, 5D theories have WZW-like topological terms involving the canonical $U(1)$, i.e. 
\begin{equation}
\begin{split}
    \label{5d CS term for KK U(1)}
  &\int_{M_5} -\frac{C_{000}}{6} A^{0} \wedge F^{0} \wedge F^{0}   + \frac{a_0}{96}A^0 \wedge \tr R^2  + ... \\
  =&\int_{M_6} - \frac{C_{000}}{3!}  F^0 \wedge F^0 \wedge F^0   - \frac{a_0}{48}F^0 \wedge p_1(M_6) + ...,
\end{split}
\end{equation}
where $\partial M_6 = M_5$. These terms are subject to the quantization condition\footnote{Note that \eqref{5d CS term for KK U(1)} are formally the same as \eqref{eq:pure5d+}. Eq. \eqref{5d CS term for KK U(1)} singles out one $U(1)$ field out of $n_V +1$, and the ellipsis stands for the missing CS terms involving other vectors. \eqref{eq:pure5d+} has $n_V= 0$.}
\begin{equation}
    \label{quantization condition for 5d CS term for KK U(1)}
  \int_{M_6} - \frac{C_{000}}{3!}  F^0 \wedge F^0 \wedge F^0   - \frac{a_0}{48}F^0 \wedge p_1(M_6) \in \bZ \ \ \Rightarrow \ \ C_{000} + \frac{a_0}{2} \in 6\bZ,  
\end{equation}
for any closed 6D spin manifold $M_6$.

A natural question is how the $S^1$ compactification relates the anomaly-free condition (\ref{6d anomaly free condtion from DF}) to the quantization condition (\ref{quantization condition for 5d CS term for KK U(1)}).
Can the local coupling \eqref{5d CS term for KK U(1)} arise from the geometric $\eta$ invariant \eqref{6d anomaly free condtion from DF}? This will be the topic of Section \ref{sec:ano-NP}. In the current subsection, we discuss the first step of this potential relation: the derivation of the topological terms (\ref{5d CS term for KK U(1)}) from (\ref{6d anomaly free condtion from DF}) via the compactification map $\varphi_{S^1}$.\footnote{This is not specific to reduction from 6D to 5D. Gravitational top-form couplings are generated in the circle compactifications of type IIB strings by the string winding contributions \cite{Antoniadis:1997eg,Liu:2010gz}. This reduction using the adiabatic $\eta$ invariant has been discussed in \cite{Dierigl:2022reg}. Differently from 5D, pure KK couplings $A^0 \wedge (F^0)^4$ in 9D are not generated, since any four-derivative coupling would break supersymmetry. }

Ref. \cite{Bonetti:2013ela} already gives the term (\ref{5d CS term for KK U(1)}) from $S^1$ compactification via one-loop calculation that sums over the massive KK modes of the 6D anomalous fields running in the loop. Just like the phase contribution to the 6D partition function from the chiral fields is given by the $\eta$ invariant $\eta_{7\text{D}, \text{grav}} + \eta_{7\text{D}, V} +\eta_{7\text{D}, H} +\eta_{7\text{D}, T}$ after $S^1$ compactification, it is the topological term (\ref{5d CS term for KK U(1)}) that, after $S^1$ compactification, contributes to the phase of the effective 5D theory (as (\ref{5d CS term for KK U(1)}) contains one time-derivative).  For the discussion of the relation between (\ref{6d anomaly free condtion from DF}) and (\ref{quantization condition for 5d CS term for KK U(1)}) in the next section,  it is useful to show how the results of \cite{Bonetti:2013ela} can be obtained from the $\eta$ invariant.\footnote{At the perturbative level, it can be verified directly that the 6D local GS terms (\ref{6d GS term}) cannot contribute to (\ref{5d CS term for KK U(1)}) after $S^1$ compactification. While this is true at the perturbative level, introduction of the topological GS terms related to non-perturbative anomalies will lead to contributions to (\ref{5d CS term for KK U(1)}), as will be discussed in subsection \ref{subsec:TGS}.}

However, as a geometric invariant, the $\eta$ invariant does not in general admit a local density description -- i.e. it is not given by an integral of a differential form. Even worse, the $\eta$ invariant depends on geometric data, and hence would not seem to give rise to topological terms like (\ref{5d CS term for KK U(1)}). What saves the day is the adiabatic $\eta$ invariant discovered by Bismut and Cheeger \cite{bismut1989eta} (see also \cite{dai1991adiabatic} for some generalizations of the results of \cite{bismut1989eta}). In short, for the fibered space $M$, i.e. $F \hookrightarrow M \to B$, the limit of the $\eta$ invariant $\eta_M$ exists as the volume of the fiber $F$ approaches zero, and this limit admits a local density description, the BC $\eta$ form.

 \subsubsection{$\eta$ invariant of circle fibrations}
 The BC adiabatic $\eta$ invariant applies to our current setup, $Y_7$ as $S^1$ fibration over a 6D spin manifold $M_6$ with radius $r$ of $S^1$ very small. The metric on $Y_7$ is of the form\footnote{For simplicity, we consider the circle compactification without Wilson lines.}
 \begin{equation}
     \label{metric for circle fibration}
     ds_{Y_7}^2 = ds_{M_6}^2 + r^2(d\theta + A^0)^2, \ \ \ \ r\to 0,
 \end{equation}
 where $A^0 = A^0_idx^i $ is the KK $U(1)$ potential. Instead of using the heat-kernel methods in the original papers of the adiabatic $\eta$ invariant \cite{bismut1989eta} to compute $\eta_{Y_7}$ with metric (\ref{metric for circle fibration}),  we follow the computation done in \cite{Diaconescu:2000wy}.\footnote{In \cite{Diaconescu:2000wy}, the computation is done for an 11D manifold, in order to check the compatibility of the K-theoretic classification of RR flux in IIA and the (differential refined) cohomology description of the $C_3$ field in M-theory. This calculation can be adapted to the cases considered in this paper.} In the $r \to 0$ limit, the Dirac operator takes the form
  \begin{equation}
  \label{Dirac operator of Y7 in adiabatic limit}
    D_{Y_7}=\left(\begin{array}{cc}
\frac{i}{r} \frac{\partial}{\partial \theta} & \bar{D}_{M_6} \\
D_{M_6} & -\frac{i}{r} \frac{\partial}{\partial \theta} 
\end{array}\right) + \mathcal{O}(r).
\end{equation}
In this limit, we can ignore the $\mathcal{O}(r)$ piece, thanks to the fact that the limit of the $\eta$ invariant exists as $r\to 0$. The 7D spinor $\p_{7\text{D}}$ decomposes as:
\begin{equation}
   \G(Y_7,S) \to  \overset{\infty}{\underset{k=-\infty}{\oplus}}\G(M_{6}, S_{6\text{D},\pm}\otimes \cL^{\otimes k}),
\end{equation}
i.e.
\begin{equation}
    \label{7d spinor decomposition}
    \p_{7\text{D}}(x_i,\theta)   = \sum_{k=-\infty}^{\infty}\left(\begin{array}{c}
 \p_{6\text{D},k+} \\
 \p_{6\text{D},k-}
\end{array}\right) \text{exp} \big\{-ik\theta \big\},
\end{equation}
where $\psi_{6\text{D},\pm}$ refers to 6D spinors with positive/negative chirality. The factor of $e^{- i k \theta}$ in (\ref{7d spinor decomposition}) indicates that the 6D spinors $\p_{6\text{D},k\pm}$ take values in $\G(M_{6},\cL^{\otimes k})$, the sections of  $\cL^{\otimes k}$ over $M_6$, where $\cL$ is the line bundle associated with the $S^1$ circle fibration. In each $\G(M_{6},\cL^{\otimes k})$ sector (level $k$ of the KK tower), (\ref{Dirac operator of Y7 in adiabatic limit}) becomes (in the $r \to 0$ limit)
\begin{equation}
    \label{adiabatic Dirac operator in k-the KK}
      D_{Y_7}\big|_{\cL^{k}}=\left(\begin{array}{cc}
\frac{k}{r}  & \bar{D}_{M_6} \\
D_{M_6} & -\frac{k}{r}
\end{array}\right),
\end{equation}
  with the Dirac operator of the 6D base
  \begin{equation}
      \label{6d dirac operator}
       \left(\begin{array}{cc}
0  & \bar{D}_{M_6} \\
D_{M_6} & 0 
\end{array}\right),
  \end{equation}
acting on 6D spinors coupled to the line bundle $\cL^{\otimes k}$. For a non-zero eigenvalue of (\ref{6d dirac operator}), we have the relation
\begin{equation}
    \left(\begin{array}{cc}
0  & \bar{D}_{M_6} \\
D_{M_6} & 0 
\end{array}\right) \left(\begin{array}{c}
 \p_{6\text{D},k+} \\
 \p_{6\text{D},k-}
\end{array}\right) =    \left(\begin{array}{cc}
0  & \bar{\lambda} \\
\lambda & 0 
\end{array}\right) \left(\begin{array}{c}
 \p_{6\text{D},k+} \\
 \p_{6\text{D},k-}
\end{array}\right),
\end{equation}
as (\ref{6d dirac operator}) is Hermitian. The $7\text{D}$ Dirac operator (\ref{adiabatic Dirac operator in k-the KK}) on this subspace becomes:
\begin{equation}
\label{7d Dirac opertor on a subspace 1st}
    \left(\begin{array}{cc}
-\frac{k}{r}  & \bar{D}_{M_6} \\
D_{M_6} & \frac{k}{r}
\end{array}\right) \left(\begin{array}{c}
 \p_{6\text{D},k+} \\
 \p_{6\text{D},k-}
\end{array}\right) =    \left(\begin{array}{cc}
-\frac{k}{r}  & \bar{\lambda} \\
\lambda & \frac{k}{r} 
\end{array}\right) \left(\begin{array}{c}
 \p_{6\text{D},k+} \\
 \p_{6\text{D},k-}
\end{array}\right).
\end{equation}
After switching to a diagonal basis, (\ref{7d Dirac opertor on a subspace 1st}) gives a pair of eigenvalues $\pm w$ and hence does not contribute to the $\eta$ invariant $\eta_{Y_{7}}$ (see the definition of (\ref{definition of eta})). 

The conclusion is that in the adiabatic limit, only zero-modes of the $6\text{D}$ Dirac operator (coupled to $\cL^{\otimes k}, k\in \bZ$) from the 6D base bundle will contribute to the eta invariant $\eta_{Y_{7}}$:
 \begin{itemize}
     \item For $k=0$, the contribution of zero-modes to $\eta_{Y_7}$ is
     \begin{equation}
         \label{k=0 contribution in eta}
        \frac{ n_{+} + n_{-}}{2},
     \end{equation}
     which $\text{mod} \ \bZ$ is the same an an index:
     \begin{equation}
     \label{6d index at k=0}
         \frac{n_+ -n_-}{2} = \frac{1}{2}\int_{M_6} \widehat{A}(M_6) \, \text{ch}(V),
     \end{equation}
     where $V$ is the vector bundle associated to additional gauge groups in the 6D $\mathcal{N} = (1,0)$ supergravity theory. As we are mainly concerned with the universal features of $6\text{D}$ $\mathcal{N} = (1,0)$ supergravity, we will ignore the gauge bundle contribution $\text{ch}(V)$. Note (\ref{k=0 contribution in eta}) in general does not admit a local density description like (\ref{6d index at k=0}), since it is not an index-like object. 
     \item The interesting case is $k\neq 0 \in \bZ$, as these sectors will contribute to the CS coupling involving the KK $U(1)$ (\ref{5d CS term for KK U(1)}). From now on we only focus on the part involving the KK $U(1)$, since this is the universal sector for $6\text{D}$ $\mathcal{N} = (1,0)$ supergravity compactified on $S^1$. Then zero modes in the $k \neq 0$ sector contribute to $\eta_{Y_7}$:
     \begin{equation}
         \label{k non zero contribution in eta}
         \frac{n_{+}}{2} \text{sgn}\Big(\frac{k}{r}\Big)+ \frac{n_{-}}{2} \text{sgn}\Big(-\frac{k}{r}\Big) = (n_{+}- n_-) \frac{\text{sgn}(k)}{2} = \frac{\text{sgn}(k)}{2}\int_{M_6}\widehat{A}(M_6)e^{k\cL},
     \end{equation}
    with $\cL$ denoting the line bundle associated with the circle fibration ($c_1(\cL) = F^0$). The integral is given by
     \begin{equation}
        \label{expansion of 6d index density}
        \int_{M_6}\widehat{A}(M_6)e^{k\cL} = \frac{k^3}{3!}(F^{0})^3 -kF^{0}\wedge \frac{p_1(M_6)}{24},
     \end{equation}
       which is of the form of the KK $U(1)$ CS terms (\ref{5d CS term for KK U(1)}).
     \end{itemize}

Summing over $k \in \bZ$ and using zeta function regularization, (\ref{k non zero contribution in eta}) gives the part of the adiabatic $\eta_{Y_{7}}$ involving the KK $U(1)$:
     \begin{equation} \label{adiabatic BC eta from a single SMW}
        \eta_{Y_{7}} =  \sum_{k\geq 0}  \int_{M_6} \frac{k^3}{6}(F^0)^3 -k F^0 \wedge \frac{p_1(M_6)}{24} = \int_{M_6} \frac{\zeta(-3)}{6}(F^0)^3 -\zeta(-1) F^0 \wedge \frac{p_1(M_6)}{24}.
     \end{equation}
This is the contribution of a single SMW fermion to the 5D CS couplings.

Using (\ref{adiabatic BC eta from a single SMW}), we can compute the contribution to the KK $U(1)$ CS term from SMW fermions, the SMW gravitini, and the (anti-)self-dual tensors multiplets. The individual contributions are given below.
\begin{itemize}
    \item The KK $U(1)$ CS terms generated by SMW fermions in the vector, tensor, and hypermultiplets are:
    \begin{equation}
    \label{adiabatic BC eta for chiral fermion}
        (V-H-T)\frac{\zeta(-3)}{6}\int_{M_6} (F^{\text{0}})^3+ (V-H-T) \frac{\zeta(-1)}{48}\int_{M_6} F^{\text{0}}\wedge \text{tr}(R\wedge R).
    \end{equation}
    \item Considering the gravitino  as a chiral fermion coupled to the vector bundle $V =TY_{7} \ominus \bR^{\oplus 2}$, its contribution to the 5D KK $U(1)$ CS term can be computed as:
    \begin{equation}
        \label{adiabatic BC eta for gravitino}
        \frac{\zeta(-3)}{6}\int_{M_6}  (F^{\text{0}})^3 (6-1) +\frac{\zeta(-1)}{48}\int_{M_6} (5-24)F^{\text{0}}\wedge \text{tr}(R\wedge R).
    \end{equation}
    \item  The KK $U(1)$ CS term from a self-dual tensor is more subtle, as it is given by $- \frac{1}{4}\eta_{\text{sig},7\text{D}}$, 
    where $\eta_{\text{sig},7\text{D}}$ is the eta invariant of the $7\text{D}$ signature Dirac operator \cite{Monnier:2018nfs}. Here the minus sign is due to the fact that (anti-)self-dual tensor fields are bosonic. The contribution to the 5D KK $U(1)$ CS terms is then given by
    \begin{equation}
        \label{adiabatic BC eta for SD tensor}
       \frac{\eta_{\text{sig},7\text{D}}}{4} = -(1-T)  \frac{\zeta(-3)}{3}\int_{M_6 }  (F^{0})^3   + (1-T)\frac{\zeta(-1)}{12}\int_{M_6} F^{0}\wedge \text{tr}(R\wedge R). 
    \end{equation}
\end{itemize}
Adding the individual contributions (\ref{adiabatic BC eta for chiral fermion}), (\ref{adiabatic BC eta for gravitino}), and (\ref{adiabatic BC eta for SD tensor}) gives:
\begin{equation}
    \label{KK U(1) CS from eta 2nd}
    (V-H + T +3 )\frac{{\zeta(-3)}}{6} \int_{M_{6}  } (F^{\text{0}})^3 + 2(V- H- 5 T-15) \frac{\zeta(-1)}{96}\int_{M_{6}} F^{\text{0}}\wedge  \text{tr}(R \wedge R).
\end{equation}
This is the same result as obtained  by integrating out the massive zero-modes of the 6D chiral fields in the loop  \cite{Bonetti:2013ela}. (The comparison requires  another reminder that the curvatures $R$ and $F$ contain hidden factors of $2\pi$.) For the reduction of anomaly-free theories in 6D, using $H+29T-V=273$ and $\zeta(-3) \rightarrow \frac{1}{120}$ and $\zeta(-1) \rightarrow -\frac{1}{12}$, one notably obtains $c_{000} = \frac{1}{4}(9-T)$ and $a^{\6d}_0=4(12 - T)$. 

\subsection{5D cubic form from 6D GS terms}
\label{subsex:TI}
We can collect all the different coefficients of the 5D CS couplings coming from an $S^1$ reduction of a 6D theory and assemble them into a triple intersection form  $c_{IJK}$ with $I = 0, i, \alpha$. As we will see, a direct comparison of the result with the integrality constraints (\ref{5d consistency condition}) is a bit premature, since it is not obvious how $c_{IJK}$ obtained from reduction is related to $C_{IJK}$ in 5D supergravity \eqref{eq:act1}, where the $U(1)$ fields are assumed to be integral (hence the different notations). We shall return to this in subsection \ref{subsec:QS1}.

The situation is similar with the gravitational CS couplings. In 5D they are specified by the vector $a_I$, with $I=0,1,..., n_V$ (\ref{eq:act1}). The gravitational couplings obtained by reduction from 6D will be denoted by $a^{\6d}_I$ with $I = 0, i, \alpha$.

From the dimensional reduction of the Green-Schwarz terms and the equations of motion, we see that the following coefficients are zero:
\begin{equation}
    c_{\alpha \beta \gamma} = c_{\alpha \beta i} = c_{\alpha i 0} = c_{\alpha 00} = 0.
\end{equation}
By comparison with \eqref{NI}, one finds that the only non-zero coefficients involving the $\alpha$ indices are 
\begin{equation}
    c_{0 \alpha \beta} = \Omega_{\alpha \beta}, \ \ \ \ \  c_{\alpha ij} =  -\Omega_{\alpha\beta} b^{\beta}_{ij}.
\end{equation}
The coefficients without an $\alpha$ index are generated at the one-loop level, as discussed previously. In subsection 
\ref{sec:cirle2}, coefficients governing the trilinear ($k_{IJK}$) and gravitational couplings ($k_{IRR}$) were determined (up to ambiguities, notably in $c_{000}$ and $a_0^{\text{6D}}$) in equations \eqref{kgrav} and \eqref{kgauge} in agreement with the loop calculation results $\cite{Bonetti:2013ela}$. Gauge fixing the symmetry under large gauge transformations such that the floor functions are set to zero we arrive at:
\begin{equation}
\label{U1 one loop cs}
\begin{split}
    &c_{ijk} = -\tfrac{1}{2}\sum_{\text{h}} q^{\text{h}}_i  q^{\text{h}}_j q^{\text{h}}_k,\\&
    c_{ij0} = -\tfrac{1}{12 }\sum_{\text{h}} q^{\text{h}}_i q^{\text{h}}_j,   \\&
    c_{i00} = 0, \\&
    c_{000} = -\tfrac{1}{120}(H - V - T - 3),
\end{split}
\end{equation}
and the $a^{\6d}_I$ coefficients are
\begin{equation}
\begin{split}
    &a^{\6d}_0 = -\tfrac{1}{6}(H - V + 5 T + 15),  \\&
    a^{\6d}_i = \sum_{\text{h}} q^{\text{h}}_i, \\&
    a^{\6d}_{\alpha} = 12 \Omega_{\alpha \beta} \tilde{a}^{\beta}.
\end{split}
\end{equation}
We can check the integrality conditions for these coefficients:
\begin{itemize}
\item The integrality conditions only involving $c_{\a\b\g}$ are satisfied:
 \begin{equation}
    c_{\alpha \alpha \alpha} + \tfrac{1}{2}a^{\6d}_{\alpha} =\tfrac{1}{2} a^{\6d}_{\alpha} =  6 \tilde{a}^{\beta} \Omega_{\alpha \beta} \in 6 \mathbb{Z}.  
\end{equation}  
    \item The integrality conditions involving $c_{I J i}$ ($I, J = \{0,\a,i \} $) are not all satisfied in general. For example, we find:
    \begin{equation}
    \label{U1 intersection in 6d basis}
    c_{iii} + \tfrac{1}{2}a^{\6d}_i = -\tfrac{1}{2}\sum_{\text{h}} [(q^{\text{h}}_i)^3  -q^{\text{h}}_i] \notin 6\mathbb{Z},
\end{equation}
\item The integrality conditions only involving $c_{\G_1\G_2\G_3}$, with $\G_{1,2,3} =(0,\a)$ are not satisfied in general:
\begin{equation}
\label{KK coeficients from cirlce reduction 1st}
    c_{0\alpha \alpha} + c_{00 \alpha} = c_{0 \alpha \alpha} =  \Omega_{\alpha \alpha} \notin 2\mathbb{Z}, \ \ \ \ 
\end{equation}
\begin{equation}\label{eq:miss3}
    c_{000} + \tfrac{1}{2}a^{\6d}_0 = -\tfrac{1}{4}(9 - T) - 2(12 - T) 
    \notin 6 \mathbb{Z}.  \ \ \ \ 
\end{equation}  
\end{itemize}
We will argue in the next section that this is not an indicator of an inconsistency, but just an artifact of the  $U(1)$ basis. 
 
\section{6D non-perturbative anomalies and 5D consistency conditions}
\label{sec:ano-NP}
As reviewed in Section \ref{sec:AT}, the anomaly of a D-dimensional theory is characterized by a bordism invariant. When the perturbative anomaly vanishes, the non-perturbative anomaly sits in  $\text{Hom}_{\bZ}\big(\Omega_{\text{D}+1,\text{tor}}^{\text{Spin}}(BG), U(1)\big)$ (see \eqref{bordism description of anomalies 1st}). In the context of circle compactification of 6D $\mathcal{N} =(1,0)$ supergravity, we want to study  the following schematic map:
\begin{equation}
    \label{schematic map of circle compactification}
    \varphi_{S^1}: \cT_{6\text{D} \ \text{SUGRA}} \to \cT_{5\text{D} \text{ SUGRA} \,  + \, U(1)},
\end{equation}
where:
\begin{itemize}
    \item $\cT_{6\text{D} \ \text{SUGRA}}$ denotes the set of $6\text{D}$ $\mathcal{N} = (1,0)$ supergravity theories.
    \item $\cT_{5\text{D} \text{ SUGRA $+ \,\, U(1)$}}$ denotes the set of $5\text{D}$ $\cN=1$ supergravity theories with the universal KK $U(1)$ vector singled out. The set $\cT_{5\text{D} \text{ SUGRA $+ \, \, U(1)$}}$ has been discussed in the example of subsection \ref{Physical interpretation of other bordism groups}.
\end{itemize}

As discussed, if the 6D theory suffers from perturbative anomalies, the 5D theory will simply fail to be invariant under gauge transformations and diffeomorphisms. In this section, we want to focus on theories with apparent $\bZ_k$ gauge symmetry and its possible anomalies discussed in \cite{Monnier:2018nfs}, as well as study the incarnation of these $\bZ_k$ gauge anomalies after circle compactification (i.e. the image of the map $\varphi_{S^1}$). Hence, we refine the setup to be
\begin{itemize}
    \item $\cT_{6\text{D} \text{ SUGRA}}$ denotes the set of $6\text{D}$ $\mathcal{N} = (1,0)$ supergravity theories with $\bZ_k$ gauge group free of perturbative anomalies.
    \item Circle compactification $\varphi_{S^1}$ is modified to have one unit of $\bZ_k$ holonomy on the compactified $S^1$.
    \item $\cT_{5\text{D}\text{ SUGRA $+ \, U(1)$}}$ denotes the set of $5\text{D}$ $\cN=1$ supergravity theories with a universal KK $U(1)$ gauge field. Due to the non-trivial $\bZ_k$ holonomy along the $S^1$, in the resulting $5\text{D}$ supergravity theories the $\bZ_k$ gauge group is broken.\footnote{In compactifications with a trivial $\bZ_k$ holonomy, this gauge group is not broken after circle compactification. Even then the $5\text{D}$ theory does not have a $\bZ_k$ gauge anomaly since $\Omega^{\text{Spin}}_{7\text{D},\text{free}}(B\bZ_k)= \Omega^{\text{Spin}}_{6\text{D},\text{tor}}(B\bZ_k)$ = 0.}
\end{itemize}
 From the bordism description (\ref{bordism description of anomalies 1st}), the potential $\bZ_k$ gauge anomaly of $6\text{D}$ supergravity sits in $\text{Hom}_{\bZ}\big(\Omega_{7\text{D},\text{tor}}^{\text{Spin}}(B\bZ_k), U(1)\big)$ and is hence non-perturbative. As pointed out in \cite{Monnier:2018nfs}, the non-perturbative $\bZ_k$ gauge anomalies indeed exist in $\mathcal{N} = (1,0)$ theories, and the condition for these theories to be anomaly-free imposes non-trivial constraints on the $\bZ_k$ charged matter spectrum. 
 
 However, in the $5\text{D}$ $\cN=1$ supergravity on the right-hand side of (\ref{schematic map of circle compactification}), there are extra consistency conditions due to non-perturbative anomalies, since $\Omega_{6\text{D},\text{tor}}^{\text{Spin}}\big(BU(1)\big) = \Omega_{7\text{D},\text{free}}^{\text{Spin}}\big(BU(1)\big) = 0$. On the other hand, as studied in the example of subsection \ref{Physical interpretation of other bordism groups}, there will be non-trivial quantization conditions for the KK $U(1)$ CS terms associated with the bordism group   $\Omega_{6\text{D}}^{\text{Spin}}\big(BU(1)\big) = \bZ^{\oplus 2}$. If a $5\text{D}$ supergravity comes from the circle compactification of a  $6\text{D}$ supergravity, then the related CS terms are determined by the matter content of the $\text{D}$-dimensional theory via one-loop effects \cite{Bonetti:2013ela,Witten:1996qb}. One of the main results of this section is that the consistency condition proposed in \cite{Monnier:2018nfs} becomes the requirement for a  well-defined KK $U(1)$ CS term (\ref{quantization condition for 5d CS term for KK U(1)}) after circle compactification. 
 
 When the $6\text{D}$ supergravity theories are obtained from F-theory compactifications, the KK $U(1)$ CS term can also be determined via the intersection numbers of the Calabi-Yau threefold.  In this way we verify that the consistency condition in \cite{Monnier:2018nfs} associated with a $\bZ_k$ gauge group is satisfied by F-theory on elliptic (or genus-one) fibered Calabi-Yau threefolds. 
 
 We will also briefly discuss possible generalization to the case when the $\bZ_k$ group acts on vector and tensor multiplets (i.e. CHL-like models).
 
\subsection{Non-perturbative anomalies in 6D}
\label{$6d$ non-perturbative anomalies}
Before turning to circle reductions, we review non-perturbative $\bZ_k$ gauge anomalies in  6D $\mathcal{N} = (1,0)$ theories free of perturbative anomalies, following \cite{Monnier:2018nfs}. Using (\ref{bordism description of anomalies 1st}), the potential anomaly is given by $\mbox{Hom}_{\bZ}\big(\Omega_{7\text{D}, \text{tor}}^{\text{Spin}}(B\bZ_k),U(1)\big)$. To draw general model-independent conclusions about the non-perturbative $\bZ_k$ gauge anomalies, we focus on one specific generator of $\Omega_{7\text{D}, \text{tor}}^{\text{Spin}}(B\bZ_k)$, the Lens space $L_{7}^{k}$ (together with proper $\bZ_k$ bundle, which we will assume from now on).

\paragraph{\textbf{7D Lens space $L_{7}^k$}:}
 The 7D sphere $S^7$ with round metric can be embedded in $\bC^4$ with standard metric:
\begin{equation}
    S^7 = \{(z_1,z_2,z_3,z_4): |z_1|^2+|z_2|^2+|z_3^2|+|z_4|^2 = 1\}. 
\end{equation}
Now consider the following $\bZ_k$ action on ${\bC^4}$:
\begin{equation}
\label{Zk action on C4}
    \sigma: (z_1,z_2,z_3,z_4) \to \bigg(\exp\bigg\{\frac{2\pi i}{k}\bigg\}z_1,\exp\bigg\{\frac{2\pi i}{k}\bigg\}z_2,\exp\bigg\{\frac{2\pi i}{k}\bigg\}z_3,\exp\bigg\{\frac{2\pi i}{k}\bigg\}z_4\bigg),
\end{equation}
where $\sigma$ is the generator of $\bZ_k$. Equation (\ref{Zk action on C4}) induces a $\bZ_k$ action on $S^7$, and the Lens space $L_{7}^{k}$ is realized as the quotient space $S^7/\bZ_k$. The $\bZ_k$ bundle is realized via $\bZ_k$ holonomy on $S^7/\bZ_k$ given by $\text{Hom}_{\bZ}\big(\pi_1(L_7^k,U(1))\big)$. We take the $\bZ_k$ bundle given by the generator of $\text{Hom}_{\bZ}\big(\pi_1(L_7^k,U(1))\big)$, i.e. the holonomy is the generator $\sigma$ when going around the loop which is the generator of $\pi_1({L_7^k})$. 

\paragraph{Remark}   $L_{7}^k$ admits an $S^1$ fibration structure:
\begin{equation}
    \label{circle fibration of lens space}
    S^1 \hookrightarrow L_{7}^{k} \to {\bC P^3}.
\end{equation}
The $\bZ_k$ holonomy now is the holonomy around the $S^1$ fiber, and $\bC P^3$ is spin. Hence $L_{7}^k$ is suitable for discussing $S^1$ compactification for 6D supergravity with $\bZ_k$ gauge group, i.e. (\ref{schematic map of circle compactification}). 

\vspace{.2cm}

As discussed in Section \ref{sec:AT}, the anomaly from a SMW fermion with $\bZ_k$ charge $s$ is given by $\exp\{2\pi i \, \eta_{7\text{D},s}\}$ and depends on the metric. Although  $\exp\{2\pi i \, \eta_{7\text{D},s}\}$  is computable for $L_7^k$ with round metric (see \cite{Monnier:2018nfs} and references therein), it is not a cobordism invariant and will change once we perturb the metric. On the contrary, the relative reduced $\eta$ invariant $\cA_s$ \cite{atiyah1975spectral2} is a cobordism invariant:
\begin{equation}
\label{relative reduced eta}
        \cA_s = \exp\{2\pi i (\eta_{7\text{D},s}-\eta_{7\text{D},0})\}.
\end{equation}
Here $\eta_{7\text{D},0}$ is the anomaly of a SMW fermion which only differs from the previous SMW fermion by carrying $\bZ_k$ charge $0$. In this way $\exp\{2\pi i(\eta_{7\text{D},s}-\eta_{7\text{D},0})\}$ does not receive any contributions from  the perturbative part and becomes a bordism invariant.

It is easy to compute both $\eta_{s}(L_{7}^{k})$ and $\eta_{0}(L_{7}^{k})$ using the round metric on $L_{7}^{k}$. By taking the difference, one obtains a bordism invariant whose physical interpretation is the non-perturbative $\bZ_k$ gauge anomaly \cite{Monnier:2018nfs}:
\begin{equation}
 \label{relative anomaly of a hyper 3rd}
   \eta_s({L_7^k}) -\eta_0(L_7^k)=  \frac{1}{24 k}\left(-2 k s+2 s^2-k^2 s^2+2 k s^3-s^4\right).  
 \end{equation}

To draw universal lessons about the $\bZ_k$ anomaly of 6D supergravity, \cite{Monnier:2018nfs} studied the following relative set up:
consider two supergravity theories $\cT_A$ and $\cT_B$ which only differ in $\bZ_k$ charged hypermultiplets, i.e. $\cT_A$ and $\cT_B$ are exactly the same if we ignore the $\bZ_k$ gauge group. Without loss of generality, let $\cT_A$ have $x_s$ hypermultiplets with charges $s\in(0,1,\dots,k-1)$, while $\cT_B$ has $y_s$ hypermultiplets with charges $s\in (0,1,\dots,k-1)$ under the $\bZ_k$ gauge group. Assuming $\cT_A$ is consistent, what is the condition for $\cT_B$ to be consistent, at least as far as the $\bZ_k$ anomaly is concerned?

As $\cT_A$ is consistent, the $\bZ_k$ gauge anomaly  for $\cT_B$ is given by
\begin{equation}
    \label{Zk anomaly for Tb}
    {\cA_{\cT_B}} = \text{exp} \bigg\{ 2\pi i\sum_{s=0}^{k-1} \Delta x_s \eta_{7\text{D},s}\bigg\}, \quad \Delta x_s \equiv y_s -x_s.
\end{equation}
From the set-up, it is clear that $\sum_{s=0}^{k-1}\Delta x_s = 0$ and ${\cA_{\cT_B}}$ is a bordism invariant. Applying (\ref{relative anomaly of a hyper 3rd}), (\ref{Zk anomaly for Tb}) becomes:
\begin{equation}
    \label{Zk anomaly for Tb 2nd}
    \begin{split}
     \sum_{s=0}^{k-1} \Delta x_s \eta_{s}(L_{7}^k)  & = \sum_{s=1}^{k-1} \Delta x_s \big[\eta_{s}(L_{7}^k)-\eta_{0}(L_{7}^k)\big]\\& =
     \sum_{s=1}^{k-1}  \frac{\Delta x_s}{24 k}\left(-2 k s+2 s^2-k^2 s^2+2 k s^3-s^4\right). 
     \end{split}
\end{equation}
For $\cT_B$ to be free of $\bZ_k$ anomalies, it is required that $\cA_{\cT_B} = 1$, and hence 
\begin{equation}
 \label{Moore Moonier eta}
\sum_{s=0}^{k-1} \frac{ \Delta x_s}{24 k}\left(-2 k s+2 s^2-k^2 s^2+2 k s^3-s^4\right)  = 0 \ \text{mod} \ \bZ,
\end{equation} 
under the condition
\begin{equation}
    \label{relative difference of two supergravity theory}
    \sum_{s=0}^{k-1} \Delta x_s = 0,
\end{equation}
e.g.
\begin{equation}
\label{Moore Moonier stronger condition}
\begin{array}{l}
\vspace{.1cm}

k=2: \qquad \frac{1}{16} \Delta x_1=0 \bmod 1, \\

\vspace{.1cm}

k=3: \qquad \frac{1}{9}\left(\Delta x_1+\Delta x_2\right)=0 \bmod 1, \\

\vspace{.1cm}

k=4: \qquad \frac{1}{32}\left(5 \Delta x_1+8 \Delta x_2+5 \Delta x_3\right)=0 \bmod 1, \\

\vspace{.1cm}

k=5: \qquad \frac{1}{5}\left(\Delta x_1+2 \Delta x_2+2 \Delta x_3+\Delta x_4\right)=0 \bmod 1, \\

\vspace{.1cm}

k=6: \qquad  \frac{1}{144}\left(35 \Delta x_1+80 \Delta x_2+99 \Delta x_3+80 \Delta x_4+35 \Delta x_5\right)=0 \bmod 1.
\end{array}
\end{equation}
As already noticed in \cite{Monnier:2018nfs}, \eqref{Moore Moonier eta} imposes conditions that are too strong and would be violated in (obviously consistent) F-theory models. In \cite{Monnier:2018nfs} a modification of GS terms by a purely torsional contribution to $\tX_4^{\alpha}$
is proposed. Denoting this $\bZ_k$ anomaly contribution  by $x_{\text{GS}}$, the anomaly-free condition for $\cT_B$ can  be modified to
\begin{equation}
    \label{Zk anomaly for TB with GS}
    \sum_{s=0}^{k-1}  \frac{\Delta x_s}{24 k}\left(-2 k s+2 s^2-k^2 s^2+2 k s^3-s^4\right) + x_{\text{GS}} = 0 \ \text{mod} \ \bZ.
\end{equation}
Due to the quadratic refinement of the GS couplings,  $x_{\text{GS}}$ can take values in $\frac{\bZ}{2k}$, and the necessary $\bZ_k$ anomaly-free condition for $\cT_B$ becomes
\begin{equation}
 \label{Moore Moonier eta with GS}
\sum_{s=0}^{k-1}  \frac{\Delta x_s}{24 k}\left(-2 k s+2 s^2-k^2 s^2+2 k s^3-s^4\right)  = 0 \ \text{mod} \ \frac{\bZ}{2k}.
\end{equation} 
As a result, the conditions \eqref{Moore Moonier stronger condition} are weakened to
\begin{equation}
\label{Moore Moonier weaker conditions}
\begin{array}{l}
\vspace{.1cm}

k=2: \qquad \frac{1}{4} \Delta x_1=0 \bmod 1, \\

\vspace{.1cm}

k=3: \qquad  \frac{1}{3}\left(\Delta x_1+\Delta x_2\right)=0 \bmod 1, \\

\vspace{.1cm}

k=4: \qquad \frac{1}{4}\left(5 \Delta x_1+8 \Delta x_2+5 \Delta x_3\right)=0 \bmod 1, \\

\vspace{.1cm}

k=5: \qquad \text { no constraints, } \\

\vspace{.1cm}

k=6: \qquad \frac{1}{12}\left(35 \Delta x_1+80 \Delta x_2+99 \Delta x_3+80 \Delta x_4+35 \Delta x_5\right)=0 \bmod 1.
\end{array}
\end{equation}

\noindent
For the rest of this section, we will study the image of these constraints under the circle compactification  $\varphi_{S^1}$, \eqref{schematic map of circle compactification}.

\subsection{$S^1$ compactifications and consistency conditions for 5D supergravities}
We can turn to the study of the relative set-up discussed in the previous subsection, i.e. a pair of 6D theories $\cT_A$ and $\cT_B$ with $\cT_A$ being consistent, in the context of circle compactification \eqref{schematic map of circle compactification}. Note that $\bZ_k$ gauge symmetry will be lost after circle compactification due to the non-trivial holonomy on the circle. The relevant bordism  groups
 \begin{equation}
     \label{6d BU(1) bordism}
     \Omega_{6\text{D}}^{\text{Spin}}\big(BU(1)\big) = \bZ \oplus \bZ, \ \ \ \ \ \Omega_{7\text{D}}\big(BU(1)\big) = 0,
 \end{equation}
 suggest that there are no relevant anomalies in 5D. Hence, an obvious question is what happens to the 6D $\bZ_k$ anomaly after the circle compactification.

As already discussed, $\Omega_{6\text{D}}^{\text{Spin}}\big(BU(1)\big) = \bZ \oplus \bZ$ suggests the possibility that  generalized WZW terms can be added to the 5D $\cN=1$ supergravity, leading to the CS couplings of the KK $U(1)$ field (\ref{5d CS term for KK U(1)}).

Well-defined KK $U(1)$ CS terms are given by $\text{Hom}_{\bZ}\big( \Omega_{6\text{D}}^{\text{Spin}}(BU(1)) = \bZ \oplus \bZ,\bZ\big)$, which implies that the CS coefficients $C_{000}$ and $a_{0}$ need to satisfy the quantization condition (see the example of subsection \ref{Physical interpretation of other bordism groups})
\begin{equation}
\label{KK u(1) quantization}
   \frac16 \Bigl( C_{000} + \frac{a_0}{2} \Bigr) \in \bZ.
\end{equation} 

On the other hand, if the 5D supergravity theory comes from the circle compactification of a 6D supergravity, then the CS term (\ref{5d CS term for KK U(1)}) is generated via one-loop effects and depends on the field content of the 6D field theory (\cite{Bonetti:2013ela}, or see (\ref{KK U(1) CS from eta 2nd}) for a derivation using the adiabatic $\eta$ invariant). A priori it is not guaranteed that the one-loop generated CS terms satisfy the quantization condition (\ref{KK u(1) quantization}), as $\cT_A$ and $\cT_B$ compactified on $S^1$ with one unit $\bZ_k$ holonomy will give different KK towers of hypermultiplets. Assume $\cT_A$ is consistent so that the one-loop generated CS terms satisfy (\ref{KK u(1) quantization}); then the quantization condition for the CS terms of $\cT_B$ after circle compactification would impose constraints on the spectrum of the 6D theory $\cT_B$. 

In this section, we argue that the necessary 6D $\bZ_k$ anomaly-free conditions, after circle compactification, become a necessary condition for ensuring the cubic form \eqref{KK u(1) quantization} is integral.

For this we will need to modify the calculation of $c_{000}$ and $a^{\6d}_0$
 (\ref{KK U(1) CS from eta 2nd}) by including the $\bZ_k$ holonomy on the circle.  The only subtlety involved in this modification is the correct accounting of the masses of KK towers.

\paragraph{Massless $U(1)_{\text{KK}}$ gauge field of $S^1$ compactification with $\bZ_k$ holonomy} We will use the intuition obtained in the F-theory context. It is argued from the UV F-theory \cite{Cvetic:2015moa}, via the Higgs mechanism, that for 6D supergravity with a $\bZ_k$ gauge field compactified on $S^1$:
\begin{itemize}
    \item Without discrete $\bZ_k$ holonomy, (part of) the 5D massless gauge group is $U(1)_{\text{KK}}\times \bZ_k$, and the KK tower of 6D hypermultiplets with $\bZ_k$ charge $s$ carries charge $(n,s) \in \bZ \times \bZ_k$ after $S^1$ compactification. 
    \item With one unit of $\bZ_k$ holonomy on $S^1$, the relevant 5D gauge group is the KK $U(1)$. The key point here is that a 6D hypermultiplet with $\bZ_k$ charge $s$ will give a KK tower of 5D hypermultiplets with $U(1)$ charge
    \begin{equation}
        \label{KK U(1) charge of hypers with holonomy}
          k \bZ +s, \quad s = 0, \dots, k-1.
    \end{equation}
    This means that the KK $U(1)$ gauge group for circle compactification with one unit of holonomy is given by the $\bZ_k$ extension of the KK $U(1)$ for circle compactification without $\bZ_k$ holonomy:
    \begin{equation}
        \label{5d KK U(1) with holonomy}
        0 \rightarrow \bZ_k \rightarrow U(1) \rightarrow U(1)_{\text{KK}} \rightarrow 0,
    \end{equation}
    where the map from $U(1)$ to $U(1)_{\text{KK}}$ is 
    \begin{equation}
        \exp\{i\theta\} \in U(1) \to \exp\{ik\theta\} \in U(1)_{\text{KK}}.
    \end{equation}
\end{itemize}
\paragraph{Remark} One can argue that this should hold in general using a supergravity-based heuristic argument:
\begin{itemize}
    \item Turning on one unit of $\bZ_k$ holonomy on the circle relies on the topological fact $\pi_1(S^1)=\bZ$ for the compactified $S^1$.
    \item Due to the completeness conjecture, any 5D supergravity should have magnetic monopole strings in its spectrum. In theories coming from a circle reduction of a 6D theory, these lift to a Taub-NUT$\times\mathbb{R}^2$, and the $S^1$ at the center of the Taub-NUT space shrinks. This space no longer has a nontrivial $\pi_1$. Hence a 5D theory obtained from $S^1$ compactification with $\bZ_k$ holonomy would not have magnetic monopole strings in the spectrum.
    \item The way out of this puzzle for $S^1$ compactification with $\bZ_k$ holonomy is that one should enlarge the $S^1$ $k$ times, absorbing the $\bZ_k$ gauge group into the KK $U(1)$, as in (\ref{5d KK U(1) with holonomy}). The effect of this enlargement is that for the $k$th KK tower of 6D hypermultiplets with $\bZ_k$ charge $s$, its dependence on $\theta \in S^1$ gets modified to
    \begin{equation}
        \phi_n(\theta + 2\pi) = \exp \Big\{2\pi i \frac{nk+s}{k} \Big\} \phi_n(\theta )\to  \phi_n(\theta + 2\pi) = \exp\{2\pi i (nk+s)\} \phi_n(\theta).
    \end{equation}
    This imposes no non-trivial topological condition on the fibered $S^1$ and becomes compatible with the existence of magnetic monopole strings for the KK $U(1)$.
    \item The above argument does not apply to 6D continuous gauge groups. In this case, the 5D monopole solution could come from a 6D Taub-NUT solution with certain gauge field configurations, see \cite{kronheimer1985monopoles}.
\end{itemize}

\vspace{.5cm}
We now return to the relative set-up and consider two 6D supergravity theories $\cT_A$ and $\cT_B$ that only differ in $\bZ_k$ charged hypermultiplets. Assuming $\cT_A$ is consistent, compactifying it on $S^1$ (with one unit $\bZ_k$ holonomy) should give well-defined KK $U(1)$ CS terms, i.e. (\ref{quantization condition for 5d CS term for KK U(1)}). The KK $U(1)$ CS terms from $\cT_B$ in this circle compactification differ from the $\cT_A$ ones by the contribution of the $\bZ_k$ charged hypermultiplets.

To calculate the difference in the KK $U(1)$ CS terms, first we need to know the contribution  from a single 6D hypermultiplet with $\bZ_k$ charge $s$ to $c_{000}$ and $a^{\6d}_{0}$.

\paragraph{Contribution to $c_{000},a^{\6d}_0$ from a 6D hyper with $\bZ_k$ charge $s$ under $\varphi_{S^1}$ }
Like for the coefficients computed in simple circle reductions, we shall use $c_{IJK}$ and $a^{\6d}_I$ for the triple intersections and gravitational coefficients respectively.
\begin{itemize}
    \item For $S^1$ compactification without $\mathbb{Z}_k$ holonomy, (\ref{KK U(1) CS from eta 2nd}) gives
    \begin{equation}
    \label{CS coefficient from 6d hyper}
    \delta c^s_{000} = \sum_{n\geq 0} n^3 \text{sgn}(n), \ \ \ \ \delta (a^{\6d}_{0})^s = -2\sum_{n\geq 0} n \, \text{sgn}(n).
\end{equation}
    \item For $S^1$ compactification with one unit $\mathbb{Z}_k$ holonomy, the contribution gets modified to
    \begin{equation}
    \label{CS coefficient from 6d hyper with Zk charge s}
    \begin{split}
    \delta c^s_{000} = \sum_{n\geq 0} (kn+s)^3 \, \text{sgn}(kn+s),\\ (a^{\6d}_{0})^s = -2\sum_{n\geq 0} (kn+s) \, \text{sgn}(kn+s).
    \end{split}
\end{equation}
\end{itemize}
Using zeta function regularization,
\begin{equation}
    \label{zeta regularization}
     \sum_{n\geq 0}   \Big(n +\frac{s}{k}\Big)^3 \text{sgn}\Big(n+\frac{s}{k}\Big) = \zeta(-3;s/k) = -\frac{1}{4}B_4(s/k),
\end{equation}
\begin{equation*}
    \sum_{n\geq 0} \Big(n +\frac{s}{k}\Big) \, \text{sgn} \Big(n+\frac{s}{k}\Big) = \zeta(-1;s/k) = -\frac{1}{2}B_2(s/k),
\end{equation*}
 where
\begin{equation}
    \label{B2 B4 polynomials}
    B_2(x) = x^2 -x + \tfrac{1}{6}, \ \ \  B_4(x) = x^4 - 2x^3 +x^2 -\tfrac{1}{30}.
\end{equation}
Equations \eqref{CS coefficient from 6d hyper with Zk charge s} become
\begin{equation}
    \label{difference CS coefficient from 6d hyper 2nd}
    \begin{split}
    \delta c^s_{000} &=-\frac{k^3}{4} B_4(s/k),\\
    \delta (a^{\6d}_{0})^s &=  k B_2(s/k),
    \end{split}
\end{equation}
and the differences in the 5D KK $U(1)$ CS terms between compactified $\cT_A$ and $\cT_B$ are given by
\begin{equation}
    \label{difference CS coefficient from 6d hypers}
    \begin{split}
     \Delta c_{000} &=\sum_{s=0}^{k-1}\Delta x_s \delta c^s_{000} =  \sum_{s= 1}^{k-1} \frac{-k^3 \Delta x_s}{4}\big[B_4(s/k)-B_{4}(0)\big],\\
     \Delta a^{\6d}_{0} &=  \sum_{s= 0}^{k-1}\Delta x_s \delta (a^{\6d}_{0})^s =\sum_{s=1}^{k-1} {k \Delta x_{s}}\big[B_2(s/k)-B_{2}(0)\big] = 0 ,
   \end{split} 
\end{equation}
 where \eqref{relative difference of two supergravity theory} is used for obtaining the second equality in the above two equations.

As the 5D supergravity from $\cT_A$ compactified on a circle with one unit $\bZ_k$ is consistent (and \eqref{KK u(1) quantization} holds), for the KK $U(1)$ CS term of $\cT_B$ upon similar circle compactification to be consistent, it is required that
\begin{equation}
    \label{well defined CS coefficient for the  2nd model}
    \Delta c_{000} + \frac{\Delta a^{\6d}_0}{2} \in 6\bZ.
\end{equation}

Using \eqref{difference CS coefficient from 6d hypers} and \eqref{B2 B4 polynomials}, equation 
\eqref{well defined CS coefficient for the  2nd model} 
can be rewritten as
\begin{equation}
\label{Consistency condition for CS term with Zk holonomy 2nd}
   \sum_{s\neq 0} \frac{ \Delta x_s}{24k}(-2ks + 2 s^2 - k^2s^2 + 2 ks^3 - s^4) = 0 \ \text{mod} \  \bZ,
\end{equation}
which is exactly the same as the 6D $\bZ_k$ anomaly-free condition, \eqref{Moore Moonier eta}! 

\subsection{5D integral cubic form
from $S^1$ compactifications}
\label{subsec:QS1}

As we saw in the last subsection, the condition of having well-defined 5D CS terms upon reduction from a 6D  theory with $\bZ_k$ symmetry \eqref{well defined CS coefficient for the  2nd model} is equivalent to the stronger set of conditions of \cite{Monnier:2019ytc} \eqref{Moore Moonier stronger condition}. However, as already mentioned, many consistent models satisfy the weaker condition \eqref{Moore Moonier weaker conditions}.

The quantization condition \eqref{5d consistency condition} is  necessary for having a well-defined 5D CS theory. However, in the derivation of \eqref{5d consistency condition}, we implicitly assumed the $U(1)$ basis to which the constants $C_{IJK}, a_{I}$ are adapted is integral, i.e. that monopole charges in this $U(1)$ basis are quantized in the unit of $\bZ$ (not $n\bZ$). More precisely,  a $U(1)$ gauge group basis is required such that the field strengths $F^I$ generate $H^2(M_5;\bZ)$, where $M_5$ is the 5D spacetime manifold. We refer to this integral $U(1)$ basis as the 5D $U(1)$ basis. 

On the other hand, the $S^1$ compactification of 6D supergravity \eqref{schematic map of circle compactification} chooses a basis, for which the CS coefficients $c_{IJK}$ are given by summing over KK towers of charged matter in the loop. We refer to this $U(1)$ basis as the (compactification) 6D basis. The 6D basis is not guaranteed to be integral. In fact, it is not an integral $U(1)$ basis even in F-theory models, as we will see in this subsection. This is the key to the question raised at the beginning of this section: the CS coefficients need to satisfy \eqref{quantization condition for 5d CS term for KK U(1)} in the 5D $U(1)$ basis, while this is not the case for $c_{IJK}$ given by one-loop calculation (i.e.\eqref{CS coefficient from 6d hyper}). This also applies to  the gravitational CS coefficients --  $a_I$ is used together with the 5D integral vector basis, while the coefficients found via circle reduction are denoted by $a^{\6d}_I$.
In order to verify the 5D consistency conditions \eqref{5d consistency condition},
the one-loop generated $c_{IJK}$ and  $a^{\6d}_I$ need to be transformed to an integral basis of vector fields.

In this subsection we will see that the 5D consistency conditions \eqref{quantization condition for 5d CS term for KK U(1)} applied to the 6D $U(1)$ basis are indeed the same as \eqref{Moore Moonier eta with GS} in the M/F-theory framework. F-theory models satisfy the consistency condition \eqref{Moore Moonier weaker conditions}.  

\paragraph{The 5D $U(1)$ basis}
Under the $S^1$ compactification \eqref{schematic map of circle compactification}, the generators of the 5D $U(1)$ basis come from:
 \begin{itemize}
     \item The KK $U(1)$, which we denote by $U(1)_0$, and its field strength $F^0$. 
     \item 6D  vector fields with gauge group $G$ 
     with field strength $F^i$, $ i = 1,\dots,\text{rk}(G)$. (When $G$ is non-Abelian, we assume a generic Wilson line on $S^1$ is turned on which breaks $G$ to $U(1)^{\text{rk}(G)}$.)
     \item 6D (anti-)self-dual tensors. We denote the associated field-strengths by $F^{\alpha}$, with  $\alpha = 1,\dots, T +1$.    
 \end{itemize}

 To apply \eqref{5d consistency condition} in the 6D $U(1)$ basis, we need to first find the transformation between the 6D $U(1)$ basis $F^{6\text{D},I}$ and the 5D $U(1)$ basis $F^I$:
 \begin{equation}
     \label{transformation between 6d and 5d basis}
     F^{6\text{D},I} = M^{I}_{\ J} F^J, \qquad I,J \in \{0,i,\a\}. 
 \end{equation}
$M^{I}_{\ J} \in \mathbb{Q}$ is  rational. Note that the 5D integral basis is denoted simply by $F^I$ ($I=0,..., n_V$) and $F^I = \{ F^0, F^i, F^{\alpha}\}$, while the (non-integral) 6D basis $F^{6\text{D},I}$ is explicitly labeled by $\6d$. 

The trilinear couplings in these two bases, $F^{\6d, I}$ and $F^I$, given by  $c_{IJK}$ and $C_{IJK}$  respectively, are related accordingly:
\begin{equation}
    \label{CS coefficients in two basis}
    c_{IJK} = (M^{-1})_{I}^{\ \, I'}(M^{-1})_{J}^{\ \, J'}(M^{-1})_{K}^{\ \, K'}C_{I'J'K'}.
\end{equation}
It is $C_{000}$ and $a_0$ that need to satisfy \eqref{quantization condition for 5d CS term for KK U(1)}, not the one-loop generated $c_{000}$ {and $a^{\text{6D}}_0$. Only by transforming \eqref{quantization condition for 5d CS term for KK U(1)} to the 6D basis using \eqref{transformation between 6d and 5d basis} can we determine the correct quantization conditions for the coefficients in that basis. Since we are interested in $c_{000}$ and $a^{\text{6D}}_6$, we do not need the entire $M^{I}_{\ J}$ matrix, but only the column $M^{0}_{\ J}$. 

\vspace{.5cm}
\noindent
To find the matrix \eqref{transformation between 6d and 5d basis}, we recall that
\begin{itemize}
    \item $c_{ijk}$ can be generated by summing over the KK tower of hypermultiplets charged under the 6D $U(1)_i$ that make up the (Cartan subgroup of the) gauge group.
    \item Similarly for $c_{000},c_{0ii},c_{00i}$.
    \item $c_{0\a\b}$ come from  6D (anti-)self-duality and the equations of motion. 
    \item $c_{00\a}$ cannot be generated by summing over the KK tower.
\end{itemize}
This suggests that $M^{0}_{\ J}$ (modulo integral transformation\footnote{Notice the solutions to \eqref{transformation between 6d and 5d basis} come in families, as there are families of 5D bases.}) in \eqref{transformation between 6d and 5d basis} is of the form
\begin{equation}\label{eq:M0J}
    M^{0}_{\ J} = \delta_{J}^{0} + x_\a \delta^{\a}_{J}, \quad \a = {1,\dots,T+1 } \,\, (\text{runs over all 6D tensors}), 
\end{equation}
subject to solving
\begin{equation}
    \label{U(1) basis of circle compactification}
    c_{00\a} = (M^{-1})_{0}^{\  I'}(M^{-1})_{0}^{\ J'}(M^{-1})_{\a}^{\ K'}C^{5\text{D}}_{I'J'K'} = 0.
\end{equation}
\paragraph{Remark} We find based on some experimental evidence that the other components of \eqref{transformation between 6d and 5d basis} take the following form:
\begin{equation}
    \label{solution of transformation matrix}
    M^I_{\ J} =  n \delta_{J}^{I}, \quad I \neq 0,
\end{equation}
where $n\in \bZ$ is an integer, needed to insure that $C_{IJK}$ is given by an integral basis. The divisors in F-theory that give rise to the $U(1)$ fields are in 
 general $\mathbb{Q}$ divisors. The fractional part comes from the inverse Cartan matrix, see \cite{Park:2011ji}. This also explains why the one-loop generated coefficients do not satisfy the integrality condition \eqref{U1 intersection in 6d basis}. Instead, if we set $n=2k > 0$,   (using string charge quantization \eqref{3.15}) \eqref{U1 intersection in 6d basis} becomes
\begin{equation}
    c_{iii} + \frac{a_i}{2} = \frac{1}{2}\sum_{\text{h}}n^3(q^{\text{h}}_{i})^3 - nq^{\text{h}}_i \in 6\bZ,
\end{equation} and the integrality condition is satisfied.

\vspace{.5cm}
It is generally not clear how to fix an integral 5D $U(1)$ basis from the IR. Here we will focus on the theories that are obtained from elliptic or genus-one CY threefolds, i.e. the  M/F-theory framework. Now an integral $U(1)$ basis is given by a basis of $H^2(\text{CY3};\bZ)$ (or the Poincaré dual $D_I \in H_{4,\text{free}}(\text{CY3};\bZ)$), and $C_{IJK}$ and $a_I$ are given by the intersection numbers of the CY3, $D_I\cdot D_J \cdot D_k, D_I \cdot c_2(\text{CY3})$. We can identify the divisors  corresponding to each $U(1)$:
\begin{itemize}
    \item $D_{0} = B$, the divisor given by the base of the elliptic/genus-one fibration. It corresponds to the $U(1)_0$ in the 5D (integral) $U(1)$ basis.
    \item $D_\a$, the divisors pulled back from the base divisors $D_\a = \pi^{-1}(C_\a), C_\a \in H_{2,\text{free}}(B;\bZ)$. These correspond to the $U(1)_\a$ gauge fields in the 5D (integral) $U(1)$ basis.
    \item $D_i$, the remaining divisors, which correspond to $U(1)_i$ in the 5D (integral) $U(1)$ basis.
\end{itemize}
\paragraph{Remark} To be more precise, the CY3 $X$ on the M-theory side is given by the resolution of the elliptic (genus-one) fibration on the F-theory side, $Y$ (if $Y$ is singular):
\begin{equation}
    \label{resolution of M/F theory}
    \sigma: X \to Y.
\end{equation}
The divisors introduced by resolution are  part of $D_i$,\footnote{Additional $D_i$ come from the free part of the Mordell-Weil  group of $Y$} and their associated Kähler moduli are (locally) parametrized by Wilson lines on the circle. The $D_0$ and $D_{\a}$ are pull-backs of the divisors on $Y$ via \eqref{resolution of M/F theory}. In the genus-one fibration case, $D_0$ is identified with the $k$-section. 

\paragraph{M/F-theory with $\bZ_k$ gauge group} We need some basic facts about the $\bZ_k$ gauge group in the M/F-theory framework,\footnote{To the best of our knowledge, the mathematical verification in general cases is still not settled.} see \cite{Cvetic:2015moa, Cvetic:2018bni} for details:
\begin{itemize}
    \item The $\bZ_k$ gauge group is realized by $H_{\text{tor}}^3(\hat{Y};\bZ)$, where $\hat{Y}$ is the Jacobian of the genus-one fibration $Y$. Genus-one fibrations that share the same Jacobian $\hat{Y}$ are classified by $H^3_{\text{tor}}(\hat{Y};\bZ)$ under some assumptions \cite{dolgachev1993ellipticthreefoldsioggshafarevich}. A subtle point regarding this is illustrated via an example with $H_{\text{tor}}^3(\hat{Y};\bZ) = \bZ_3$ in  \cite{Cvetic:2015moa}.
    \item F-theory on a genus-one fibration gives the same physical theory as on its associated Jacobian fibration.
    \item The differences appear on the M-theory side, where elements of $ H_{\text{tor}}^3(\hat{Y};\bZ)$ are identified with the $\bZ_k$ holonomy on the circle compactification. With the chosen $\bZ_k$ holonomy, the physics is identified with M-theory on the corresponding genus-one fibration. 
\end{itemize}
    
\subsubsection{Solving $c_{0 0 \alpha} =0$ in M/F-theory} 
\label{Solving in $M/F$ thoery}
 As mentioned before, the CY3 in M-theory provides a natural integral $U(1)$ basis, labeled by $D_I \in  H_{4,\text{free}}(\text{CY3};\bZ)$. In this integral $U(1)$ basis, we have:
\begin{equation}
    \label{5d CS coefficients from M theory}
    C_{IJK} = D_I\cdot D_J \cdot D_{K}, \quad  a_I = D_I \cdot c_2(\text{CY3})\,.
\end{equation}
We already showed in Section \ref{sec:cubic} that \eqref{5d consistency condition} are satisfied by \eqref{5d CS coefficients from M theory}. 

\vspace{.5cm}
With the knowledge of a 5D (integral) $U(1)$ basis and the CS coefficients $C_{IJK}$ in this basis, we can verify \eqref{eq:M0J} and \eqref{CS coefficients in two basis}. Equation \eqref{U(1) basis of circle compactification} gives the 6D $U(1)$ basis with one-loop induced CS coefficients. According to \eqref{solution of transformation matrix}, $U(1)_0$ in the 6D $U(1)$ basis is different from $U(1)_0$ in the 5D $U(1)$ basis. The relation between the two is
\begin{equation}
    \label{6d U(1) KK generator}
    B^{6\text{D}} = B+ D_f = B + x^\a D_\a,
\end{equation}
and $D_f = x^\a D_\a$ ($x^\a$ are rational constants) is constrained by  \eqref{U(1) basis of circle compactification} as
 \begin{equation}
    \label{vertical divisor ambiguity}
   c_{00\a} = 0 \ \ \Rightarrow \ \   B^{6\text{D}}\cdot B^{6\text{D}}\cdot D_{\a} = (B+ D_f)(B+D_f)(D_{\alpha}) = 0, \quad \forall \a,
\end{equation}
i.e. $D_f$ satisfies
\begin{equation}
    \label{vertical divisor ambiguity 2nd}
    B\cdot B\cdot D_{\alpha} = -2 D_f \cdot B\cdot D_{\alpha}, \quad \forall \a.
    \end{equation}

\paragraph{Example 1: Elliptically fibered CY3}
This case corresponds to a reduction of 6D supergravity, obtained from F-theory on a CY3, on $S^1$ without discrete holonomies.

Applying \eqref{vertical divisor ambiguity 2nd} to the elliptically fibered CY3 gives
\begin{equation}
    \label{vertical divisor ambiguity in elliptic case}
    B\cdot B\cdot D_{\alpha} = -c_1(B)\cdot \pi (D_\alpha) = -2 D_\alpha  \cdot   \pi(D_f),
\end{equation}
where $\pi: \text{CY3} \to B$.  The solution is $\pi(D_f) = \frac{1}{2} c_1(B)$, and in the 6D $U(1)$ basis, $U(1)_0$ corresponds to $B^{6\text{D}} = B+ \frac{1}{2} \pi^{-1}(c_1(B))$. As a result, the one-loop generated $c_{000}$ and $a^{\6d}_0$  in the 6D $U(1)$ basis should be $B^{6\text{D}}\cdot B^{6\text{D}}\cdot B^{6\text{D}}$ and $B^{6\text{D}}\cdot c_2(\text{CY3})$, respectively.  Explicit one-loop calculation shows this is indeed the case \cite{Bonetti:2013ela}. Hence, \eqref{U(1) basis of circle compactification} provides an alternative way of deriving the results in \cite{Bonetti:2013ela}. 

\paragraph{Example 2: Genus-one fibered CY3}
This case corresponds to 6D supergravity, obtained from F-theory after reduction on a CY3 and further compactified on $S^1$ with discrete $\bZ_k$ holonomies. In the genus-one fibered CY3 case, $B$ will be the $k$-section and $B\cdot D_{\alpha}\cdot D_{\beta} = k \pi(D_\a)\cdot \pi(D_\b) \in k\bZ$.  The $U(1)_0$ in the 6D basis corresponds to $B^{6\text{D}} = B + D_{f} = B+ \frac{D_v}{2k}$ ($D_v$ is an integral divisor), and \eqref{vertical divisor ambiguity 2nd} yields
\begin{equation}
    \label{vertical divisor ambiguity in genus one case}
    B\cdot B\cdot D_{\alpha} = -\frac{2}{2k} D_v \cdot B\cdot D_{\alpha} = -\frac{1}{k} \pi(D_v) \cdot \pi(D_\a) \qquad \forall \a.
\end{equation}
 
  \paragraph{Remark} One peculiar fact: the divisor corresponding to $U(1)_0$ in the 6D basis, $B^{6\text{D}}= B + \frac{D_v}{2k}$,  is the same as the divisor basis adopted in \cite{Cota:2019cjx}. 
  This basis has the property that the flat coordinates of the Kähler moduli space in the large-base limit have $\Gamma_1(k)$ modular behavior. This behavior follows from the relative conifold transformation, and it makes the topological string partition function (partially) solvable via the modular bootstrap ansatz. It would be interesting to understand better whether this is a coincidence. 

Unlike the elliptic case above, there in general does not exist a universal solution to \eqref{vertical divisor ambiguity in genus one case} in terms of divisors. Here we check the solution $B^{6\text{D}}= B + \frac{D_v}{2k}$ by matching the one-loop induced $c_{000}$ with $B^{6\text{D}}\cdot B^{6\text{D}}\cdot B^{6\text{D}}$ for a family of genus-one CY threefolds discussed in \cite{Dierigl:2022zll}.  
\begin{table}[h]
\label{table of genus one CY3}
    \centering
    \begin{tabular}{|c|c|c|c|c|c|c|c|c|}
        \hline
        $Y_i$ & $x_{1} + x_2$ & $x_0$ &   $C_{111}$ & $C_{112}$ & $C_{122}$ & $C_{222}$ & $c_{2}\cdot D_1$ & $c_{2}\cdot D_2$   \\
        \hline
        1 & 177 & 96 & 14 & 7 & 3 & 0 & 68 & 36  \\
        2 & 180 & 93 & 21 & 9 & 3 & 0 & 78 & 36  \\
        3 & 186 & 87 & 5 & 5 & 3 & 0 & 50 & 36  \\
        4 & 186 & 87 & 11 & 7 & 3 & 0 & 62 & 36   \\
        5 & 189 & 84 & 0 & 3 & 3 & 0 & 36 & 36   \\
        6 & 195 & 78 & 2 & 5 & 3 & 0 & 44 & 36   \\
        7 & 198 & 75 & 8 & 7 & 3 & 0 & 56 & 36  \\
        8 & 198 & 75 & 15 & 9 & 3 & 0 & 66 & 36   \\
        9 & 204 & 69 & 5 & 7 & 3 & 0 & 50 & 36 \\
        10 & 207 & 66 & 12 & 9 & 3 & 0 & 60 & 36  \\
        11 & 213 & 60 & 2 & 7 & 3 & 0 & 44 & 36 \\
        12 & 216 & 57 & 9 & 9 & 3 & 0 & 54 & 36  \\
        13 & 225 & 48 & 6  & 9 & 3 & 0 & 48 & 36   \\
        \hline
    \end{tabular}
    \caption{Sample Data Table}
    \label{tab:sample_data}
\end{table}
 
 Table \ref{tab:sample_data}    lists 13 genus-one fibered smooth CY threefolds, $Y_i, i =1,\dots,13$ with the base $B= \bC P^2$ from \cite{Dierigl:2022zll}. The triple intersections $C_{111},C_{112},C_{122}, $ and $C_{222}$ have already been computed there.
 We computed the last two columns, $c_{2}\cdot D_1$ and $c_{2}\cdot D_2$. 
 
 F-theory on $Y_i$ gives 6D supergravity with gauge group $\bZ_3$, $n_T=0$, and $n_H = x_0+ x_1+x_2 = 273$. Here $x_n$ ($n=0,1,2$) are the numbers of hypermultiplets with $\bZ_3$ charge $n$. $H_4(Y_i;\bZ) = \bZ[D_1]\oplus \bZ[D_2]$, where $D_1 =B$ is the divisor given by the base $B=\bC P^2$ and $D_2 = \pi^{-1}[H]$ is the pull-back from the base of the divisor $[H]$. $C_{ijk}$ in Table \ref{tab:sample_data} give the triple intersection data. Note $C_{122} = 3$ for all $Y_i$, indicating that these are genus-one fibered threefolds with a $3$-section. 

 Solving \eqref{U(1) basis of circle compactification} for the $Y_i$'s gives
 \begin{equation}
     \label{6d U(1)0 for genus one fibered Yi}
     B^{6\text{D}} = B -\frac{C_{112}}{6}D_2.
 \end{equation}

As a result, the one-loop induced CS coefficient $c_{000}$ should be
\begin{equation}
    \label{C000 from one loop for genus one Yi}
   c_{000} =  B^{6\text{D}}\cdot B^{6\text{D}}\cdot B^{6\text{D}}= C_{111} -\frac{C^2_{112}}{4},
\end{equation}
and $a^{\6d}_0$ should be:
\begin{equation}
    \label{a0 for genus one Yi}
    a^{\6d}_0 = B^{6\text{D}} \cdot c_2(Y_i) = \Big(B -\frac{C_{112}}{6}D_2\Big)\cdot c_2 (Y_i).
\end{equation}
Using the one-loop calculation \cite{Bonetti:2013ela} with properly normalized KK tower (as there is a discrete $\bZ_3$ holonomy along $S^1$), it is checked that \eqref{C000 from one loop for genus one Yi}  and \eqref{a0 for genus one Yi} indeed give the one-loop generated $c_{000}$ and ${a^{\text{6D}}_0}$ for all $Y_i$'s in Table \ref{tab:sample_data}.

\paragraph{($B^{6\text{D}}\cdot B^{6\text{D}} \cdot B^{6\text{D}}$, $B^{\6d}\cdot c_2(\text{CY3})$)$_{\textbf{M-theory}}$  $=$ (One loop generated $c_{000},a^{6\text{D}}_0$)$_{\textbf{F-theory}}$} Notice in the above discussions that $c_{00\a} =0$ is enough to solve for $B^{6\text{D}}$, while the fact that its intersection numbers are precisely the same as the  CS terms generated at one-loop from circle compactification, can be considered a result of M/F-theory duality:
 \begin{itemize}
     \item The 5D CS terms from the F-theory side are given by one-loop effects after circle compactification.
     \item The 5D CS terms from the M-theory side are given by triple intersection numbers on the CY3 from $-\frac{1}{6}C\wedge G \wedge G + C\wedge X_8$ in the $11$D supergravity Lagrangian.
     \item As M- and F-theory give the same 5D supergravity, the 5D CS terms should be the same up to a basis change \eqref{transformation between 6d and 5d basis}. For 6D $U(1)_{{0}}$, the basis change can be determined by solving $c_{00\a}=0$.
 \end{itemize}

\paragraph{Remark} The above discussion also resolves the contradiction in \eqref{KK coeficients from cirlce reduction 1st}. 

Here we want to illustrate this point in the M/F-theory framework, as in this case  the $({0,\a,\b})$ interaction ($\a,\b$ indicate that the 5D vectors originate from 6D tensors) can be computed both from the 6D Lagrangian and from the triple intersection of the underlying CY3 manifold. The former is denoted by $c_{0\a\a}$ and the latter by $C_{0\a\a}$. As mentioned, $C_{0\a\a}$ corresponds to the 5D integral basis, and hence should satisfy \eqref{5d consistency condition}:
\begin{equation}
    \label{cosnsitency condition in a integral base 1st}
    C_{0\a\a}+ C_{00\a} = B\cdot D_\a\cdot D_\a + B\cdot B\cdot D_\a\in 2 \bZ,
\end{equation}
where $B$ is the base divisor. However, the $c_{0\a\a}^{6\text{D}},c_{00\a}^{6\text{D}}$ are not given in  an integral basis. The $U(1)_0$ in the 6D basis is given by a fractional shift of the base divisor $B$, i.e. ${B}^{\text{6D}} = B+ \pi^{-1}(\frac{1}{2}c_1(B))$ (here we take the elliptic fibration for simplicity). In the 6D basis the consistency condition \eqref{cosnsitency condition in a integral base 1st} is no longer valid. However, we could first translate the consistency condition in the integral 5D basis to the fractionally shifted one and then check if it is satisfied by \eqref{KK coeficients from cirlce reduction 1st}. As the relation of the two bases is given by $(D_{0},D_\a ) =\big(B+ \frac{1}{2}c_1(B),  D_\a\big)$, we have:
\begin{equation}
    c_{0\a\a} = \big(B+\pi^{-1}\big(\tfrac{1}{2}c_1(B)\big)\big)\cdot D_\a\cdot D_\a, 
\end{equation}
\begin{equation}
    c_{00\a} = (B+\tfrac{1}{2}c_1(B))\cdot  (B+\tfrac{1}{2}c_1(B))\cdot D_\a = C_{00\a} + c_1(B) \cdot \pi(D_{\a})=0.
\end{equation}
Here $c_1(B)\cdot \pi(D_\a)$ is the intersection on the base. Now the consistency condition becomes
\begin{equation}
\label{consistency condition in shifted basis 1st}
    c_{0\a\a} + c_{00\a}  - c_1(B) \cdot \pi(D_{\a}) =  C_{0\a\a}+ C_{00\a} = 0 \ \text{mod} \ 2,
\end{equation}
 which gives:
\begin{equation}
    c_{0\a\a} = c_1(B) \cdot \pi(D_{\a}) \,\,\,  \text{mod} \  2.
\end{equation}
It is this consistency condition we should apply in \eqref{KK coeficients from cirlce reduction 1st}. On the base $B$, the adjunction formula  yields:
\begin{equation}
K\cdot \pi(D_\a) +  \pi(D_\a)\cdot \pi(D_\a) = -c_1(B)\cdot \pi(D_\a) +  \pi(D_\a)\cdot \pi(D_\a) = 2g(\pi(D_{\a}))-2  \in 2\bZ,
\end{equation}
where $g(\pi(D_{\a}))$ is the arithmetic genus of base divisor $\pi(D_{\a})$. Together with
\begin{equation}
    \Omega_{\a\a} = \pi(D_\a)\cdot \pi(D_\a) = c_1(B)\cdot \pi(D_\a) \ \ \text{mod} \ \, 2,
\end{equation}
we see that \eqref{KK coeficients from cirlce reduction 1st} indeed satisfies the condition \eqref{consistency condition in shifted basis 1st}, i.e.
\begin{equation}
    c_{0\a\a} + c_{00\a} = c_{0\a\a}= \Omega_{\a\a} = c_1(B)\cdot \pi(D_\a)  \ \ \text{mod} \  2.
\end{equation}
Note that this argument uses the M/F-theory framework and  the key fact that $c_1(B)$ is the characteristic vector in the lattice of $H_2(B)$. When translated to the supergravity data, the lattice $H_2(B)$ maps to the BPS charge lattice $\Lambda$, and $c_1(B)$ maps to the anomaly coefficient $\ta \in \Lambda$. That $\tilde{a}$ should be a characteristic vector of $\Lambda$ is proven in  from the supergravity point of view in \cite{Monnier:2018nfs}.

The contradiction in equation \eqref{eq:miss3} is resolved in a similar fashion.

Although the discussion above uses the M/F-theory framework, it seems natural to expect that the consistency condition \eqref{5d consistency condition} in the basis picked by circle reduction will be shifted to $ c_{00\a} + c_{0\a\a} = \Omega_{\a \b }{\tilde{a}^{\b} (\text{6D})}$ $\text{mod} $ $2$ (notice that $\Omega_{\a\a}+ \Omega_{\a \b}\ta^\b \in 2\bZ $, since $\ta$ is a characteristic vector of $\Lambda$) for any consistent supergravity theory. This claim hinges on finding an integral basis from the IR perspective (the (co)homology of the CY3 provides such a basis in the M/F-theory framework). We hope to return to this question in the future.

\paragraph{Example: model with $T=1$, $V=0$, $H=244$} In the supergravity with such a field content, the anomaly is given by $\frac{1}{16} (\tr \tR^2)^2$, and there are seemingly two inequivalent anomaly factorizations -- a symmetric one with $\tilde{a} = (2,2)$ and one with $\tilde{a} = (4,1)$. The second is inconsistent since $\tilde{a} = (4,1)$ is not characteristic \cite{Monnier:2017oqd, Monnier:2018nfs}. It is not hard to see that for this choice of $\tilde{a}$, $ c_{00\a} + c_{0\a\a} = \Omega_{ \a \b} \tilde{a}^{\b}$ $\text{mod} $ $2$ is not even for all $\a$}, and hence the circle reduction to 5D detects the inconsistency of the 6D theory.

For the consistent choice $\tilde{a} = (2,2)$, it is not hard to compute the intersections 
\beq \label{eq:244F}
-\frac{1}{3} \int_{M_6}  (F^{\6d,0})^3 + 3 F^{\6d,0} \wedge F^{\6d,1} \wedge F^{\6d,2},
\eeq
with only $c_{000} =2 $ and $c_{012} = 1$ (with permutations)  non-vanishing and with  $a_I^{\6d} = (44, 24, 24)$.

It is not hard to see that a linear transformation to a 5D basis $F^{\6d,0} \mapsto F^0$,   $F^{\6d,1} \mapsto F^2 + F^0$,  $F^{\6d,2} \mapsto F^1 + F^2 + F^0$ transforms \eqref{eq:244F} to 
\beq
-\frac{1}{6} \int_{M_5} 8 (F^0)^3 + 6 (F^0)^2 (F^1 + 2 F^2) + 6F^0 (F^1 \wedge F^2 + (F^2)^2)
\eeq
with gravitational couplings given by $a^I = (92,24,48)$.
These are the intersection numbers of CY3 $X_{24}(12, 8, 2, 1, 1)$ with Hodge numbers $h^{1,1} = 3$ and $h^{2,1} = 243$  (see e.g. \cite{Hosono:1995bm}), as expected by the duality between the reduction of the 6D model and the CY compactification of M-theory.

\subsubsection{M/F-theory duality and constraints on $\bZ_k$ charges in 6D}
\label{subsec:TGS}

Recall that due to the topological GS mechanism, the necessary consistency condition for 6D supergravity $\cT_B$ is:
\begin{equation}
\label{necessary 6d Zk anomaly free condition}
    \sum_{s=0}^{k-1}  \Delta x_s\frac{1}{24 k}\left(-2 k s+2 s^2-k^2 s^2+2 k s^3-s^4\right)  \in  \frac{\bZ}{2k}.
\end{equation}
In the previous section, we interpreted the left-hand side as the difference of one-loop generated KK $U(1)$ CS terms between $\cT_A$ and $\cT_B$ (with $\cT_A$ well-defined) under circle compactification with one unit holonomy:
\begin{equation}
    \label{one loop CS difference}
    \frac{1}{6} \Bigl(\Delta c_{000} + \tfrac{1}{2}\Delta a^{6\text{D}}_0 \Bigl).
\end{equation}
There are two ways to understand why \eqref{Moore Moonier stronger condition} is too strong:
 \begin{itemize}
     \item In 6D \cite{Monnier:2018nfs}, it does not take into account the contribution from topological GS terms. 
     \item In 5D, we have already argued that the quantization condition should apply to the 5D integral basis. The one-loop generated couplings do not need to satisfy it.
 \end{itemize}  
In order to reconcile the weak condition \eqref{necessary 6d Zk anomaly free condition} with the consistency of 5D supergravity, we need to show that the quantization condition for $\cT_B$ (after circle compactification) in the 6D basis is indeed compatible with \eqref{necessary 6d Zk anomaly free condition}, i.e. is given by
\begin{equation}
    \label{one loop CS consistency in 6d basis}
    \Delta c_{000} + \tfrac{1}{2} \Delta a^{6\text{D}}_0 \ \in \ \frac{3\bZ}{k}.
\end{equation}

\vspace{.5cm}
Due to the fact that it is hard to find a 5D basis from the IR for 5D supergravity theories, here we will once more focus on the M/F-theory case. As already discussed in Section \ref{sec:cubic}, the 5D supergravities coming from the compactification of M-theory on a CY3 automatically satisfy the quantization condition in the 5D basis \eqref{5d consistency condition}. The relation between $U(1)_0$ in the 6D basis and the 5D basis is solved by \eqref{6d U(1) KK generator}. In this section, we show that \eqref{one loop CS consistency in 6d basis} is satisfied in general M/F-theory cases.\footnote{We take the CY3 to be a genus-one fibered manifold admitting a crepant resolution, if singular. Shortly, we will discuss one possible way to drop this assumption.} The strategy is as follows:

 \begin{itemize}
     \item According to the M/F-theory duality, we can relate the one-loop induced CS coefficients $(c_{000},a^{\6d}_{0})$ in the 6D basis with the intersection numbers of the fractional divisor $(B^{6\text{D}}\cdot B^{6\text{D}}\cdot B^{6\text{D}}, B^{6\text{D}}\cdot c_2(\text{CY}3))$ (see previous subsection).
     \item The left-hand side of \eqref{necessary 6d Zk anomaly free condition} is the same as $(\Delta c_{000}+ \tfrac{1}{2}\Delta a^{6\text{D}}_{0})/6$ generated by one loop.
     \item The above two points establish the identification of the left-hand side of \eqref{necessary 6d Zk anomaly free condition} and the intersection data of the ($\mathbb{Q}$) divisor $B^{6\text{D}}$ up to a factor of 6. Hence we can study the intersection data of $B^{6\text{D}}$ and compare with \eqref{necessary 6d Zk anomaly free condition}.
 \end{itemize}

\paragraph{The relative set-up in M/F-theory} 
\begin{itemize}
    \item Consider two 6D $\mathcal{N} = (1,0)$ supergravity  theories  $\cT_A$ and $\cT_B$ which only differ in matter content with respect to the $\bZ_k$ gauge group (the differences of hypermultiplets with $\bZ_k$ charge $i$ is $\Delta x_i$). Assuming $\cT_A$ is well-defined, the consistency of $\cT_B$
  will impose conditions on $\Delta x_i$  \eqref{Moore Moonier weaker conditions}.
    \item In the M/F-theory framework, this translates into 6D theories $\cT_A$ and $\cT_B$ arising from F-theory on two different genus-one CY3 manifolds, $Y_{A}$ and $Y_B$ respectively, that have the same base\footnote{We believe this is the general case. One hint is that $\cT_A$ and $\cT_B$ have the same perturbative anomalies, which are known to be tied to the geometry of the base $B$, see \cite{Cheng:2021zjh, Katz:2020ewz, Kumar:2009ac, Kumar:2010ru}.} but non-equivalent Jacobian fibrations. Compactifying both on $S^1$ with one unit $\bZ_k$ holonomy and no Wilson lines for any continuous gauge group is the same as compactifying M-theory on $Y_A$ and $ Y_B$ respectively.
\end{itemize}
The one-loop-induced CS coefficients $c_{000, A(B)}$ and $a^{6\text{D}}_{0,A(B)}$ for $\cT_{A(B)}$ compactified on $S^1$ with one unit $\bZ_k$ holonomy should be given by
\begin{equation}
\begin{split}
\label{fraction C000 from KK summation}
    c_{000,A(B)} = B^{6\text{D}}_{A(B)}\cdot  B^{6\text{D}}_{A(B)} \cdot  B^{6\text{D}}_{A(B)} &= \Big(B_{A(B)}+ \frac{1}{2k}D_{v,A(B)}\Big)^3 \\& = B_{A(B)}^3 + \frac{3}{4k} B_{A(B)}^2 \cdot D_{v,A(B)},
\end{split}
    \end{equation}
     \begin{equation}
         \label{fractional a0 in 6d basis}
        a^{6\text{D}}_{0,A(B)} =  B^{6\text{D}}_{A(B)}\cdot \frac{1}{2}c_2 = B_{A(B)}\cdot \frac{1}{2}c_2 + \frac{12}{4k} \pi(D_{v,A(B)})\cdot c_1(B),
     \end{equation} 
     where $B_{A(B)}$ is the $k$-section of $Y_{A(B)}$ and $D_{v,A(B)}$ is the divisor in $Y_{A(B)}$ which satisfies \eqref{U(1) basis of circle compactification}:
     \begin{equation}
         \label{definition of Dv}
         B_{A(B)}\cdot B_{A(B)}\cdot D_{\alpha,A(B)} = -\frac{2}{2k} D_{v,A(B)} \cdot B_{A(B)}\cdot D_{\alpha,A(B)} = -\frac{1}{k} \pi(D_{v,A(B)}) \cdot \pi(D_{\a,A(B)}),  
     \end{equation}
     for $D_{\a,A(B)}$ any integral divisor in $Y_{A(B)}$ coming from the pull-back of the integral divisors in the base $B_{A(B)}$.
    $B_{A(B)},D_{v,A(B)},$ and $D_{\a,A(B)}$ are integral divisors inside the CY3, and they automatically satisfy (see Section \ref{sec:cubic})
     \begin{equation}
         \label{CY3 intersection condition}
          D_{A(B)}^3 + \frac{D_{A(B)}\cdot c_2(Y_{A(B)})}{2} \in 6\bZ,  \quad D = B,D_v,D_\a.
     \end{equation}
     Using \eqref{fraction C000 from KK summation}, \eqref{fractional a0 in 6d basis}, and \eqref{CY3 intersection condition}, we find:
     \begin{equation}
         \label{fractional intersection in 6d basis}
         \begin{split}
         &\ \ \ \ \ \ \ \ \ \  \ \ \ \ \ \  B_{A(B)}^{6\text{D}}\cdot B_{A(B)}^{6\text{D}} \cdot B_{A(B)}^{6\text{D}} + \frac{1}{2}B_{A(B)}^{6\text{D}}\cdot c_2(Y_{A(B)}) = \\& -\frac{3}{4k}\big(\pi(D_{v,A(B)})\cdot \pi(D_{v,A(B)})\big) +\frac{3}{k}\pi(D_{v,A(B)})\cdot c_1(B_{A(B)}) \ \, \text{mod} \ 6.
    \end{split}
     \end{equation}
The second line is written in terms of the intersection numbers of the base divisor $B$. Since $Y_{A}$ and $Y_B$ share the same base, we can directly compare the difference \eqref{one loop CS difference}:
\begin{equation}
    \label{diference between two consistent combination}
  \Delta c_{000} + \frac{1}{2}\Delta a^{6\text{D}}_{0} =  -\frac{3}{4k}\big(\pi(D_{v,A})^2-\pi(D_{v,B})^2\big) + \frac{3}{k}\big(\pi(D_{v,A} -D_{v,B})\cdot c_{1}(B)\big) \ .
\end{equation}
 To gain further information from \eqref{diference between two consistent combination}, note that $D_{v,A(B)}$ is characterized by \eqref{vertical divisor ambiguity in genus one case}:
\begin{equation}
     \label{characterize vertical divisor}
        -k B_{A(B)}^2 \cdot D_{\alpha,A(B)} = B_{A(B)}\cdot D_{v,A(B)} \cdot D_{\alpha,A(B)},
\end{equation}
for all $D_{\alpha,A(B)}$ divisors of $Y_{A(B)}$ pulled-back from the divisors in the base $B_{A(B)}$. Since $B_{A(B)},$ $D_{\alpha,A(B)}$, and $D_{v,A(B)}$ are integral, we find\footnote{Although the following is proven for smooth CY threefolds, it is also true for singular $Y_{A,B}$ with crepant singularities, as resolving the singularities will not change this specific set of intersection numbers. \eqref{even condition of CY3 intersection} may also be generalized to $Y_{A,B}$ with non-crepant singularities. One possible way to do so is to deform the singular CY3 to a smooth one (physically this corresponds to moving to a generic point of the Higgs branch). Then all results mentioned here go through for a CY3 with non-crepant singularities, provided the related intersection numbers remain invariant under deformation.}
    \begin{equation}
        \label{even condition of CY3 intersection}
        k B_{A(B)}^2 \cdot D_{\alpha,A(B)} = kB_{A(B)} \cdot D_{\alpha,A(B)}^2 \ \  \text{mod} \ \, 2k  \, .
    \end{equation}
    Equations \eqref{characterize vertical divisor} and \eqref{even condition of CY3 intersection} give $B_{A(B)}\cdot (D_{v,A(B)}-kD_{\a,A(B)})\cdot D_{\a, A(B)} \in 2k\bZ$. As $B$ is a $k$-section, we have
    \begin{equation}
        \label{characterize vertical divisor in terms of base}
       \big( \pi(D_{v,A(B)})-k\pi(D_{\a})\big)\cdot \pi(D_{\a}) \in 2\bZ.
    \end{equation} 
    Note this is a condition in terms of the base and holds for both $D_{v,A}$ and $D_{v,B}$.  We neglect the subscript $A(B)$ on $\pi(D_\a)$, since the divisor on the base $B$ is the same.
    Applying \eqref{characterize vertical divisor in terms of base} to $D_{v,A}$ and $D_{v,B}$ and taking the difference gives
    \begin{equation}
        \big(\pi(D_{v,A})-\pi(D_{v,B})\big) \cdot \pi(D_{\a}) \in 2\bZ \quad \ \ \ \ \forall D_{\alpha} \in \pi^{-1}(H_2(B)).
    \end{equation}
     Self-duality of $H_2(B;\bZ)$ implies that $D_{v,A} - D_{v,B}$ is valued in $2H_2(B;\bZ)$ and $\pi(D_{v, A})^2-\pi(D_{v,B})^2 \in 4\bZ$. Hence,
    \begin{equation}
        -\frac{3}{4k}\big(\pi(D_{v,A})^2-\pi(D_{v,B})^2\big) + \frac{3}{k}\big(\pi(D_{v,A} -D_{v,B})\cdot c_{1}(B)\big) \in \frac{3}{k} \bZ.
    \end{equation}  
As a result, \eqref{diference between two consistent combination} becomes
\begin{equation}
    \label{5d consistent condition in 6d basis from M/F}
    \Delta c_{000} + \tfrac{1}{2} \Delta a^{6\text{D}}_{0} = 0 \text{ mod } \frac{3}{k} \, ,
\end{equation}
i.e. exactly the weaker consistency condition \eqref{necessary 6d Zk anomaly free condition} derived in \cite{Monnier:2018nfs}.

To conclude:
\begin{itemize}
    \item The 6D $\bZ_k$ anomaly-free condition (with topological GS mechanism included) is related to the quantization condition of the 5D KK $U(1)$ CS term after circle compactification (as proven at least in the general M/F-theory context).
    \item The physical consistency condition \eqref{necessary 6d Zk anomaly free condition} proposed in \cite{Monnier:2018nfs} is indeed satisfied in general M/F-theory compactifications. 
\end{itemize}
\paragraph{Remark} In \cite{Monnier:2018nfs} it is proposed  (using differential cohomology) that the full GS term will have a (torsion) part in the $\bZ_k$ gauge group (related to the fact that $H^4(B\bZ_k;\bZ) = \bZ_k$). The coefficient of the torsion part is hard to determine. The analysis in this section indicates that this coefficient should be strongly related to the $0$th entry $M^{0}_{\ J}$ in \eqref{transformation between 6d and 5d basis}. Circle compactification may provide a way of computing this coefficient -- this problem is currently under our investigation. 

\subsection{Consistency conditions for CHL-like models}
String theory can give rise to interesting 6D supergravities where $\bZ_k$ is not just a simple gauge group, but also acts on a continuous gauge group $G^k$ or tensor multiplets. One can also find anomaly-free supergravity models that may display some apparent $\bZ_k$ symmetries (without knowing their stringy realization). If the symmetry is anomaly-free, the circle reduction of the 6D theory with $\bZ_k$ holonomy should lead to a consistent 5D theory. Hence, the examination of 5D consistency conditions is a way of determining if the 6D theory does indeed have an anomaly-free $\bZ_k$ symmetry.

We shall discuss this class of theories based on the prominent set of examples of the  $\bZ_2$ CHL action\footnote{Higher-dimensional anomalies in the CHL action were discussed e.g. in \cite{Debray:2023rlx, Basile:2023knk}.} of the $E_8\times E_8$ heterotic string compactified on K3 with symmetric instanton embeddings (i.e. (12,12) with instanton profiles for both $E_8$ identical) and consider two possibilities:
\begin{enumerate}
    \item[(1)] Symmetric instanton configuration with remaining gauge group $E_7\times E_7$: the CHL $\bZ_2$ swaps the two $E_7$ gauge groups \cite{Duff:1996rs}.\footnote{This theory is obtained by choosing $SU(2)$ instantons. Other choices of instantons will lead to different 6D gauge groups $G^2$, which can be completely Higgsed (in which case they can equally be thought of as the result of turning on two $E_8$ instantons with $c_2 = (12,12)$). The fully Higgsed theory, where $\bZ_2$ becomes an ordinary gauge symmetry, is probably the most studied $\mathcal{N} = (1,0)$ supergravity and has $T=1$ and $H=244$,  of which 20 hypermultiplets are neutral and 112 hypermultiplets are charged under $\bZ_2$. In all models related to each other by Higgsing, the difference in the number of hyper- and vector multiplets coming from each $E_8$ stays the same and is 112.} By taking linear combinations, we can make one $E_7$ vector multiplet neutral and one $E_7$ vector multiplet charged under $\bZ_2$. This is compatible with the spectrum, as the instanton configuration is symmetric. 
    \item[(2)] Symmetric instanton configuration with $n$ point-like instantons in each $E_8$: the supergravity theory has $T = 1 + n + n$, and the CHL action swaps the two sets of $n$ tensors given by the $n$ point-like instantons in the respective factors of $E_8$. Again, by taking linear combinations, we can make one sector neutral and make one carry charge 1 under $\bZ_2$. The corresponding multiplet numbers will be denoted as $T_1 = n$, $V_1$, and $H_1$.
\end{enumerate}

\noindent
In both cases, the $\bZ_2$ acts not just on the hypermultiplets and hence cannot be viewed as an ordinary $\bZ_2$ gauge group. However, using the same reasoning, the following should still apply:
\begin{itemize}
    \item In 6D, an arbitrarily chosen $\bZ_2$ action can be anomalous. Demanding the absence of non-perturbative anomalies imposes restrictions on how the $\bZ_2$ acts on the multiplets of the 6D supergravity theory.
    \item For circle compactification with one unit CHL $\bZ_2$ holonomies, the one loop generated KK $U(1)$ CS terms will depend on the $\bZ_2$ action on the 6D spectrum. The quantization condition \eqref{quantization condition for 5d CS term for KK U(1)} is a priori not guaranteed and needs to be verified. 
\end{itemize}
We already saw that the $6\text{D}$ $\bZ_k$ gauge anomaly is closely tied with the quantization condition for KK $U(1)$ CS terms after circle compactification. Could this relation also be generalized to the CHL $\bZ_k$-like action? In this section we speculate about this possibility.

\paragraph{Set-up} Here we will mainly focus on the two types of examples mentioned before. We know from construction that in both cases, the $\bZ_2$ action should be anomaly-free and that it acts on the matter multiplets in a specific way. We will slightly modify the relative setup:
\begin{itemize}
    \item $\cT_A$ refers to the theory compactified on $S^1$ without the CHL $\bZ_2$ holonomy. It is well-defined after $S^1$ compactification.\footnote{The resulting theory will have $n_V = 2$ and $n_H = 132$. We can compute the intersection numbers and gravitational couplings. Comparing with the reduction without $\bZ_2$ holonomy discussed in subsection \ref{Solving in $M/F$ thoery}, the intersection numbers do not change, and the gravitational couplings are given by $a_I^{\6d} = (32, 24, 24)$.}
     \item $\cT_B$ refers to the same theory compactified on $S^1$ with the CHL $\bZ_2$ holonomy. What is the condition for $\cT_B$ to be well-defined? 
\end{itemize} 
As we saw before, the quantization condition of KK $U(1)$ CS terms in the 6D basis is expected to take the form
\begin{equation}
    \label{CHL consistent in 6d}
    \Delta c_{000} +\frac{a_{0}^{6\text{D}}}{2} \in \frac{3\bZ}{k}.
\end{equation}
We will assume this is the case also for the CHL $\bZ_2$ action. Repeating the calculation  for the CHL $\bZ_2$ action, we find
\begin{equation}
    \label{Conjecture CHL consistency condition}
    \frac{3}{8}(V_1-H_1-3 T_1) \in \frac{3\bZ}{2} \qquad \Rightarrow \qquad (V_1-H_1-3 T_1) \in 4\bZ.
\end{equation}
Here, $V_1$, $H_1$, and $T_1$ are $\bZ_2$ charged vectors, hypers, and tensors in 6D. Equation \eqref{Conjecture CHL consistency condition} tells us:
\begin{itemize}
    \item When $V_1=T_1=0$, this is the same as the standard consistency condition for freedom from $\bZ_2$ gauge anomalies \eqref{Moore Moonier weaker conditions}.
    \item It is the combination $(V_1-H_1)$, invariant under Higgsing, that appears in \eqref{Conjecture CHL consistency condition}. It is compatible with the anomaly matching condition for the CHL $\bZ_2$ (the  anomaly remains unchanged under Higgsing unless  broken explicitly).
\end{itemize}
It is  easy to check that \eqref{Conjecture CHL consistency condition} is satisfied by the first type of model and not the second:
\begin{itemize}
    \item For the first type, $T_1 = 0$. $H_1 - V_1$ (or simply $H_1$ in a completely Higgsed theory) = 112, satisfying \eqref{Conjecture CHL consistency condition}.
    \item For the second type, note that the phase transition via small instantons \cite{Witten:1996qb} can be realized by exchanging $29$ hypers with one tensor, i.e. $(29T_1 + H_1)$ remains unchanged. Hence, before phase transition, \eqref{Conjecture CHL consistency condition} gives $(V_1-H_1) \in 4\bZ$. After phase transition, we have:
     \begin{equation} \label{eq:fail}
         V_1 -(H_1-29T_1) -3T_1  \neq 0 \ \ \text{mod} \ \ 4 \, .  
     \end{equation}
\end{itemize}
Seemingly, the consistency conditions fail for the second type of example, where $\bZ_2$ acts non-trivially on anti-self-dual tensor fields. One possible reason for this is that the holonomy action on {anti-}self-dual tensors is more subtle, and they should not be treated in the same way as chiral fermions. Notably, {anti-}self-dual tensors are involved in the GS mechanism, and the holonomy action on the topological GS terms should also be taken into account, correcting \eqref{eq:fail}. We hope to come back to this question in the future. 

We point out some features of this example that may apply to general (1,0) theories with $\bZ_k$ action:
\begin{itemize}
    \item In the first type of example, i.e. $T=1$,  $H_1-V_1 =112 = 0 \mod 16$. The even  stronger set of conditions \eqref{Moore Moonier stronger condition} is satisfied. We believe this is due to the absence of subtleties associated with the self-duality of tensor fields (the anti-self-dual tensor combines with the self-dual tensor in the gravity multiplet).
    \item For the second type of example,  anti-chiral tensor fields appear after phase transition. The number of $\bZ_2$ charged fermions becomes $H_1 - V_1  = 112 - (29-1)H_1 =0 \mod 4$, which satisfies the weak condition \eqref{Moore Moonier weaker conditions}, thanks to extra tensor fields and the quadratic refinement.
    \item Note that in the second type of model, the $\bZ_2$ also acts on anti-self-dual tensors. However, the fermion spectrum satisfies the $\bZ_2$ anomaly-free condition, which means that the anomaly from the $\bZ_2$ action on anti-self-dual tensors, if non-trivial, should contribute $\frac{1}{4}\bZ$. It would be interesting to directly check this. Currently we seem to have no framework for dealing with the anomaly of a $\bZ_k$ action that is non-trivial on the anti-self-dual tensors of 6D $\mathcal{N} = (1,0)$ supergravity. One possible option is to promote the 6D $(1,0)$ BPS charge lattice $\Lambda$ in \cite{Monnier:2018nfs} to a local system $\tilde{\Lambda}$ and perform a similar analysis.
    \item The one-loop generated 5D CS terms \eqref{KK U(1) CS from eta 2nd} seem to be in tension with the phase transitions, i.e. $H - T,H + 5T$ in the coefficients of \eqref{KK U(1) CS from eta 2nd} do not accommodate the fact that $29\Delta T = -\Delta H$ after phase transition. This provides an additional hint as to why the one-loop computation performed in this section does not work for the examples when $\bZ_2$ acts on tensors. Circle compactification with discrete holonomy for anti-self-dual tensor fields needs extra care. 

\end{itemize}

\section{Discussion}
\label{sect: discussion}
The main subject of this paper was to study the relations between anomalies and the GS mechanism in 6D and the consistency of supergravities with eight supercharges in 5D.
\begin{enumerate}
    \item Upon circle compactification, the perturbative anomalies lead to bad, i.e. gauge or diffeomorphism non-invariant, local couplings. Perhaps the right interpretation here is that although the 5D theory can be made well-defined by adding local counterterms, these cannot be lifted to 6D as local couplings. In other words, in making the 5D theory consistent, one obstructs the lift to 6D. This is compatible with the fact that the one-loop perturbative anomaly only exists in even dimensions.
    \item We proposed a new consistency test of 6D non-perturbative anomalies (characterized by $\text{Hom}_{\bZ}\big(\Omega^{\text{Spin}}_{7\text{D},\text{tor}}(BG),U(1)\big)$) via circle compactification:  the one-loop induced WZW terms  are well-defined (characterized by $\text{Hom}_{\bZ}(\Omega_{6\text{D}}^{\text{Spin}}(BG'),\bZ)$, where usually $G'\subset G$) after the circle compactification with corresponding discrete holonomies. We argued that the $\bZ_k$ 6D gauge anomalies contribute to the coefficients and hence the quantization condition of certain universal CS terms (\eqref{quantization condition for 5d CS term for KK U(1)} in a suitably chosen basis) after circle compactification. When the five- and six-dimensional theories arise from CY compactifications, this can be checked using the M/F-theory framework. Assuming M/F-theory duality, we found that the consistency condition proposed in \cite{Monnier:2018nfs} is satisfied for generic F-theory models.  
    \item Based on CHL-like examples (twisted compactifications)  and the anomaly matching, we found that the consistency condition of circle compactifications may be generalized to study the $\bZ_k$ part of semi-product $\bZ_k\ltimes G$ gauge groups.
\end{enumerate}
It is curious that although the non-perturbative $\bZ_k$ anomaly cancellation mechanism is rather complicated (it typically involves a particular quadratic refinement), after a circle compactification it transforms into the integrality of a 5D cubic form, i.e. a set of quantization conditions. It would be of interest to carry out the same analysis from the IR perspective in order to find a 5D $U(1)$ basis. 

\vspace{0.3cm}
Comparing six- and five-dimensional non-perturbative anomalies, one finds:
\begin{itemize}
    \item[] In 6D, the $\bZ_k$ anomalies are determined by the relative reduced eta-invariant \cite{atiyah1975spectral2}, which in general does not admit a differential form (like index density) expression.
    \item[] After circle compactification, the $\bZ_k$ anomalies given by the relative reduced eta-invariant are related to the contributions to the partition function by the topological WZW-like terms in 5D (the CS couplings involving the KK $U(1)$ field), which can be expressed as local differential forms.
\end{itemize}
These two statements do not contradict each other and are reconciled by the adiabatic eta invariant of Bismut and Cheeger \cite{bismut1989eta}: when the circle fiber is extremely small, there exists a limit for the eta invariant\footnote{Usually the limit is valued in $\bR/\bZ$, but in many cases it can be promoted to being valued in $\bR$ \cite{dai1991adiabatic}.} that can be expressed in terms of local differential forms on the base. Hence, within the Dai-Freed framework of anomalies, the theory of the adiabatic eta invariant is a powerful tool for the study of anomalies in compactifications. We hope to  return to applications of the BC theory in the future.

One may wonder if and how much one can do beyond the adiabatic limit. As mentioned, the topological terms, computed in the adiabatic limit, are not the only local terms that arise in the reduction of anomalous theories. There are also explicitly non-invariant terms that involve the scalar fields in the 5D vector multiplets or the graviphoton field strength. The latter vanish in the strict limit of the circle radius $r \rightarrow 0$ and may be thought of as subleading contributions to the adiabatic limit. The expectation that a simple (without $\bZ_k$ holonomies) circle reduction of an anomaly-free supergravity should be consistent and \eqref{eq:Delta-total} should be reproduced suggests that these subleading contributions can be determined by individual $\Delta_5$ contributions up to local terms that do not obstruct the lift to 6D.

There are a number of questions that we have not addressed in this paper. Notably,
\begin{itemize}
     \item[] Can the relation between 6D non-perturbative $\bZ_k$ anomalies and 5D quantization conditions be generalized to other adjacent dimensions? In Appendix \ref{app:W} we revisit the Witten $SU(2)$ global anomaly and find similar relations between the 4D theory and the circle-reduced 3D counterpart. To what extent can these considerations be generalized to compactifications of higher dimensional theory on internal manifolds other than $S^1$? 
    \item[] The topological GS terms  play an important role in our discussion. However, their coefficients remain undetermined, and a systematic way of determining them is lacking.
    \item[] $\bZ_k$ action on {anti-}self-dual tensor fields clearly needs to be better understood. The action of non-trivial    $\bZ_k$ holonomies on topological GS terms is clearly worth  studying further.
    \item[] In this paper we only considered spacetimes $M_5$ with Spin structure. Given that M-theory is consistent on Pin$^+$ manifolds, it is natural to expect that its CY3 compactifications should be as well. Can our analysis be extended to Pin$^+$ structure? Can there be anomalies associated with non Spin tangential structures in 5D?  Can any of these 5D theories be lifted to 6D and how?
\end{itemize}

Finally we remark that while the relation between the 6D and 5D consistency conditions is rather intricate, and it is not straightforward to verify that anomaly-free 6D theories indeed reduce to consistent 5D supergravities, the 5D conditions by themselves are not very restrictive. Indeed the integral cubic form can exist for an infinite number of solutions of conditions \eqref{eq:CC1}-\eqref{eq:CC3}. On the other hand, the number of consistent 6D theories is claimed to be finite \cite{Kim:2024eoa}. Some extra restrictions due to the consistency of BPS strings in 5D have already been studied in \cite{Katz:2020ewz} and in 5D reductions of 6D theories in \cite{Cheng:2021zjh}. It will be interesting to understand how the non-perturbative anomalies and circle reductions with $\bZ_k$ holonomy will modify these considerations.


\subsubsection*{Acknowledgments}

PC is partially supported by the DFG Excellence Strategy EXC-2094 390783311. 
Useful discussions with  F. Bonetti, M. Dierigl, J. Gray, T. Grimm, P.K. Oehlmann, I. Melnikov, M. Montero, Y. Proto, W. Taylor, S. Theisen, C. Vafa,  and P. Yi are gratefully acknowledged. RM would like to thank Arnold Sommerfeld Center, LMU, and Simons Center for Geometry and Physics (Physics Summer Workshop) for hospitality during the course of this work.


\appendix


\section{Details of the circle reduction}
\label{app:reduce}

The circle reduction of 6D GS terms and four forms $\tX^\a_4 ( \tilde{R}, \mathcal{\tilde{F}}_m, \tilde{F^i}) $  \eqref{eq:tX4} was discussed in subsection \ref{sec:circle}. Here we give some details and explicit formulae, as well as spell out the form of 5D non-invariant terms discussed in subsection \ref{sec:cirle2}.

We consider U(1) fibered backgrounds:
$U(1) \hookrightarrow {\tilde M} \stackrel{\pi}{\longrightarrow}M_B$. There exists on $\tilde M$ a globally-defined smooth one-form $e$ such that its curvature is a horizontal two-form.
For deriving  local formulae when considering a circle reduction, we shall  take a
six-dimensional metric of the form
\begin{equation}\label{eq:fibr}
\dd {\tilde s}^2=\eta_{\alpha\beta}e^\alpha e^\beta+r^2 (\dd \theta+ A^0)^2 = \dd s^2 + (e^{\circ})^2,
\end{equation}
with 
\begin{equation}
e^{\circ}= r (\dd\theta+ A^0),\qquad \dd e^{\circ} = \dd \ln(r) \wedge e^{\circ} + rT \, .
\end{equation}
We define $e = \big(\tfrac{1}{r}e^{\circ}\big) = (\text{d} \theta + A^0)$ such that $\dd e = \pi^* T$ and $T \in H^2(M_B, \mathbb{Z})$. In the following, we omit the pull-back signs.

For generic radius $r$, horizontal-vertical decompositions have the schematic form
$$
\tX_4=X_4+  X_3\wedge e.
$$
For $\tilde{X}_4$ closed, the 5D forms $X_4$ and $X_3$ are subject to \eqref{eq:h-v}.

\paragraph{Curvatures} The discussion concerning the gravitational part of $\tX_4$ follows closely \cite{Liu:2013dna}.  We start by reducing the curvature $\tR(\tome)=\dd \tome +\tome\wedge
\tome$. Once more, all quantities with tildes are in $\text{D}=4n+2$, and quantities without tildes are in $\text{D}=4n+1$ (notably $R(\omega) = \dd \omega + \omega \wedge \omega$). Flat indices in $\text{D} = 4n+ 2$ are denoted by $a = (\alpha, \circ)$, and curved indices in $\text{D} = 4n+ 2$ by $\mu = (i,\theta)$, where in each case the final item denotes the circle direction, and $\alpha$ and $i$ are the $\text{D} = 4n + 1$ indices. The spin connection is now
\begin{eqnarray}
\tome^{\alpha\beta}&=&\omega^{\alpha\beta}-\tfrac{r}{2}T^{\alpha\beta}e,\nonumber\\
\tome^{\alpha \circ}&=&- \tfrac{r}{2}T^\alpha{}_\gamma e^\gamma - \partial^\alpha r \, e,
\end{eqnarray}
where we have used ${\tilde e} = \{ e^{\alpha}, e^{\circ} \equiv re \}$. The reduction of the Riemann curvature two-form is
\begin{equation}
\begin{aligned}
\tilde{R}^{\alpha \beta} (\tilde{\omega}) & \ \ = \ \ \big[R^{\alpha \beta} (\omega) - \tfrac{r^2}{4} T^{\alpha}\wedge T^{\beta}  - \tfrac{r^2}{2}T^{\alpha \beta}T \big] - \big[\tfrac{1}{2} \nabla(r^2T^{ab}) + r\partial^{[\alpha}r T^{\beta]} \big]\wedge e, \\
\tilde{R}^{\alpha \circ}(\tilde{\omega}) & \ \ = \ \ \tfrac{1}{2} \nabla (r T^{\alpha}) - \partial^{\alpha}r T + \big[\tfrac{r^3}{4} T^{\alpha}_{\,\,\,\,\, \beta} T^{\beta} - \nabla \partial^{\alpha}r \big] \wedge e,
\end{aligned}
\end{equation}
where we have introduced the one-form $T^\alpha \equiv T_\gamma{}^\alpha e^\gamma$. In components, the Riemann tensor is

\begin{equation}
\begin{aligned}
\tilde{R}_{\alpha \beta}{}^{\gamma \delta}(\tilde{\omega})& \ \ = \ \ R_{\alpha \beta}{}^{\gamma \delta}(\omega) - \tfrac{r^2}{2}T_{\alpha}{}^{[\gamma} T_{\beta}{}^{\delta]}- \tfrac{r^2}{2} T_{\alpha \beta}T^{\gamma \delta}, \\
\tilde{R}_{\alpha \circ}{}^{\gamma \delta}(\tilde{\omega}) & \ \ = \ \ -\tfrac{r}{2} \nabla_{\alpha} T^{\gamma \delta} - \partial_{\alpha} r T^{\gamma \delta} + T_{\alpha}{}^{[\gamma}\partial^{\, \delta]}r, \\
\tilde{R}_{\alpha \beta}{}^{\gamma \circ}(\tilde{\omega}) & \ \ = \ \ r \nabla_{[ \alpha} T_{\beta]}{}^{\gamma} + \partial_{[\alpha}r \, T_{\beta]}{}^{\gamma} - T_{\alpha \beta} \partial^{\gamma}r,\\
\tilde{R}_{\alpha \circ}{}^{\gamma \circ}(\tilde{\omega}) & \ \ = \ \ \tfrac{r^2}{4} T_{\alpha \delta}T^{\gamma \delta} - \tfrac{1}{r}\nabla_{\alpha}\partial^{\gamma}r.
\label{eq:riemanns2}
\end{aligned}
\end{equation}
The components of the Ricci tensor are given by
\begin{eqnarray}
\tilde{R}_{\alpha\beta}(\tome)&=&R_{\alpha\beta}(\omega)-\tfrac{r^2}{2}T_{\alpha\gamma} T_{\beta}{}^\gamma - \tfrac{1}{r} \nabla_{\alpha} \partial^{\gamma}r,\nonumber\\
\tilde{R}_{\alpha \circ}(\tome)&=&-\tfrac{r}{2} \nabla^\gamma T_{\gamma\alpha} - \tfrac{3}{2} \partial^{\gamma}r \, T_{\gamma \alpha},\nonumber\\
\tilde{R}_{\circ \circ}(\tome)&=&\tfrac{r^2}{4} T_{\alpha\beta}T^{\alpha\beta} - \tfrac{1}{r} \partial_{\alpha} \partial^{\alpha} r,
\label{eq:riccis1}
\end{eqnarray}
and the Ricci scalar is
\begin{equation}
    \tilde{R}(\tome) = R(\omega) - \tfrac{r^2}{4} T_{\alpha \beta} T^{\alpha \beta} - \tfrac{2}{r} \partial_{\alpha} \partial^{\alpha} r.
\end{equation}

The decomposition of $\tr (\tilde{R} \wedge \tilde{R})$ (for $r=1$) was derived in \cite{Liu:2013dna}. We will give the explicit formulae shortly.

\paragraph{Gauge fields } When a Wilson line breaks the gauge group $\mathcal{G}$ to $U(1)^{\mbox{rk}(\mathcal{G})}$ (where the Wilson line is valued in the Cartan subalgebra of $\mathfrak{g}$) it is not hard to see that
\beq
\label{abelian}
\tr \tF^2 \, \mapsto \, \tr [(F+ \phi T) \wedge (F + \phi T)] + \dd \left[ \tr \phi (2 F + \phi T)  \right] \wedge e \,,
\eeq
where we use the ansatz for the reduction of abelian gauge fields $\tilde{A} = A + \phi e$, and the 5D abelian field strength is $F = \text{d} A$. All fields in \eqref{abelian} are abelian, but the trace is kept to denote the sum over the ${\mbox{rk}(\mathcal{G})}$ $U(1)$ fields. In the language of Section \ref{sec:AT}, the global two-form that is invariant under $\delta_{\epsilon}$ is $\mathfrak{X}_2 =  \text{tr} [\phi (2 F + \phi T)]$, and the global and closed four-form is $[X_4 - \mathfrak{X}_2 T] = \text{tr} F^2$, which is constructed out of 5D curvatures. In this case, direct reduction of the usual 6D Chern-Simons form, $\tilde{\omega}_3^{(0)} = \text{tr} [\tilde{A}\wedge \tilde{F}]$ gives
\begin{equation}
\omega_3^{(0)} = \text{tr} [A\wedge(F + \phi T)], \ \ \ \ \ \ \ \ \ \omega_2^{(0)} = \text{tr}[A \wedge \text{d}\phi + \phi F + \phi^2 T],
\end{equation}
and we see explicitly that $\mathfrak{X}_2$ differs from $\omega_2^{(0)}$ by a total derivative:
\begin{equation}
    \mathfrak{X}_2 - \omega_2^{(0)} = \text{d} \big(\text{tr} [\phi A] \big).
\end{equation}

Simple reduction of a non-abelian gauge group (without Wilson line) preserving the gauge group gives
\beq
\tilde{\mathcal{A}} = \mathcal{A}  + \varphi e \qquad \mbox{and} \qquad \tilde{\mathcal{F}} = [ \mathcal{F} + \varphi T] + D \varphi \wedge e,
\eeq
with $D = \dd + [\mathcal{A}, .]$ and $\mathcal{F} = D\mathcal{A}$. Then the reduction of $\text{tr}\tilde{\mathcal{F}}^2$ is given by
\beq
\tr \tilde{\mathcal{F}}^2 = \tr (\mathcal{F} + \varphi T)^2 + 2 \tr \left[ ( \mathcal{F} + \varphi T) \wedge D \varphi \right ] \wedge e.
\eeq
The global two-form can be found in the form analogous to the Abelian case:  $\mathfrak{X}_2 = 2 \tr\left( \varphi \mathcal{F} \right) + \tr(\varphi^2 )\, T$ by the manipulation $\dd  \left( \tr\left[ \varphi \mathcal{F} \right] +  \frac12 \tr[\varphi^2] \, T \right) =   \tr \left[ (\mathcal{F} + \varphi T) \wedge D \varphi \right ]$, and the shifted closed four-form, $[X_4 - \mathfrak{X}_2 T]$ is $\tr \mathcal{F}^2$. The general Wilson line (in the Cartan subalgebra $\mathfrak{h} \subset \mathfrak{g}$) yields a mixture of the two cases above.

\paragraph{The reduction of $\tX_4$} 
We now perform the explicit reduction for $\tilde{X}_4$:
\begin{equation}
\tX_4(\tome, \tilde{\mathcal{A}}_m,\tA_i)\equiv \tfrac{1}{8} a \, \tr \tR(\tome)\wedge \tR(\tome)+ \tfrac{1}{2} b_m \, \tr \tilde{\mathcal{F}}_m \wedge \tilde{\mathcal{F}}_m + \tfrac{1}{2} b_{ij} \tilde{F}^i \wedge \tilde{F}^j.
\label{eq:X4Omega+}
\end{equation}
where sums over $m, i,$ and $j$ are implied.
We can now construct 
\begin{eqnarray}\label{eq:expl}
 X_4-  \mathfrak{X}_2\wedge T  &=& \tfrac{1}{8} a \big[R^{\alpha\beta}(\omega )\wedge R_{\beta\alpha}(\omega)
-\tfrac{r^2}{2} R^{\alpha\beta}(\omega)\wedge T_{\beta} \wedge T_{\alpha}\nonumber 
-\tfrac{1}{2} \nabla (r T^\alpha) \,
\nabla (r T_{\alpha})\big] \nn \\
&&\ + \tfrac{1}{2}b_m\ \tr \mathcal{F}_m \wedge \mathcal{F}_m + \tfrac{1}{2} b_{ij} F^i \wedge F^j
,\nonumber\\
X_3&=&\dd \mathfrak{X}_2,\nonumber\\
 \mathfrak{X}_2 &=& \tfrac{1}{8} a \big[\mathfrak{R}^{\alpha \beta}T_{\alpha\beta} + 2 \partial_{\alpha} r \nabla (r T^{\alpha}) - 2 (\partial r)^2 T\big] \nn \\
&&+ \tfrac{1}{2} b_m \big[ 2\tr (\varphi_m \mathcal{F}_m) + \text{tr} (\varphi^2_m) T\big] + \tfrac{1}{2} b_{ij} \big[2 \phi^i F^j + \phi^i \phi^j T \big],
\label{eq:X4X2}
\end{eqnarray}
where we have introduced the useful symbol $\mathfrak{R}^{\alpha \beta} \equiv  R^{\alpha \beta} - \tfrac{r^2}{4} T^{\alpha \beta} T - \tfrac{r^2}{4} T^{\alpha} \wedge T^{\beta}$.

We recall once more that since $X_2^{(0)}$ is defined as an inverse derivative of the closed form $X_3$ (formally $X^{(0)}_2 = \text{d}^{-1} \big[\iota_v \tX_4\big] $, where $\iota_v$ denotes the contraction with the isometry vector $v$), it is ambiguous up to closed forms. Similar ambiguity is present in the definition of the closed four-form $ [X_4- X_2^{(0)}\wedge T]$. Demanding that the latter is globally defined  fixes the ambiguity.  In this sense, \eqref{eq:X4X2} gives the unique
expressions for the globally-defined $ \mathfrak{X}_2$ and  $ [X_4- \mathfrak{X}_2\wedge T]$. 

We can now compute the difference between these expressions and the reduction of the usual CS term, $\tilde{\omega}_{3}^{(0)} = \omega_{3}^{(0)} + \omega_{2}^{(0)}e$:
\beq
\label{eq:ambigA}
\mathfrak{X}_2 - \omega_2^{(0)} = \dd \Big[ \tfrac{1}{8} a \big(\tfrac{r^2}{2}\omega^{\alpha \beta} T_{\alpha \beta} + r\partial_{\alpha} r  T^{\alpha}\big) + \tfrac{1}{2}b_m \, \text{tr}(\varphi_m \mathcal{A}_m) + \tfrac{1}{2} b_{ij} \phi^i A^j \Big].
\eeq

\paragraph{5D non-invariant terms}
As discussed in the main text, the reduction of the anomaly gives rise to non-invariant 5D Chern-Simons terms. We record the explicit forms of $\Delta_5(\tilde{R})$ and $\Delta_5(\tilde{\mathcal{F}})$ below for the choice $X_2^{(0)} \equiv \omega_2^{(0)}$. We spell out only the pure gauge terms and the pure gravitational terms. There are, of course, mixed non-invariant terms as well, but their form is analogous and easily derivable, and we will only be interested in the non-Abelian pure-gauge terms in the following. 
\begin{equation}
    \begin{split}
    \label{eq:Delta}
        &\Delta_5(\tilde{\mathcal{F}}) = \tfrac{1}{8}(b\cdot b)\big\{ -X_3^{(0)}(\tilde{\mathcal{F}}) \mathfrak{X}_2 (\tilde{\mathcal{F}})+ x_1(\tilde{\mathcal{F}}) [X_4(\tilde{\mathcal{F}}) - \mathfrak{X}_2(\tilde{\mathcal{F}})T] \big\}  \\& \ \ \ \ \ \ \ \ \  = \,\,  \tfrac{1}{8}(b\cdot b)\Big\{ \big(\text{tr}[2\phi F] + \text{tr}[\phi^2] T \big) \big(\text{tr}[A F] - \tfrac{1}{3} \text{tr}[A^3] \big) \\&  \ \ \ \ \ \ \ \ \ \ \ \, + \text{tr}[\phi A] \big( \text{tr}[F^2] + \text{tr}[2 \phi F] T + \text{tr}[\phi^2] T^2 \big) \Big\},\\&
        \Delta_5(\tilde{R}) = \tfrac{1}{128} (a \cdot a) \big\{ -X_3^{(0)}(\tilde{R}) \mathfrak{X}_2 (\tilde{R}) + x_1(\tilde{R}) [X_4(\tilde{R}) - \mathfrak{X}_2(\tilde{R})T] \big\} \\& \ \ \ \ \ \ \ \ \ = \,\, \tfrac{1}{128}(a \cdot a) \Big\{ \text{tr}[T_0 \mathfrak{R}] \big(\text{tr}[\omega R]  - \tfrac{1}{2} \text{tr}[T_1 \nabla T_1] \big) \\& \ \ \ \ \ \ \ \ \ \ \  +  \text{tr}[T_0 \omega] \big( \text{tr}[R^2]  - \tfrac{1}{2}\text{tr}[T_1^2 R] - \tfrac{1}{4}\text{tr}[\nabla T_1 \nabla T_1] + \tfrac{1}{2}\text{tr}[T_0 \mathfrak{R}]T \big)\Big\}.
        \end{split}
\end{equation}

\subsection{Fermionic terms}
\label{app:B}
In the following appendix, we perform a loop computation and calculate the simplest of the non-invariant terms in $\Delta_5(\tilde{\mathcal{F}})$. In preparation for this, we present the dimensional reduction of the fermion kinetic term in this subsection. We expect the chiral anomaly in 6D to correspond to mass terms in 5D with a sign dependence corresponding to the higher dimensional chirality. 

Our conventions for indices are the same as before. The components for one choice of vielbeins in metric \eqref{eq:fibr} (as well as their inverses) are given by
\begin{equation}
\tilde{e}_{ \, i}^{ \, \alpha } = e_{\, i }^{\, \alpha}, \ \ \ \ \ \ \tilde{e}_{\, \theta}^{\, \alpha } = 0, \ \ \ \ \ \ \tilde{e}_{\, i}^{\, \circ}  = r A^0_{i}, \ \ \ \ \ \ \tilde{e}_{\theta}^{\,\circ} = r,
\end{equation}
\begin{equation}
\tilde{e}^{\, \, i}_{ \, \alpha} = e^{\, \, i}_{\, \alpha}, \ \ \ \ \ \ \tilde{e}^{\, \, \theta}_{\, \alpha} = -A^{0}_{\alpha}, \ \ \ \ \ \ \tilde{e}^{\, \, i}_{\,\circ}  = 0 ,\ \ \ \ \ \ \tilde{e}^{\, \, \theta}_{\, \circ} = \tfrac{1}{r}.
\end{equation}
In a particularly convenient basis, the 6D gamma matrices are given by
\begin{equation}
\Gamma^{i} = \left(
\begin{array}{cc}
0 & \gamma^i \\
\gamma^i & 0 \end{array}
\right), \ \ \ \Gamma^{\circ} = i \left(
\begin{array}{cc}
0 & -\mathbbm{1}_4 \\
\mathbbm{1}_4 & 0 \end{array} \! \right), \ \ \ \Gamma^{*} = i\Gamma^{0}\Gamma^1\cdots \Gamma^{\circ}  = i\left(
\begin{array}{cc}
\mathbbm{1}_4 & 0 \\
0 & -\mathbbm{1}_4 \end{array} \! \right),
\end{equation}
where $\gamma^i$ are $4 \times 4$ matrices and generate the 5D Clifford algebra. Using this basis, it is easy to see that chiral fermions in 6D reduce to Dirac fermions in 5D. Now we consider the reduction of the 6D fermionic action coupled to a gauge field $\tilde{A}_{\mu}$ and to gravity:
\begin{equation}
	\mathcal{L}_{\Psi} = \overline{\Psi} \Gamma^{a} \tilde{e}_a^{\, \, \, \mu}(\partial_{\mu} + i\tilde{\omega}_{\mu}^{\,\,\, ab} \Gamma_{ab} - i q A_{\mu}) \Psi.
\end{equation}
Since $\theta$ is periodic, we can expand the fermions as
\begin{equation}
	\Psi = \sum_{n \in \mathbb{Z}} \psi_n e^{-i n \theta}.
\end{equation}
For the vector $\tilde{A}$, we choose $\tilde{A} = A + \phi (\text{d}\theta + A^0)$. And finally, for the spin-connection, we have components
\begin{equation}
\begin{split}
	& \tilde{\omega}_{i}^{\, \, \alpha \beta} = \omega_i^{\, \,  \alpha \beta} - \tfrac{r^2}{2} T^{\alpha \beta} A^0_{i} \ \ \ \ \ \ \ \tilde{\omega}_{\theta}^{\, \, \alpha \beta} = - \tfrac{r^2}{2} T^{\alpha \beta}, \\&
	\tilde{\omega}_i^{\, \, \alpha \circ} = -\tfrac{r}{2} T^{\alpha}_{\,\,\, i} - \partial^{\alpha} r A^0_i \ \ \ \ \ \ \ \ \tilde{\omega}_{\theta}^{\, \, \alpha \circ} = - \partial^{\alpha} r.
	\end{split}
\end{equation}
We are now ready to expand the action. We find
\begin{equation}\label{eq:fermi}
	\mathcal{L}_{\Psi} = \sum_{n \in \mathbb{Z}_n} \overline{\psi}_{-n}\Big[\gamma^i D_i + in \gamma^i A^0_i \pm \big( \tfrac{n}{r} + \tfrac{1}{r}\phi + i \tfrac{r}{2} T_{\alpha \beta}\gamma^{\alpha \beta}\big)+ 2 \gamma^{i} \partial_i \ln{r}  \Big] \psi_n,
\end{equation}
where $D_i$ is the 5D covariant derivative, $D_i = \partial_i  - i q A_i + \omega_i^{\, \, \alpha \beta} \gamma_{\alpha \beta}$, and where the plus/minus corresponds to the chirality of the 6D chiral fermion. We thus find the usual Kaluza-Klein tower of massive fermions, as well as interactions between the 5D fermions and $\phi, T_{\alpha \beta},$ and $r$. One subtlety is that after reduction, the fermion needs to be redefined as $\psi_\text{D} = \psi_{\text{D}-1}\times \sqrt{r}$  to compensate for the dimension change. We ignore this point in writing the above action, as it seems to add extra terms proportional to $r{'}(x)$  but does not depend on the chirality in one dimension higher (i.e. the term with $\pm$ ambiguities). 

Note that in \eqref{eq:fermi} $T$ plays a role of a mass term, just like the holonomy of the gauge group. The similarity of the roles of $T$ and gauge scalars was already noticed in \cite{Cheng:2021zjh}.

\section{Cancellation of 5D non-invariant couplings}
\label{app:FDC}
As shown in Section \ref{sec:pert_an}, the reduction of a 6D (1,0) theory with GS terms leads to 5D non-invariant couplings of the form
\bea
	\Delta_5(\tilde{R},\tilde{\mathcal{F}}_m, \tilde{F}^i) &= (H-V+T) \Delta_5^{\psi}(\tilde{R}) - \Delta_5^{\psi_{\mu}}(\tilde{R}) + (T-1) \Delta_5^{B_{\mu\nu}}(\tilde{R}) + \Delta_5^{\psi}(\tilde{R},\tilde{\mathcal{F}}_m, \tilde{F}^i) \nn \\
    &= \tfrac{1}{2}\big[- X_3^{(0)}\cdot \mathfrak{X}_2 + \text{x}_1 \cdot  [X_4 -  \mathfrak{X}_2 T]\big],
\eea
where different contributions are determined by anomalies of the individual fields  as 
\begin{equation}\label{eq:deltaPhi+}
	\Delta_5^{\Phi} = \text{d}^{-1} [\iota_v \text{d}^{-1}\tilde{I}_8^{\Phi}  - \mathfrak{D}^{\Phi}_6 ],
\end{equation}
for  $\Phi = (\psi, \psi_{\mu}, B_{\mu \nu})$.

This form of $\Delta_5$ should facilitate the comparison with the results of the loop calculation. The reduction of the 6D anomaly for abelian fields was discussed in \cite{Corvilain:2020tfb}. Below we consider the non-invariant terms in the reduction of the anomaly for a single non-abelian gauge field, $\Delta_5^{\psi}(\tilde{\mathcal{F}})$.

In Appendix \ref{GS cancellation of perturbative anomalies is not affected by discrete holonomy} we argue that the cancellation of the 5D non-invariant terms in the reduction of anomaly-free 6D (1,0) theories will hold when $\bZ_k$ holonomies are turned on.

\subsection{Non-abelian loop computation}
\label{app:C}
\noindent In this section, we show how the non-invariant terms appear at the level of a 5D loop computation for non-abelian gauge fields by computing the simplest term in $\Delta^{\psi}_5(\tilde{\mathcal{F}})$, up to an overall factor. The usual Chern-Simons term in the 5D action is given by 
\begin{equation}
	\mathcal{L}_{\text{CS}} = k_{\mathcal{A}\mathcal{F}\mathcal{F}} \, \text{tr}[\mathcal{A} \mathcal{F}^2]\propto k_{\mathcal{AFF}} \epsilon^{\mu \nu \rho \sigma \tau} \mathcal{A}^a_{\mu} \partial_{\sigma} \mathcal{A}^b_{\nu} \partial_{\tau} \mathcal{A}^c_{\rho} \text{tr}[t^a t^b t^c],
\end{equation}
which corresponds to the following term in the evaluation of the diagram depicted in Figure \ref{Figure2}:
\begin{equation}
- (i 3!) k_{\mathcal{AFF}}  \epsilon^{\mu \nu \rho \sigma \tau} p_{\sigma} q_{\tau}  (e_1)_{\mu} (e_2)_{\nu} (e_3)_{\rho} \text{tr}[t^a t^b t^c]
\end{equation}
where the $e_{\mu}$'s are the polarization vectors for the three external gauge bosons. For the $\varphi$-dependent term, we have similarly
\begin{equation}
\begin{split}
	\mathcal{L}_{\text{CS}} &= k^{\varphi}_{\mathcal{AFF}} \big[ \, \text{tr}[\mathcal{A} \varphi]\text{tr}[\mathcal{F}^2] + 2\text{tr}[\varphi \mathcal{F}] \text{tr}[\mathcal{A} \mathcal{F}]\big] \\&
	\propto k^{\varphi}_{\mathcal{AFF}} \epsilon^{\mu \nu \rho \sigma \tau} \varphi^d\mathcal{A}^a_{\mu}  \partial_{\nu} \mathcal{A}^b_{\rho} \partial_{\sigma} \mathcal{A}^c_{\tau} \big(\text{tr}[t^a t^d] \text{tr}[t^b t^c] + 2\text{tr}[t^a t^b] \text{tr}[t^c t^d] \big).
\end{split}
\end{equation}
This term in the 5D effective action corresponds to the following term in the amplitude:
\begin{equation}
	\propto - (i 3!) k^{\varphi}_{\mathcal{AFF}} p_{\tau} q_{\lambda} \epsilon^{\mu \nu \rho \sigma \tau} (e_1)_{\mu} (e_2)_{\nu} (e_3)_{\rho} \varphi^d \big( \text{tr}[t^a t^d]\text{tr}[t^b t^c]+ 2\text{tr}[t^a t^b] \text{tr}[t^c t^d] \big).
\end{equation}
We now extract the Chern-Simons coefficient from a loop diagram in 5D, shown in Figure \ref{Figure2}. Using the reduced Feynman rules, we write
\begin{equation}
	\mathcal{M}^{\mu \nu \rho}_{abc} = g^3 \sum_n \int \! \frac{d^5 \ell }{(2\pi)^5} \, \text{Tr}\bigg\{\gamma^{\mu} t^a\bigg[\frac{\slashed{\ell} + i m_n}{\ell^2 + m_n^2} \bigg]\gamma^{\nu}t^b\bigg[\frac{\slashed{\ell} + \slashed{p} + im_n}{(\ell + p)^2 + m_n^2} \bigg] \gamma^{\rho}t^c\bigg[\frac{\slashed{\ell} + \slashed{p} + \slashed{q} +im_n}{(\ell + p + q)^2 + m_n^2} \bigg]\bigg\}.
\end{equation}
\begin{center}
\label{Figure2}
\includegraphics[scale=.3]{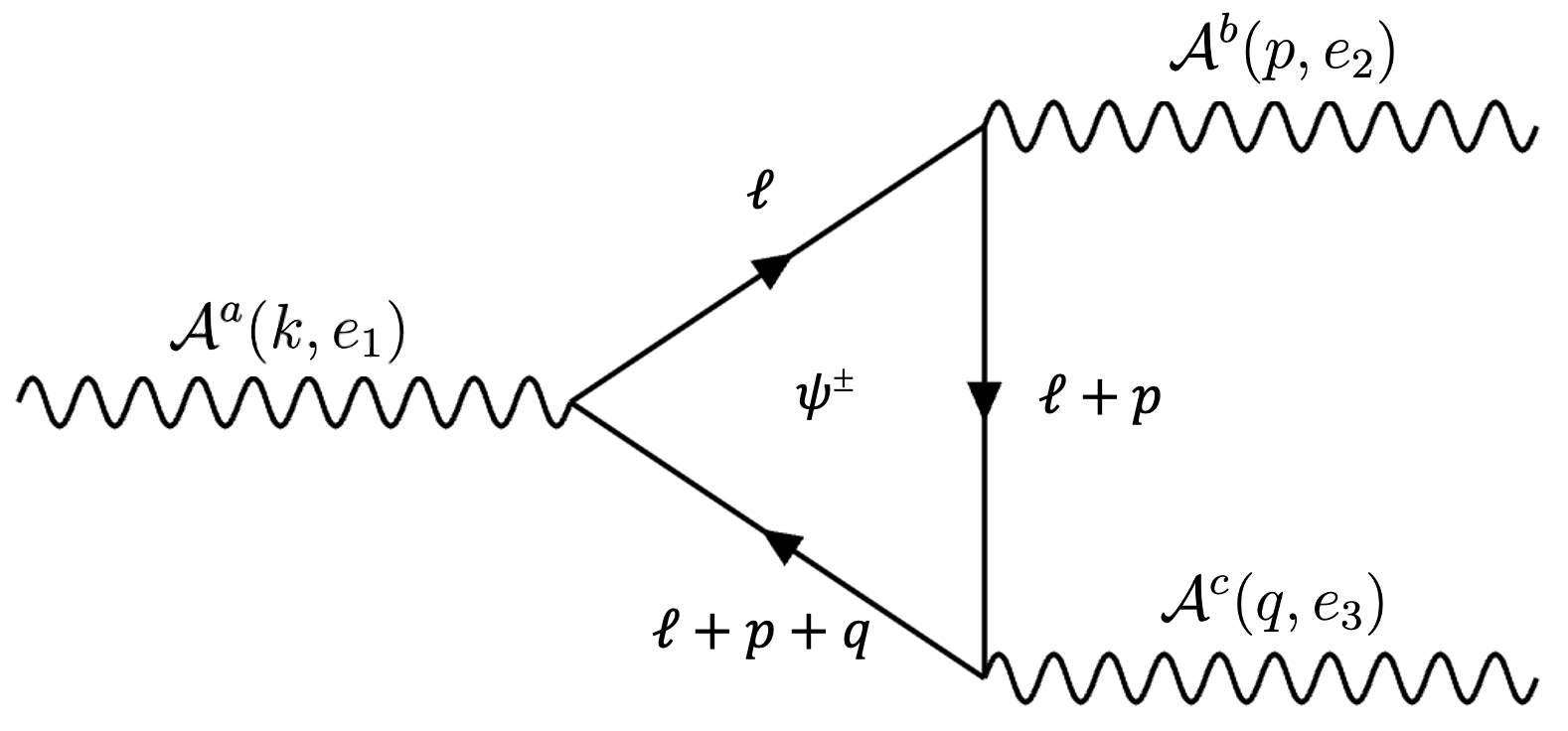}
\end{center}
\begin{quote}
\textbf{\  \ \ \ \ \ \ \ \ \ \ \ \ \ \ \ \ \text{Figure B.1}} 5D Non-abelian three-vertex diagram
\end{quote}
In the non-abelian case, $m_n$ is a matrix given by
\begin{equation}
	m_n = \tfrac{1}{r} \big(n\mathbbm{1} + \varphi^d t^d\big).
\end{equation} The only term that survives the Dirac trace and integration (and which will depend on $\epsilon_{\mu \nu \rho \sigma \tau}$), is
\begin{equation}
g^3  i p_{\tau} q_{\lambda} \epsilon^{\mu \nu \tau \rho \lambda} \int \frac{\text{d}^5\ell}{(2\pi)^5}\text{Tr} \bigg[t^a \frac{m_n}{\ell^2 + m_n^2} t^b \frac{1}{(\ell+p)^2 + m_n^2} t^c \frac{1}{(\ell + p + q)^2 + m_n^2} \bigg].
\end{equation}
We cannot immediately rearrange denominators, since now $m_n$ is Lie-algebra valued. Consider the simplest case of $SU(2)$ and expand in the Pauli matrices. The mass matrix is then given by
\begin{equation}
	m_n  = \frac{1}{r}\left(
\begin{array}{cc}
 n + \tfrac{\varphi^3}{2} & \tfrac{\varphi^1}{2} - i \tfrac{\varphi^2}{2} \\
\tfrac{\varphi^1}{2} + i \tfrac{\varphi^2}{2} & n - \tfrac{\varphi^3}{2} \end{array}
\right).
\end{equation}
If we diagonalize this matrix, we find
\begin{equation}
	m_n  = \frac{1}{r}\left(
\begin{array}{cc}
n + \tfrac{1}{2}|\varphi| & 0 \\
0 & n - \tfrac{1}{2}|\varphi|)  \end{array}
\right),
\end{equation}
where $|\varphi|$ is just the norm of the vector $\varphi^a$: $|\varphi| = \sqrt{\varphi_1^2 + \varphi_2^2 + \varphi_3^2}$. Using this diagonalization, we can do the integral, and using the fact
\begin{equation}
	\int \frac{d^5 \ell}{(2\pi)^5} \frac{1}{\ell^2 + m^2}\frac{1}{(\ell + p)^2 + m^2}\frac{1}{(\ell + p + q)^2 + m^2} = \frac{1}{|m|},
\end{equation}
we have the result
\begin{equation}
	\sum_n \text{Tr} \Bigg[ t^a t^b t^c\left(
\begin{array}{cc} \text{sgn}(n + \tfrac{1}{2}|\varphi|) & 0 \\
0 & \text{sgn}(n - \tfrac{1}{2}|\varphi|)  \end{array}
\right) \Bigg].
\end{equation}
Regularizing each element individually gives
\begin{equation}
	\text{Tr} \Bigg[ t^a t^b t^c\left(
\begin{array}{cc} \tfrac{1}{2}
 +  \tfrac{1}{2}\lfloor |\varphi| \rfloor -  \tfrac{1}{2}|\varphi| & 0 \\
0 & \tfrac{1}{2}
 -  \tfrac{1}{2}\lfloor |\varphi| \rfloor + \tfrac{1}{2} |\varphi|  \end{array}
\right) \Bigg].
\end{equation}
The second term is proportional to the Cartan generator $t^3 = \frac{\sigma^3}{2}$, while the last terms can be returned to non-diagonal form, and we find
\begin{equation}
	\tfrac{1}{2}\text{Tr} [ t^a t^b t^c ] + \lfloor |\varphi| \rfloor \text{Tr} [ t^a t^b t^c t^3] - \varphi^d\text{Tr} [ t^a t^b t^c t^d].
\end{equation}
Inserting into the amplitude, we find
\begin{equation}
	-i g^3  p_{\tau} q_{\lambda} \epsilon^{\mu \nu \rho \tau \lambda}   \big\{ \tfrac{1}{2}\text{Tr} [t^a t^b t^c] + \lfloor |\varphi| \rfloor \text{Tr}[t^a t^b t^c t^3] - \varphi^d \text{Tr}[t^a t^d t^b t^c]  \big\}.
\end{equation}
The second term may look strange, but it is necessary, since it becomes the usual floor function coefficient of the CS term if we break the gauge group to its Cartan sub-algebra. Generically there should be a term like this for each Cartan generator. For a generic Lie group, the diagonalization we have performed in order to do the integral is always possible, but the form of the diagonal entries will generally not be as simple as the $SU(2)$ case. Finally, we use the trace identity
\begin{equation}
	\text{Tr}_{\mathcal{R}}[t^a t^d t^b t^c] = B_{\mathcal{R}} \text{tr}[t^a t^d t^b t^c]+ \frac{C_{\mathcal{R}}}{3} \big[\text{tr}[t^a t^d] \text{tr}[t^b t^c]+ \text{tr}[t^a t^b] \text{tr}[t^c t^d]  +\text{tr}[t^a t^c] \text{tr}[t^b t^d]\big],
\end{equation}
to write this in the expected form.
After breaking the trace using the representation theory identities, the loop computation seems to select  the CS form $\omega_2^{(0)}$; for this choice, the relative coefficients between $\text{tr}(\varphi A) \text{tr}(F^2)$ and $\text{tr}(\varphi F) \text{tr}(AF)$ match those of $\Delta_5^{\psi}(\tilde{\mathcal{F}})$ \eqref{eq:Delta}.  Considering the gauge group to be $U(1)$, we also recover the zeta function regularization field-dependent term of $k_{AFF}$ which is linear in $\phi$. 

\subsection{Discrete holonomy and 5D non-invariant couplings}
\label{GS cancellation of perturbative anomalies is not affected by discrete holonomy}
In this section, we argue that upon circle reduction of a 6D supergravity with $\bZ_k$ symmetry, for which the anomalies are canceled via the GS mechanism, turning on discrete holonomy will not affect the cancellation of non-invariant 5D terms.

\paragraph{Dai-Freed anomaly point of view}
 As reviewed in Section \ref{sec:AT}, the anomalies from chiral fields in 6D supergravity can be captured by the eta invariant of a suitable 7D Dirac operator $\exp\{2\pi i \eta_{7\text{D}}\}$. Using the APS index theorem, the perturbative anomaly is given by $\exp\{2\pi i \eta_{7\text{D}}\}$ on $Y_{7}$, the structures of which can be extended to a bulk $Z_{8}$:
 \begin{equation}
     \label{anoamly with GS 6d}
     \exp\{2\pi i \eta(Y_7)\} = \exp \bigg\{\int_{Z_8} I_{8\text{D}} \bigg\},
 \end{equation}
where $I_{8\text{D}}$ is the anomaly polynomial from chiral fields in 6D supergravity. Freedom from perturbative anomalies via the GS mechanism implies
\begin{equation}
    \label{anoamly with GS 6d 2nd}
     \exp\{2\pi i \eta(Y_7) + \cA_{\text{GS}} \} = 1 \ \ \ \Rightarrow \ \ \ I_{8\text{D}} = I_{\text{GS}}=\frac{1}{2  }\tX_4\wedge \tX_4.
\end{equation}
To perform a circle compactification with discrete $\bZ_k$ holonomy, one assumes that $Y_7$ is a circle fibration with a nontrivial flat $\bZ_k$ bundle over it. A $\bZ_k$ anomaly may obstruct extending $Y_7$ to $Z_8$. Consider a continuous family of geometric structures, e.g. metric on $Y_7$, $g_t(Y_7),$ with $t \in [0,1]$. Applying the APS index theorem to
\begin{equation}
\label{a family of geometric structure on Y7}
   Z_8 = Y_{7} \times [0,1], \ \ \ \ \ \  g({Z_8}) = dt^2 + g_{t}(Y_7),
\end{equation}
gives
\begin{equation}
\label{anoamly with GS 6d 3rd}
   \exp \big\{2\pi i  \eta(Y_{7,t=1})- 2\pi i\eta(Y_{7,t=0})\big\} = \exp \bigg\{\int_{Z_8} I_{\text{GS}}\bigg\}.
\end{equation}
%
The equality is given by the APS theorem, and since $I_{\text{GS}}$ is a differential form, the expression does not depend on the flat structure.

Now consider circle compactification without discrete holonomy, i.e. the same $Y_7$ but without a non-trivial flat $\bZ_k$ bundle. We refer to this case as $Y_{7,0}$. We can also view $Y_{7,0}$ the same as $Y_{7}$ and modify the Dirac operator to be the one without coupling to the $\bZ_k$ bundle. Applying the APS index theorem and repeating for $Y_{7,0}$ the same steps as for $Y_{7}$ yields
\begin{equation}
    \label{anomaly 6d without discrete holonomy}
      \exp \big\{2\pi i  \eta(Y_{7,0, t=1})- 2\pi i\eta(Y_{7,0,t=0}) \big\} = \exp\bigg\{\int_{Z_8} I_{\text{GS}}\bigg\},
\end{equation}
with the same $I_{\text{GS}}$ appearing on the right-hand side as in \eqref{anoamly with GS 6d 3rd}. Equations \eqref{anoamly with GS 6d 3rd} and \eqref{anomaly 6d without discrete holonomy} suggest the relative reduced eta invariant in \cite{atiyah1975spectral2} is independent of the geometric structure (is actually a cobordism invariant):
\begin{equation}
\label{rigid relative eta}
    \exp \big\{2\pi i  \eta(Y_{7,t=1}) - 2\pi i  \eta(Y_{7,0, t=1}) \big\} = \exp \big\{2\pi i  \eta(Y_{7,t=0}) - 2\pi i  \eta(Y_{7,0, t=0}) \big\}.
\end{equation}
Note that for the circle compactification without holonomy, the  one-loop anomaly  from chiral fields $\exp\{2\pi i \eta(Y_{7,0})\}$ is canceled by the reduction of the perturbative GS term. Eq. \eqref{rigid relative eta} states that the  one-loop anomaly $\exp\{2\pi i \eta(Y_7)\}$ is the same as $\exp\{2\pi i \eta(Y_{7,0})\}$, up to a rigid cobordism invariant (which is related to the non-perturbative anomaly).  Hence, no new continuous local non-invariant terms should appear upon circle compactification with discrete holonomy, provided the 6D supergravity is perturbative-anomaly-free via the GS mechanism. Notice the above discussion works for any $Y_7$ with circle-fibration structure (which is a requirement of the circle compactification setup). 

\paragraph{An example of a loop calculation}
Consider 6D supergravity with $\bZ_k$ gauge group and without a perturbative gauge anomaly, i.e. the one-loop generated 5D local non-invariant terms cancel the non-invariant terms from the reduction of the GS term. For illustration, we shall consider only the gauge part:
\begin{itemize}
    \item The relevant bad terms $\Delta_5(\tilde{F})$ are generated by integrating out the KK tower of charged hypers, which is:
    \begin{equation}
    \label{one loop bad term in 5d}
       \Delta_5(\tilde{F}) \propto \sum_{i}\sum_{n\geq 0} n^i_{H} \text{sgn}(n+\phi)= \sum_{i} n^i_{H}  \Big(\phi -\frac{1}{2}\Big) ,
    \end{equation}
    and the non-invariant terms come from the $\phi$ dependent part of $\zeta$ function regularization of \eqref{one loop bad term in 5d}. $n^i_{H}$ is the number of hypermultiplets with $\bZ_k$ charge $i$. 
    \item Turning on one unit of $\bZ_k$ holonomy changes the mass of the KK tower of hypermultiplets \eqref{KK U(1) charge of hypers with holonomy}. In this case \eqref{one loop bad term in 5d} is modified to:
    \begin{equation}
    \label{one loop bad term in 5d 2nd} 
        \Delta_5(\tilde{F}) \propto \sum_{i}\sum_{n\geq 0} n^i_{H} \text{sgn}(kn+i+k\phi)  = \sum_{i}  n^{i}_{H}  \Big(\phi + \frac{i}{k} -\frac{1}{2}\Big).
    \end{equation}
    \end{itemize}
    Equations \eqref{one loop bad term in 5d} and \eqref{one loop bad term in 5d 2nd} give the same one-loop local non-invariant terms in 5D, as suggested by the general argument given above. The difference between the two is a constant shift that affects the coefficient of the CS coupling. This shift is expected. For example, in M/F-theory the theories with different holonomies correspond to M-theory on two different CY3 manifolds.  
    
 \paragraph{GS term reduction with $\bZ_k$ automorphism} Above we consider $\bZ_k$ as a gauge group, i.e. it does not affect the perturbative part of the GS term after circle compactification. We may also discuss the case where $\bZ_k$  acts on 6D tensors and vectors and ask about the effect of $\bZ_k$ holonomy on the perturbative part of the GS terms. 
 \begin{itemize}
     \item Perturbatively and locally, a GS term is a counterterm in 6D supergravity and can be viewed as a local functional $\cL_{\text{GS}}$  of the field configuration space $\cM =$ ($B,A, \phi$, etc.).
     \item Turning on discrete holonomy of CHL-like $\bZ_k$ reduces the field configuration from $\cM$ to the $\bZ_k$ invariant subspace $\cM^{\bZ_k} \subseteq \cM$.
     \item {The reduction} of $\cL_{\text{GS}}$ in this case is the circle compactification of $\cL_{\text{GS}}$ restricted to the invariant subspace $\cM^{\bZ_k} \subseteq \cM$.
     \item An example is given by the CHL string with $\cM=({A_{E_8,1},A_{E_8,2}})$. The invariant subspace is $\cM^{\bZ_2}=({A_{E_8,1}=A_{E_8,2})}$. Without CHL $\bZ_2$ holonomy, the GS term is
     \begin{equation}
         B\wedge \big[\text{tr}(\mathcal{F}^4_{E_8,1})+ \text{tr}(\mathcal{F}^4_{E_8,2})\big].
     \end{equation}
     With CHL $\bZ_2$ holonomy, it requires $\mathcal{F}_{E_8,1}= \mathcal{F}_{E_8,2}$, and the GS term is:
      \begin{equation}
         B\wedge \big[\text{tr}(\mathcal{F}^4_{E_8})+ \text{tr}(\mathcal{F}^4_{E_8})\big]= 2 B\wedge \text{tr}(\mathcal{F}^4_{E_8}).
     \end{equation}
     Reducing this term on the circle gives the corresponding term in the 9D CHL string.
     \item As another example, consider $\bZ_2$ acting on two tensor multiplets (labeled by $1$ and $2$) and a part of the GS couplings given by 
     \begin{equation}
         \cL_{\text{GS},6} = \tfrac{1}{2}\big[B^{-}_1 \wedge (X^1) + B^{-}_{2}\wedge (X^2)\big],\quad  X^{1,2} = \tfrac{1}{2}b^{1,2} \text{tr}(\mathcal{F}^2_{G}).
     \end{equation}
     The existence of a potential $\bZ_2$ automorphism swapping the two anti-self-dual $B^{-}$  requires $b^1 = b^2=b$, i.e. $X^1 =X^2 =X$. Here $\cM= (B_1,B_2,\mathcal{F}_{G})$, and $\cM^{\bZ_2} = (B_1=B_2 =B,\mathcal{F}_G)$. The reduced coupling in 5D is the circle reduction of the following term:
     \begin{equation}
         \cL_{\text{GS},6} = \tfrac{1}{2} \big[B^-_1 \wedge (X^1) + B^-_{2}\wedge (X^2) \big]  = \tfrac{1}{2} \big[ B^- \wedge X + B^- \wedge X \big] = B^-\wedge X.
     \end{equation}
 \end{itemize}
The above arguments are very sketchy and only apply to the reduction of perturbative GS terms.

\section{Bordism group calculations}
\label{Some bordism group calculations}

In this section we present some bordism calculations used in the main text:
\begin{itemize}
    \item $\Omega^{\text{Spin}}_{6\text{D}}(BU(1)) = \bZ \oplus \bZ$.
    \item $\Omega^{\text{Spin}}_{6\text{D}}(B\bZ_{k}) = 0 $.
\end{itemize}
In the following discussion, we omit the $\text{D}$ in the subscripts for brevity. The main techniques for computing these groups are the Atiyah-Hirzebruch spectral sequence (AHSS) and the Adams spectral sequences; see \cite{davis2001lecture, beaudry2018guide} for nice introductions. We will use the notations from \cite{beaudry2018guide} for the Adams spectral sequence calculations.
\begin{itemize}
    \item For AHSS, consider the fibration:
\begin{equation}
    X \hookrightarrow Z \to Y.
\end{equation}
 Then the AH spectral sequence gives:
 \begin{equation}
     \label{AHSS 1st}
     E_{2}^{p,q} = H_{p} \big(Y;\Omega^{\text{Spin}}_q(X) \big), d_2 \to \Omega^{\text{Spin}}_{*}(Z)) \  ,
 \end{equation}
 where $d_2$ is
 \begin{equation}
     d_2: E_{2}^{p,q} \to E_{2}^{p-2,q+1}.
 \end{equation}
 For $q=0,1$, $d_2$ is given by the dual of the Steenrod cohomology operator $Sq^2$.
 \item One key ingredient of the Adams spectral sequence is the cohomology of the Steenrod algebra $A_{p}$ with $p$ a prime number. Here we will only need the $p=2$ case, which we simply call $A$.  For computing $\Omega^{\text{Spin}}_{*}(Z)$ for $*\leq 6$, the computation can be reduced to the cohomology of $A_1$, the subalgebra of $A$ generated by $Sq^1,Sq^2$. This is our case, and the Adams spectral sequence gives\footnote{$ \Omega_{*=(t-s)}^{\text{Spin}}(Z)\big|_2^{\wedge}$ stands for the 2-completion of $\Omega_{*=(t-s)}^{\text{Spin}}(Z)$, see \cite{beaudry2018guide}.}:
 \begin{equation}
     \label{ASS}
     E_{2}^{s,t} = \text{Ext}^{s,t}_{A_1}(H^*(Z;\bZ_2);\bZ_2), d_2 \to \Omega_{*=(t-s)}^{\text{Spin}}(Z)\big|_2^{\wedge}, \quad \text{for } *\leq 7. 
 \end{equation}
\end{itemize}
\paragraph{$\underline{\Omega^{\text{Spin}}_{6}\big(BU(1)\big) = \bZ \oplus \bZ}$}

We use the Adams spectral sequence for this computation.\footnote{A $\bZ_2$ factor survives the second page when using the AH spectral sequence to compute $\Omega^{\text{Spin}}_{6\text{D}}\big(BU(1)\big)$, which makes the analysis harder.}
As an $A_1$ module for degree $\leq 7$, $H^*(BU(1);\bZ_2)$ is given by (using the notation of \cite{beaudry2018guide})
\begin{equation}
    \label{BU(1) as A1 module}
    H^*\big(BU(1);\bZ_2\big) = \bZ_2 \oplus \Sigma^2 \cA_{1}/\cE_1 \oplus \Sigma^6 \cA_{1}/\cE_1.
\end{equation}
The corresponding Adams chart is:
 \begin{center}
\begin{sseqdata}[name=MMMM,Adams grading,classes = fill,xrange = {0}{7},yrange = {0}{7}]
	 
    \tower(0,0)
	\class(1,1)
    
    \class(2,2)
    \structline(0,0)(1,1)
    \structline(1,1)(2,2)
    \tower[red](2,0)
    \tower(4,3)
    \tower[red, classes = {insert =2}](4,1)

	\tower[red](6,2)
    \tower[blue, classes = {insert =2}](6,0)

\end{sseqdata}

\printpage[name = MMMM,page = 2]
\end{center}

The above chart gives:
\begin{table}[h]
    \centering
    \begin{tabular}{c|cccccccc}
        $*$ & 0 & 1 & 2 & 3 & 4 & 5 & 6 & 7 \\ \hline
        $\Omega_{*}^{\text{Spin}}\big(BU(1)\big)$ & $\mathbb{Z}$ & $\mathbb{Z}_2$ & $\mathbb{Z}_2 \oplus \mathbb{Z}$ & 0 & $\mathbb{Z} \oplus \mathbb{Z}$ & 0 & $\bZ \oplus \bZ$ & 0 \\
    \end{tabular}
    \caption{Spin Bordism Groups of $BU(1)$}
    \label{tab:spin_bordism BU(1)}
\end{table}

Hence we have $\Omega_{6}^{\text{Spin}}\big(BU(1)\big) = \bZ \oplus \bZ$.

\paragraph{$\underline{\Omega_{6}^{\text{Spin}}(B\bZ_k) = 0} $}

{Note for any $k \in \bZ_{\geq 0}$, we have $k= 2^n l$ with $l$ odd.} $\text{Ext}^{1}_{\bZ}(Z_{2^n},\bZ_l) = 0$ gives $\bZ_k = \bZ_{2^n} \times \bZ_{l}$. 

\paragraph{n=0:} Applying the AH spectral sequence to the fibration
\begin{equation}
    pt \to B\bZ_k \to B\bZ_k,
\end{equation}
 gives:
 \begin{equation}\label{omegaBSU}
	\begin{array}{c}
	E^2_{p,q}=H_p\big(B\bZ_k;\Omega_q^{\text{Spin}}\big)\\
	\begin{array}{c|cccccccc}
		6 &0&0&0&0&0&0&0& 0\\
		5 &0&0&0&0&0&0&0&0\\
		4 & \bZ &  \bZ_k &0&\bZ_k& 0 &\bZ_k& 0 & \bZ_k\\
		3 &0&0&0&0&0&0&0&0 \\
		2 & \bZ_2 &0&0& 0 &  0 &0&  0 &0\\
		1 & \bZ_2 &0&0&0& 0 &0& 0&0\\
		0 & \bZ&  \bZ_k &0&\bZ_k& 0 &\bZ_k& 0 & \bZ_k\\
		\hline
		& 0 & 1 & 2 & 3 & 4 & 5 & 6 & 7
	\end{array}
	\end{array}
\end{equation}

The above chart gives $\Omega_{6}^{\text{Spin}}(B\bZ_k) = 0$ for $k$ odd.

\vspace{.5cm}

The Künneth formula and the AH spectral sequence give $\Omega_{6}^{\text{Spin}}(B\bZ_{k = 2^n l}) = \Omega_{6}^{\text{Spin}}(B\bZ_{2^n})$, and the second page of the AH sequence for $\Omega_{6}^{\text{Spin}}(B\bZ_{2^n})$ gives
 \begin{equation}\label{omegaBSU+}
	\begin{array}{c}
	E^2_{p,q}=H_p\big(B\bZ_k;\Omega_q^{\text{Spin}}\big)\\
	\begin{array}{c|cccccccc}
		6 &0&0&0&0&0&0&0& 0\\
		5 &0&0&0&0&0&0&0&0\\
		4 & \bZ &  \bZ_{2^n} &0&\bZ_{2^n}& 0 &\bZ_{2^n}& 0 & \bZ_{2^n}\\
		3 &0&0&0&0&0&0&0&0 \\
		2 & \bZ_2 &\bZ_2&\bZ_2& \bZ_2 &  \bZ_2 &\bZ_2&  \bZ_2 &\bZ_2\\
		1 & \bZ_2 &\bZ_2&\bZ_2&\bZ_2& \bZ_2 &\bZ_2& \bZ_2&\bZ_2\\
		0 & \bZ&  \bZ_{2^n} &0&\bZ_{2^n}& 0 &\bZ_{2^n}& 0 & \bZ_{2^n}\\
		\hline
		& 0 & 1 & 2 & 3 & 4 & 5 & 6 & 7
	\end{array}
	\end{array}
\end{equation}
\paragraph{n$>$1:} First we focus on $n>1$, as the $A_1$ module structures of $n=1$ and $n>1$ are different. In this case $d_2$ gives isomorphisms in the following maps:
 \begin{equation}
     \begin{split}
         d_2: E_{5,1}^2 \to E^2_{3,2},\\
         d_2: E_{5,0}^2 \to E^2_{3,1},\\
          d_2: E_{4,1}^2 \to E^2_{2,2}.
     \end{split}
 \end{equation}

 Together with the fact that $E_{*,0}^2$ survives in every page due to the fact that $\text{id}: pt \to pt$ can be factorized to $pt \to B\bZ_{2^n} \to pt$, we get the following results:
  
\begin{table}[h]
    \centering
    \begin{tabular}{c|cccccc}
        $*$ & 0 & 1 & 2 & 3 & 4 &5 \\ \hline
        $\Omega_{*}^{\text{Spin}}(B\bZ_{2^n})$ & $\mathbb{Z}$ & $\mathbb{Z}_2\oplus \bZ_{2^n}$ & $\mathbb{Z}_2 \oplus \mathbb{Z}_2$ & $\bZ_{2}^{\oplus2}\oplus\bZ_{2^n}(\text{or } \bZ_{2}\oplus\bZ_{2^{n+1}} \text{ or } \bZ_{2^{n+2}})$ & $\mathbb{Z} $ & $ G$\\
    \end{tabular}
    \caption{Spin Bordism Groups of B{$\mathbb{Z}_{2^n}$}}
    \label{tab:spin_bordism B2n(1)}
\end{table}
with $G$ a torsion group and  $|G|=2^{2n-2} \text{ or } 2^{2n-1} $. Now we use the Adams spectral sequence to compute $\Omega_{6}^{\text{Spin}}(B\bZ_{2^n})$. The $A_1$-module of $H^*(B\bZ_{2^n})$ has the following structure:
\begin{equation}
    \label{A1 module of BZ2n}
    \bZ_2 \oplus \Sigma^1 \bZ_2 \oplus \Sigma^2\cA_1/\cE_1\oplus \Sigma^3\cA_1/\cE_1\oplus \Sigma^6\cA_1/\cE_1 \dots \ \ .
\end{equation}
The Adams chart is:
 \begin{center}
\begin{sseqdata}[name=MMM,Adams grading,classes = fill,xrange = {0}{7},yrange = {0}{7}]
	\tower(0,0)
	\class(1,1)
    \class(2,2)
    \structline(0,0)(1,1)
    \structline(1,1)(2,2)
     
    \tower(4,3)
     \tower[green](1,0)
	\class[green](2,1)
    
    \class[green](3,2)
    \structline(1,0)(2,1)
    \structline(2,1)(3,2)
     
    \tower[green](5,3)

    \tower[red](2,0)
    \tower[red](4,1)
    \tower[red](6,2)

    \tower[blue](3,0)
    \tower[blue](5,1)

	\tower[purple](6,0)

\end{sseqdata}

\printpage[name = MMM,page = 2]
\end{center}
Together with the results from the AH spectral sequence, Table \ref{tab:spin_bordism B2n(1)}, we have:
 \begin{itemize}
     \item The green tower of the 1st column is killed by the red tower of the 2nd column in the $n$th page by $d_n$, to match the result $\Omega_{1}^{\text{Spin}}(B\bZ_{2^n}) = \bZ_2 \oplus \bZ_{2^n}$  and $\Omega_{2}^{\text{Spin}}(B\bZ_{2^n}) = \bZ_2 \oplus \bZ_2$ from the AH spectral sequence.
     \item The red tower from 4th column kills part of the blue tower in the 3rd column in the $n$th page by $d_n$ and gives $\Omega_{3}^{\text{Spin}}(B\bZ_{2^n}) = \bZ_2 \oplus \bZ_{2^{n+1}}$. Note this result is hard to get purely using the AH spectral sequence.
     \item The two towers (in red and purple) in the 6th column kill (part of) the two towers (green and blue) in the 5th column (at least one of them is killed in the $n$th page by $d_n$). It also gives refined information of $G$, i.e. $G = \bZ_{2^p}\oplus \bZ_{ 2^q}$ . 
 \end{itemize}
 The last point of the above discussion gives $\Omega_{6}^{\text{Spin}}(B\bZ_{2^n}) = 0$ for $n>1$.
 
\paragraph{n=1:}The fact that $\Omega_{6}^{\text{Spin}}(B\bZ_2) = 0$ can be found in \cite{Kapustin:2014dxa}. It can also be computed using similar techniques mentioned above.

Hence we reach the conclusion $\Omega_{6}^{\text{Spin}}(B\bZ_k) = 0$ for $k>1 \in \bZ$.

\section{Witten (global) anomaly and circle reduction}
\label{app:W}

In this appendix, we discuss the circle compactification of Witten's $SU(2)$ global anomaly \cite{Witten:1982fp}: 4D $SU(2)$ gauge theory with an odd number of fundamental Weyl fermions has a global anomaly characterized by the non-trivial element of $\text{Hom}_{\bZ}\big(\Omega_{5\text{D}}^{\text{Spin}}(BSU(2)),U(1)\big)= \bZ_2$.\footnote{The perturbative anomaly vanishes. Some recent discussions are in \cite{Garcia-Etxebarria:2017crf, Jia:2024kbl, Saito:2025idl, Okada:2025kie}.} The key differences compared with the 6D/5D examples discussed in the main text are:
\begin{itemize}
    \item The gravity is not dynamical, i.e. we are considering a quantum field theory. As a result, we do not have the KK $U(1)$ gauge group after circle compactification. The 4D theory is also free of gravitational anomalies.
    \item The 4D fermion compactified on a circle gives a full KK tower (labeled by $n \in \bZ$) of 3D Dirac Fermions, whereas in 6D the fermions are SMW, and the reality condition halves the full KK tower to $n \in \bZ_{n\geq 0}$.
\end{itemize}
Putting the 4D $SU(2)$ theory on $\bR^{2,1}\times S^1$, we can study the 3D footprint of the 4D non-perturbative anomaly.

\subsection{Global $SU(2)$ Witten anomaly and instanton numbers} Following the adiabatic techniques in \cite{Witten:1982fp}, it is clear that the $\bZ_2$ global anomaly is related to the odd instanton number sector:
\begin{equation}
    \label{odd SU2 instanton number sector}
   \int_{S^4} \frac{1}{2}\tr_f( \mathcal{F}^2) = 1 \ \text{mod} \ 2.
\end{equation}
Generalizing to Weyl fermions in the spin-$\frac{m}{2}$ representation of $SU(2)$, the contribution to the global $SU(2)$ anomaly as
\begin{equation}
    \label{spin m/2 weyl}
     \int_{S^4} \frac{1}{2}\tr_{m/2}(\mathcal{F}^2) = \frac{m(m+1)(m+2)}{6} \int_{S^4} \frac{1}{2}\tr_f(\mathcal{F}^2) \, ,
\end{equation}
i.e.
 \begin{equation}
    \label{spin m/2 weyl 2nd}
     m=4k+1 \ \  \ \Rightarrow \ \ \  \ \frac{m(m+1)(m+2)}{6} \neq 0 \ \text{mod} \ 2 \,.
\end{equation}
Only Weyl fermions in spin-$(2k+\frac{1}{2})$ representations contribute to global $SU(2)$ anomalies. Hence, a more general statement regarding the global $SU(2)$ anomaly is that 4D $SU(2)$ gauge theory with an odd number of Weyl fermions in spin-$(2k+\frac{1}{2})$ representations is anomalous \cite{Witten:1982fp}.

\subsection{Global $SU(2)$ anomaly and circle compactification}
 To analyze circle compactification,\footnote{We only consider the case without Wilson lines here.} we first recall the general principle of fermionic anomalies \cite{Witten:2015aba}: a gappable fermion does not contribute to anomalies. Here gappable refers to a fermion for which a mass term compatible with all symmetries under consideration can be written down. In this case the theory can be regularized using Pauli-Villars regularization without spoiling the relevant symmetries.

 Following the general principle, the KK tower except for $n=0$ will not be a cause of any inconsistencies. {The contribution from the $n$th KK mode is canceled by that of the $(-n)$th KK mode}: 
 \begin{equation*}
     \sum_{n<0}\frac{\text{sgn}(n)+ \text{sgn}(-n)}{2} \text{CS}(A_{SU(2)}) = 0.
 \end{equation*}
 We therefore only need to analyze the 3D gauge theory with massless 3D Dirac spinors in the fundamental representation on $\bR^{1,2}$ parametrized by $(x_0,x_1,x_2)$. Now note that 3D Dirac spinors admit a mass term compatible with $SU(2)$ gauge symmetry. The theory is consistent! However, there is one subtlety:
 \begin{itemize}
     \item Such a mass term breaks 3D parity symmetry, e.g. $x_2 \to -x_2$, explicitly.
     \item The 3D parity symmetry comes from rotation symmetry in 4D: $(x_2,x_3) \to (-x_2,-x_3)$, where $x_3$ labels the circle. This is the remaining symmetry of $SO(2)$, the rotation group of the $(x_2,x_3)$ {plane, after making the $x_3$ direction a circle}.
 \end{itemize}
 If we choose to preserve the parity symmetry after circle compactification, which is natural from the 4D perspective, then a 3D mass term cannot be turned on and a potential anomaly can exist. 
 
 The following analyses are not new and can be found in \cite{Redlich:1983kn}:
 to see the potential anomaly, following \cite{Witten:1982fp}, consider the following mapping tori:
\begin{equation}
    \label{3d mapping tori}
    S^3 \times [0,1]_t/\sim, A_t = (1-t)A + t A^g,
\end{equation}
where $A$ is the $SU(2)$ connection on $S^3$ and $g$ is the large gauge transformation with $[g] = n \in \pi_3(SU(2)) = \bZ$. The $SU(2)$ bundle over $ S^3 \times [0,1]_t/\sim$ has instanton number $n$. Equation \eqref{3d mapping tori} gives the change of the fermionic measure of a 3D Dirac fermion in the fundamental representation (more generally in the spin-$(2k+\frac{1}{2})$ representation) under the large gauge transformation $g$:
\begin{equation}
\label{3d large gauge transformation}
    \prod \bar{\psi}\psi \to (-)^n \prod \bar{\psi}\psi.
\end{equation}
This is the same as a 3D $SU(2)$ CS term with level $\frac{1}{2}$:
\begin{equation}
\label{1/2 CS}
    \frac{1}{2}\text{CS}(A_{SU(2)}) \ \to \ (-)^n  \frac{1}{2} \text{CS}(A_{SU(2)}).
\end{equation}
 
On the other hand, we can add a 3D mass term to preserve gauge invariance (\eqref{1/2 CS} will be introduced after using Pauli-Villars regularization \cite{Redlich:1983kn}), but this would break parity symmetry. 

This analysis leads to:
\begin{itemize}
    \item The global $SU(2)$ anomaly does not go away after circle compactification; it enters as a mixed anomaly between parity symmetry and 
    large gauge transformations.
    \item If the 4D theory is free of the global $SU(2)$ anomaly, the 3D theory after circle compactification can preserve both parity symmetry and gauge invariance under LGTs.
\end{itemize}

 Notice that well-quantized 3D WZW terms are given by $\text{Hom}_{\bZ}\big(\Omega^{\text{Spin}}_{4\text{D}}(BSU(2));\bZ\big)$, which require the $SU(2)$ CS level to be an integer. So the potential anomaly can be characterized by ill-defined 3D WZW terms, as in the 6D/5D cases.

\subsection{A puzzle for $SU(N)$ gauge theory with $N>2$}
In this subsection we want to discuss a puzzle for $SU(N>2)$ gauge group in the 4D/3D context:
 \begin{itemize}
     \item The 3D analysis given in \cite{Redlich:1983kn} also works for $SU(N>2)$, and the anomaly of large gauge transformations \eqref{3d large gauge transformation} also exists for {$SU(N>2)$} if a parity symmetry preserving regularization scheme is chosen. This anomaly is equivalent to an ill-quantized 3D $SU(N>2)$ CS term.
     \item  There are no $SU(N>2)$ global anomalies in 4D, since $\Omega_{5\text{D}}^{\text{Spin}}\big(BSU(N>2)\big) = 0$.
 \end{itemize}
 It seems that a well-defined 4D $SU(N>2)$ theory after circle compactification may give a 3D theory with an anomaly equivalent to an ill-quantized 3D $SU(N>2)$ CS term when choosing a parity-preserving regularization scheme. 

\vspace{.5cm}
 
To solve this puzzle, notice the following facts:
 \begin{itemize}
     \item  The 3D anomaly $(-)^n$ is again associated with instanton numbers on the mapping torus $X$:
     \begin{equation}
          \int_{X} c_2(SU(N)) = n,
     \end{equation}
     where here the $c_2(SU(N))$ depends on the spectrum of fermions. If the spectrum is such that
     \begin{equation}
          \int_{X} c_2(SU(N)) = 0 \ \text{mod} \ 2,
     \end{equation}
     then there is no anomaly.
     \item On the other hand, if we start with an anomaly-free 4D $SU(N>2)$  gauge theory, i.e. with $c_3(SU(N>2)) = 0$, due to the following fact:\footnote{This fact is the reason why $\Omega_{5\text{D}}^{\text{Spin}}\big(BSU(N>2)\big) = 0$ while $\Omega_{5\text{D}}^{\text{Spin}}\big(BSU(2)\big) = \bZ_2$}
     \begin{equation}
     \label{K theory relation}
         c_3(SU(N>2)) = Sq^2 c_2(SU(N>2)) \mod 2,
     \end{equation}
     the freedom from perturbative anomalies gives
     \begin{equation}
      Sq^2 c_2\big(SU(N>2)\big)=0 \ \text{mod} \ 2 \ \ \ \Rightarrow \ \ \ c_2 \big(SU(N>2)\big)=0 \ \text{mod} \ 2.
     \end{equation}
 \end{itemize}
 So for a well defined 4D $SU(N>2)$ gauge theory, the 3D theory after circle compactification will be automatically free of large gauge transformation anomalies using a parity preserving regularization scheme.\footnote{The fact that $c_3 =0 \ \Rightarrow \ c_2 \text{ is even }$ for $SU(N>2)$ also appeared in \cite{Jia:2024kbl} by directly comparing the Dynkin index.}
  

\bibliographystyle{JHEP}
\bibliography{biblio}

\end{document}